\documentclass{birkmult}
\usepackage{amsmath}
\usepackage{amsfonts}
\usepackage{latexsym}
 \usepackage{amssymb}
 \usepackage{graphicx}
          \newtheorem{lemma}{Lemma}[section]
           \newtheorem{theorem}{Theorem}[section]
           
           \newtheorem{definition}{Definition}[section]

\title[A solvable model for scattering on a
junction \ldots ]{A solvable model for scattering on a junction
and a modified analytic perturbation procedure.}
\author{B. Pavlov}
\dedicatory{The  paper  is dedicated  to   Mihail  Samoilovich
Livshits, who  was  first to  consider \\ a   nonselfadjoit
operator as a  part  of   an  extended  sefadjoint   scattering
system. }
\begin{document}
\address{
V.A. Fock Institute for Physics of  St.-Petersburg
University,\\
 Petrodvorets, 198905, Russia.}%
 \email{pavlovenator@gmail.com}

\begin{abstract}
We consider  a one-body spin-less electron spectral  problem  for
a resonance scattering  system constructed of a quantum well
weakly connected to a noncompact exterior reservoir, where the
electron is free. The simplest kind of the resonance scattering
system is a quantum network, with the reservoir composed of  few
disjoint cylindrical  quantum wires, and the Schr\"{o}dinger
equation on the network, with the real bounded potential on the
wells and  constant potential  on the wires. We propose a
Dirichlet-to-Neumann - based analysis to reveal the resonance
nature of  conductance across the star-shaped element of the
network (a junction), derive an approximate formula for the
scattering matrix of the junction, construct  a fitted zero-range
solvable model of the junction and interpret  a phenomenological
parameter arising in Datta-Das Sarma boundary condition, see
\cite{DattaAPL}, for T-junctions. We also propose using of the
fitted zero-range solvable model as the first step in a modified
analytic perturbation procedure of calculation of the
corresponding scattering matrix.
\end{abstract}
\maketitle

\vskip0.3cm \begin{center}{\bf Outline}\end{center}
\begin{enumerate}
\item Introduction.
\item Scattering in Quantum Networks and Junctions via  DN-map.
\item Krein formulae for the intermediate DN-map and ND-map,
with the compensated singularities.
\item Approximate Scattering matrix and  the  boundary
condition at the  vertex of the quantum graph.
\item A solvable model of a thin junction.
\item Fitting of the solvable model.
\item A solvable model as a jump-start in the analytic perturbation
procedure.
\item Acknowledgement.
\item  Appendix: symplectic operator extension procedure.
\end{enumerate}
\vskip0.3cm

\section{Introduction}
A typical quantum  resonance scattering system  is composed of a
compact inner region surrounded  by barriers and  an exterior
reservoir, where  the quantum  dynamics is free. These components
are weakly connected due to  tunneling across the barriers or via a
narrow  connecting channels. Non-compact quantum networks (QN) are
typical resonance scattering systems. Manufacturing of QN with
prescribed transport properties is now a most challenging problem of
computational nano-electronics. While physical laws defining
transport properties of the QN are mostly represented in form of
partial differential equations, the direct computing can't help
optimization of design of the QN, because it requires expensive and
resource  consuming scanning over the space of physical and
geometrical parameters  of the network. The domain of scanning could
be essentially reduced in the case when there exist an approximate
explicit formula connecting directly the transport characteristics
with the parameters defining the geometry and the physical
properties of the network.

We derive an explicit approximate formula for the scattering matrix
of a simplest QN- a junction,- consisting of a vertex domain - a
quantum well - connected  to the outer reservoir {\it decomposed
geometrically into a sum of cylindrical leads}. The corresponding
model Hamiltonian is obtained based on Glazman splitting,
\cite{Glazman}, ${\mathcal{L}}\to L_{\Lambda}\oplus l_{\Lambda}$ of
the original Hamiltonian, depending on the Fermi level $\Lambda$,
into the sum of two operators with complementary branches of the
continuous spectra. The model proves to be fitted because the
corresponding model Dirichlet-to-Neumann map ( DN-map, see
\cite{SU2} ) serves a rational approximation of the DN - map of the
non-trivial component $ L_{\Lambda}$ of the split system.

\quad  In  an important alternative  class  of  the  resonance
scattering systems, represented by the Helmholtz resonator, the
reservoir can't be decomposed into simple components  similar to the
cylindrical leads, but  the finite leads connecting the compact
subsystem - the resonator - with the reservoir, admit a similar
decomposition. Then again, a fitted solvable model can be
constructed, see \cite{Kirchhoff_constants_08}, based on the
splitting of the spectral channels in the leads. The model obtained
can serve again as a first step - a jump start - of the
corresponding analytic perturbation procedure. We postpone the
discussion of  the  Helmholtz  resonator and other systems with
nontrivial reservoirs to oncoming publications.

Main difficulty  of  analysis  of resonance scattering systems of
both above kinds on the  networks  is defined  by  presence of the
eigenvalues of the isolated compact subsystem, embedded  into the
continuous spectrum of the reservoir, separated from the compact
subsystem. Indeed, for a selfadjoint operator $A_0$ in the Hilbert
space $E$, {\it with discrete spectrum}, and small $\varepsilon V$
,the self-adjoint perturbation, $\parallel \varepsilon V\parallel
\leq \varepsilon$, defines, for each simple isolated eigenvalue
$\lambda_s^0$ of $A_0$
     \[
2 \varepsilon < \mbox{min}_{t\neq s} |\lambda_s - \lambda_t|\equiv
\rho_s
  \]
a  branch of eigenvalues $\lambda_s^{\varepsilon}$ of the perturbed
operator $A_{\varepsilon}:= A_0 + \varepsilon V$  represented in
form of a geometrically convergent series
  \[
\lambda_s^{\varepsilon} = \lambda_s^{0} + \varepsilon \,
\lambda_s^{0}(1) + \varepsilon^2 \, \lambda_s^{0}(2)+ \varepsilon^3
\, \lambda_s^{0}(3) + \dots,
\]
and  the  corresponding  branch of eigenfunctions, see \cite{Kato}.

{\it This standard analytic perturbation approach  is not
applicable, generally, to operators with  eigenvalues embedded  into
the continuous spectrum}, in  particular to non-compact QN where the
``spacing'' $\rho$ is  zero. Development of  radio-location during
WWII required  analysis  of scattering problems  on the networks of
electromagnetic wave guides, in particular on junctions. The
scattering on the junction is a  typical perturbation problem  for
embedded  eigenvalues. The perturbation of the problem causes the
transformation of  real eigenvalues on the vertex domain of the
junction into complex resonances. This problem can't be solved  by
methods of the standard spectral  theory of selfadjoint operators.
In the paper \cite{Livshits62} M.S. Livshits proposed an elegant
approach to the problem of transmission of electro-magnetic signals
across the junction, taking into account only  oscillatory
electro-magnetic modes in the wave-guides and neglecting the
``evanescent''- exponentially decreasing  modes. He reduced the
calculation of the scattering matrix  to  calculation of the
corresponding characteristic function and  found a real wave
conductance for the oscillatory  modes and  pure imaginary  wave
conductance for evanescent modes. The discovery, based on
M.S.Livshits ideas,  of the connection between the scattering matrix
and the characteristic function of the corresponding
non-self-adjoint operator, see \cite{{AA_1965}}, was  an
extraordinary achievement and became eventually a source/basement of
a series of important results  in the theory  of the functional
models of the dissipative operators, see
\cite{Lax,Sz_N_F_1970,NK_87}. The approach to  the perturbation
theory  developed  in  these  papers permitted to understand the
spectral  nature of the resonances, but it does not help  practical
problem  of  optimization of design of quantum or electro-magnetic
networks with prescribed  transport properties. One more detail in
the above  paper \cite{Livshits62} was  important in this respect.
M.S. Livshits completely disregarded the evanescent waves in the
wave-guides, which  was usual in  the papers  of engineers and
physicists  on  the electro-magnetic wave-guides. But he  emphasized
in \cite{Livshits62} importance of accurate elimination of the
evanescent waves. This was not done till now. We  see  now, that the
absence of  analysis of the evanescent waves prevented M.S. Livshits
from establishing, at that time,  an effective connection between
the geometry of the junction and the transmisson/reflection
coefficients, see also our comments below, section 3.

In the case of  operators with continuous or dense discrete spectrum
one can substitute  the Hamiltonian $A_0$ of an unperturbed system
by a fitted solvable model $A^{\varepsilon}$, and then develop  an
analytic perturbation procedure  between  the perturbed Hamiltonian
$A_{\varepsilon}$ and the model   $A^{\varepsilon}$. This two-steps
idea $A_0 \to A^{\varepsilon}\to A_{\varepsilon}$ of the modified
analytic perturbation procedure was suggested, in implicit form, by
H. Poincare for relevant problems of celestial mechanics, see
\cite{Poincare}, and formulated in an explicit  form  in 1972 by I
Prigogine. In 1972  I. Prigogine, \cite{Prigogine73}, declared
importance of the  search of a general practical algorithm  for  the
two-step analytic perturbation procedure
\[
A_0 \longrightarrow A^{\varepsilon}\longrightarrow A^{\varepsilon}
\]
implementing the  above Poincare idea. Prigogine attempted to find
an  Intermediate operator $ A^{\varepsilon}$  as a  function of the
unperturbed  operator $ A^{\varepsilon} = \Phi (A_0 )$, and  he
wanted to have  the  above two step  analytic  perturbation
procedure on the whole Hilbert space. The search of the
corresponding ``intermediate operator''$A^{\varepsilon}$ continued
for almost 20 years, but did not give any results. Finally
 Prigogine  declared that  the intermediate operator with the
 expected   properties does not exist and  can't be  constructed.

We guess now, that I.Prigogine's suggestion  based on the
Intermediate Operator $A^{\varepsilon}$ was very close to success.
The idea of Prigogine was commonly used  by physicists in form of
effective Hamiltonian of a complex quantum systems, and, after
essential modification, in \cite{Marsden2003} for ``geometrical
intergration'' in dynamical problems of classical mechanics.

In our recent papers \cite{BMPPY,MathNach07,HPY07}, see also  an
extended  list of references below, we suggested a method of
accurate elimination of the evanescent waves based  on the idea of
the intermediate Hamiltonian. Our method also permits to accurately
eliminate  the evanescent waves  in  the case  studied by M.S.
Livshits. In this paper we provide, following the quoted papers, a
review of  the corresponding  modified  approach to  the analytic
perturbation procedure and describe, based on \cite{Pavlov_C07}, an
algorithm of construction of the solvable model and  the  procedure
of fitting. We  developed the corresponding  general approach to the
spectral problems with embedded eigenvalues  in the series of papers
\cite{BMPPY,Hadr05,NZMJ05,HPY07,PY07,MathNach07,Pavlov_C07,PP08}.
Contrary to the original Prigogine's idea, we  do  a couple of
changes:

1. We  search for the  Intermediate  Operator - the ``jump start'',
see \cite{PA05}- $A^{\varepsilon}$, on  the first step of the above
mentioned  two-step procedure,  not among functions $\Phi(A_0)$ of
$A_0$, as I. Prigogine suggested,  but among weak
(finite-dimensional) perturbations of the non-perturbed Hamiltonian
$A_0$, which is close to the method suggested by  Livshits.

2. We  restricted  our  analysis  to the part of the unperturbed
operator  on a spectral subspace which corresponds to  some
``essential spectral interval'', contrary to I. Prigogine who
attempted  to find a global  Intermediate Operator  on the whole
space. Thus we develop our modified  analytic perturbation procedure
{\it locally}. A similar  requirement of locality is applied in
\cite{Marsden2003} on the space  of  initial data of the
structure-preserving model.

We  begin with  two classical examples of the  resonance  scattering
systems, to reveal typical difficulties arising from the very
beginning when considering perturbations of systems with eigenvalues
embedded into the continuous spectrum, and discuss nearest prospects
of the perturbative analysis of these systems. \vskip0.3cm

{\bf Example 1: Helmholtz Resonator.}

 Helmholtz resonator was probably the first  resonance scattering
system  discussed  mathematically, see \cite{Rayleigh}. It is
composed of  the typical details: the inner domain $\Omega_{int}$,
the shell $\Omega_{shell}$, and the reservoir $\Omega_{out}$
separated  by the shall  from $\Omega_{int}$. Consider  the
Helmholtz equation
  $-\Delta u = \lambda u$ in a  domain  $\Omega\in R_3$
 represented as a sum of two disjoint parts $\Omega = \Omega_{int}
 \cup \Omega_{out}$ and  a  shell  $\Omega_{shell}$ with a small opening.
 Kirchhoff suggested  to  substitute the problem by the model where
 the opening is pointwise, so that there exist only one common point
 $a \in \bar{\Omega}_{int}\cup \bar{\Omega}_{out}\cup\bar{\Omega}_{shell}$.
Kirchhoff  suggested  an Ansatz for  the  Green-function
$G_{\lambda} (x,y)$ of the Helmholtz  equation
 in  $\Omega$ with Neumann  boundary condition
\[
- \Delta G_{\lambda}(x,y) - \lambda G_{\lambda}(x,y) = \delta
(x-y),\,\, \frac{\partial G}{\partial n_x} \bigg|_{\partial\Omega} =
0,\, x,y \in {\bar{\Omega}}.
\]
in the form of  a linear  combination of the Green functions
 $G^{int}_{\lambda} (x,y),\,G^{out}_{\lambda} (x,y)$ of the  inner
 and the outer  problems:
\[
- \Delta G^{in, out}_{\lambda} (x,y) - \lambda G^{in,
out}_{\lambda}(x,y) = \delta (x-y),\,\, \frac{\partial G}{\partial
n_x} \bigg|_{\Omega_{in, out}} = 0.
\]

\begin{figure} [ht]
\begin{center}
\includegraphics [width=4in] {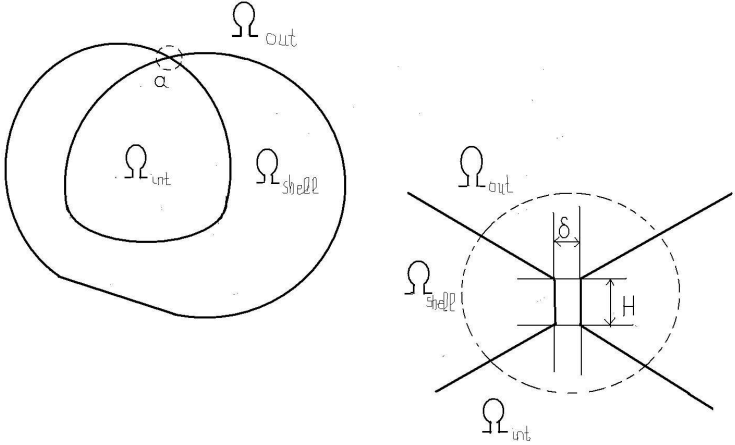}
\end{center}
\caption{Hemlholtz Resonator with a point-wise opening at the
point a and the enlarged  detail of the resonator  with a narrow
short channel, $\delta << H << \lambda^{-1/2}$} \label{F:figure 0}
\end{figure}

\[
G_{\lambda}(x,y)  - \left\{
\begin{array}{ccc}
G^{ out}_{\lambda} (x,y) & + & A^{out}\,G^{out}_{\lambda} (x,a)
,\,\, x,y \in \Omega_{out}\\
 A^{int}\,\,G^{in}_{\lambda} (x,a), & \mbox{if} &  y \in
 \Omega_{out},\,\,  x \in \Omega_{int},
\end{array},
\right.
\]
with  undefined  constants -the  Kirchhoff coefficients $ A^{out},\,
A^{int}$. This  {\it Kirchhoff Ansatz} satisfies the equation and
the Neumann boundary conditions everywhere on  $\partial \Omega$,
except the point $a$, where  the Ansatz is  singular. The problem of
choice of the Kirchhoff constant and other interesting problems
concerning the resonator, see \cite{Rayleigh} remained open until
recent time, see in this connection the  preprint \cite{BP07}.

\vskip0.3cm {\bf Example 2: Zero-range potential}

In  1933  E. Fermi, \cite{Fermi}, considered  the problem of
scattering of neutrons $n$ by the nucleon $S$ of Sulphur. He
suggested  to choose for this problem the model Hamiltonian in the
form of Laplacian in $L_2 (R_3)$ defined on smooth functions $u\in
L_2 (R_3)$ with a singularity at the  origin
 \[
u(x ) = \frac{A^u}{4\pi |x|} + B^u + \dots,
 \]
and a  special boundary condition imposed  on the asymptotic
boundary values  $A^u,B^u$:
\[
A^u = \gamma B^u,\,\, \gamma = \bar{\gamma}.
\]
The  Laplacian is  symmetric and even self-adjoint with this
boundary condition, and  admits explicit calculation of the
eigenfunctions: this model is  ``solvable''. Fermi  suggested to
``fit'' this model choosing  $\gamma = - 4\pi p_0^{-1}$, if $-
p_0^2$ is a small negative eigenvalue  in the  system  $n,S$.

The model  can be  extended to the case when $\gamma >0$, and fit to
the  purely imaginary resonance $ p_0 = i \gamma$. The resulting
mysterious ``zero-range potential'' suggested  by Fermi was
interpreted  by  F. Berezin and L. Faddeev  \cite{BF61} in terms of
von Neumann Operator Extensions Theory, \cite{Neumann}. Later this
``zero-range potential'' was used in numerous physical and
mathematical papers and books, see for instance \cite{AK00}.

In both above examples the reservoirs are either  a large  exterior
domain, or  the whole space with single  point $ x=0$ removed. The
first  example was also treated by the operator extension methods in
\cite{Opening}, producing a {\it zero-range solvable model} of the
resonator immersed into  3D  space.  The role of the unperturbed
operator in \cite{Opening}  played an orthogonal sum of the Neumann
Laplacian $L_{int}$ in $L_2 (\Omega_{int})$ and $L_{out}$ in $L_2
(\Omega_{out})$. The basic difficulty of  the original perturbation
problem, with a thin short channel, is caused by presence  of the
eigenvalues of $L_{int}$ embedded into the continuous spectrum of
$L_{out}$. The standard self-adjoint spectral theory is {\it
generally} unable to treat the problem of embedded eigenvalues, by
observing  transformation of them into the corresponding complex
resonances. The elegant Lax-Phillips approach to resonance
scattering problems reveals the spectral nature or resonances, see
\cite{Lax}, but does not help to calculate them. In \cite{Opening}
the resonances  can be easily calculated via solving an algebraic
equation, but yet the {\it fitting} of the suggested zero-range
model remained a problem. In \cite{BP07}  an approach to the problem
of fitting of the model is  suggested based on an explicit formula
connecting the ``full'' scattering matrix of the Helmholtz Resonator
with the Neumann-to-Dirichlet map, see \cite{DN01}, and a subsequent
rational approximation of the Neumann-to-Dirichlet map for the inner
domain  of the resonator (the cavity $\Omega_{int}$). Fortunately
the problem of search of the resonances, in the case of small
opening, becomes finite-dimensional after replacement of the
Neumann-to-Dirichlet map of the cavity by the corresponding rational
approximation, see more details in \cite{BP07}.

In the  second  example just a  self-adjoint operator
$-\Delta_{\gamma}$  is suggested, with only parameter  $\gamma$,
which can be  interpreted in spectral terms. This operator plays a
role of an effective Hamiltonian of the original  scattering problem
for the neutrons and the  nuclei, see \cite{Fermi}. Yet  again, the
substitution of the original perturbed ({\it full}) Hamiltonian  by
the effective solvable Hamiltonian $-\Delta_{\gamma}$  requires
fitting of the model, at least on some essential interval of energy.

Note that the role of the effective Hamiltonian is played, in the
second example, by  a  self-adjoint extension of the unperturbed
Hamiltonian $-\Delta$. Numerous effective Hamiltonians in quantum
mechanics are  constructed as  zero-range solvable models of quantum
systems  see for instance \cite{Extensions} and our recent papers
quoted above.

In  this review we represent some results of our recent papers
quoted above  (see the  text preceding the  Example 1)  where the
effective Hamiltonians are constructed as zero-range solvable
models. To make the text easily readable, we omit  some proofs and
most of technical details, which can be found in the original
publications. But  we pay additional attention to the
interconnections of our constructions previously spread in different
publications.

\section{ Scattering on Quantum Networks\\ and Junctions via  DN-map.}

The basic idea  of  analysis of partial differential equations on
quantum networks is that the corresponding Schr\"{o}dinger problem
can be divided in two parts: a Schr\"{o}dinger equation on  the
region surrounded  by  barriers ( e.g. a quantum well ) and one on
the reservoir the two being weakly coupled by tunneling, see for
instance \cite{PS96} or by a thin channel. It is noticed in
\cite{PS96} that this decomposition ``corresponds to the
schematization of the transport process as a coherent process on the
quantum well, fed by the exterior reservoirs - quantum wires''. On
the reservoirs, assumed to be homogeneous and neutral, the
electron–electron interaction is neglected, and the single electron
is  free.  But the resonance properties  of  the quantum well and
the tunneling on the contacts define the transport properties of the
whole network. It is  a  common belief that thin quantum network can
be modeled by a  1d graph, see \cite{Novikov_99}, with either
Kirchhoff boundary conditions, or just non-specified selfadjoint
boundary conditions at the vertices. There exist an extended
bibliography concerning one-dimensional models of quantum networks,
see  for instance \cite{Datt,Kuch01,Kuch02,Har2_00,StrSeba03}.
Notice that even sharp resonance  effects  on 2d  wave-guides and
networks were studied theoretically mainly by numerical methods, see
for instance \cite{Gramotnev_2000, ScattXu02}.

\begin{figure}
[ht]
\begin{center}
\includegraphics [width=3in] {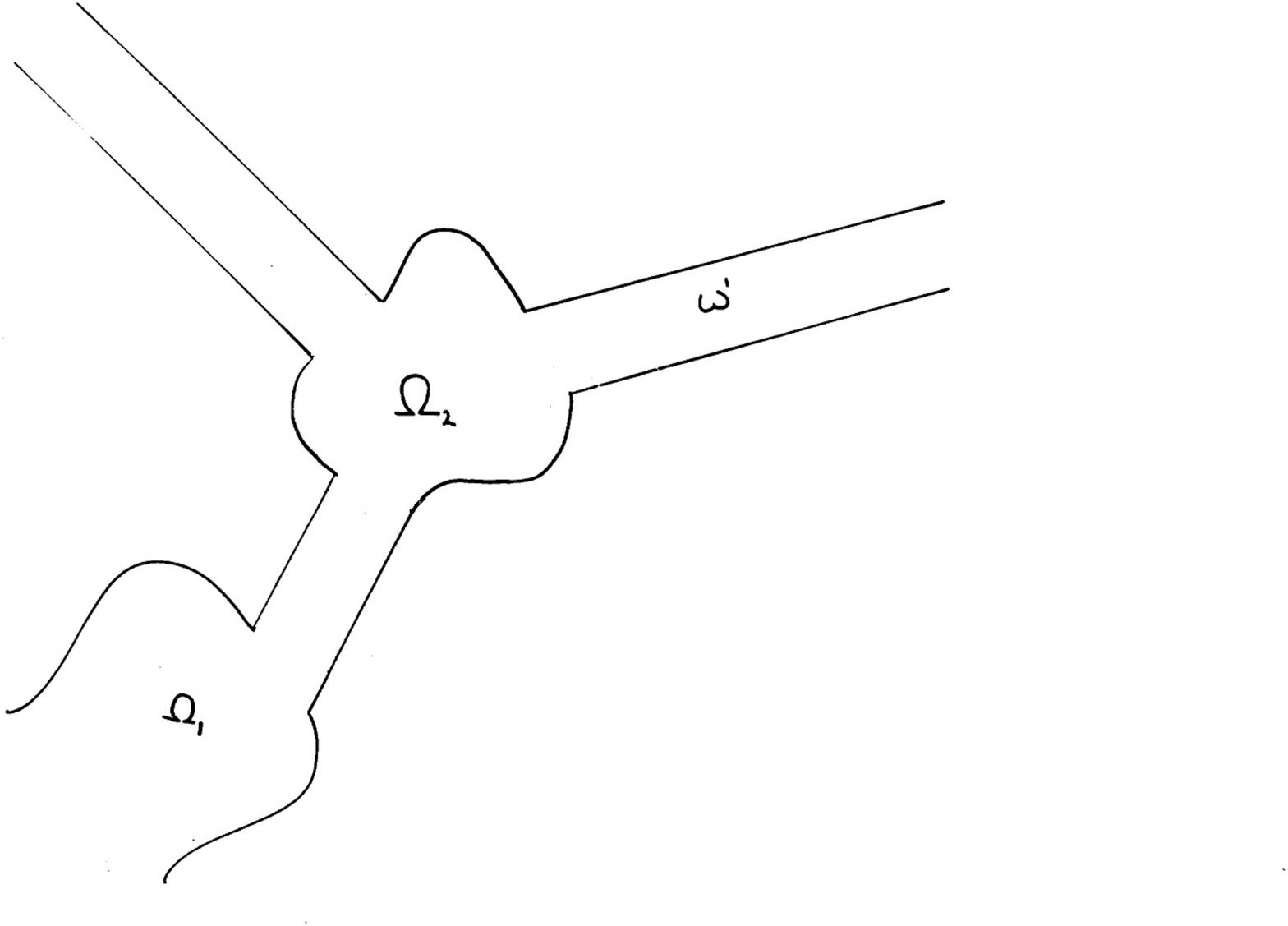}
\end{center}
\caption{Quantum Network: a detail} \label{F:figure 2}
\end{figure}

\quad Quantum network  $\Omega$  which  is being studied in this
paper, is composed of straight  leads (quantum wires)  width
$\delta$, some of them semi-infinite, and vertex domains $\Omega_s$
( quantum wells), see Fig.\ref{F:figure 2}.  An important basic
detail of the quantum network is a junction, see Fig. \ref{F:figure
3}.
The junction is a non-compact quantum  network composed of  a
quantum well and few semi-infinite  quantum wires, of constant
width, attached to it.
\begin{figure}
[ht]
\begin{center}
\includegraphics [width=4.5in] {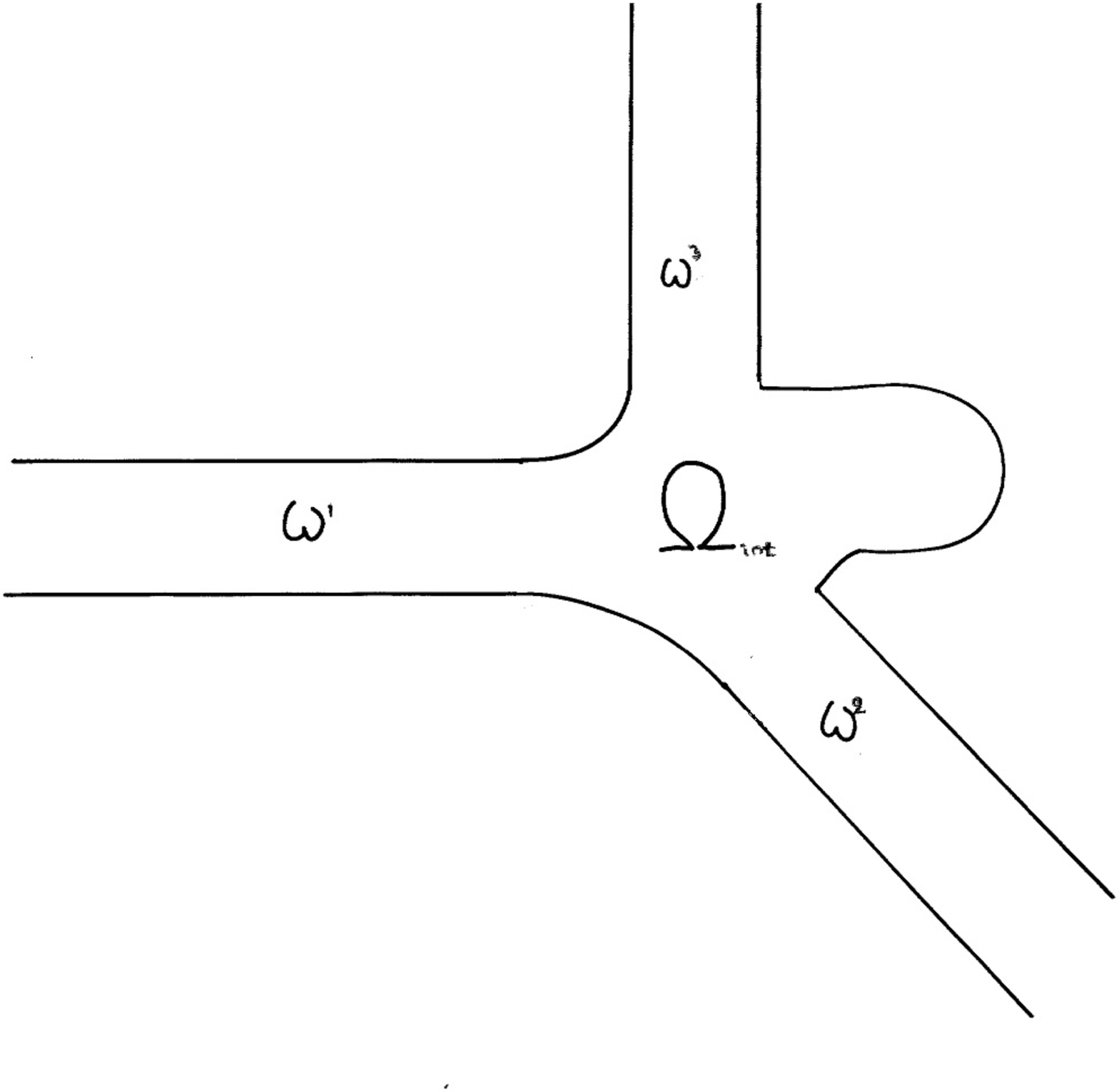}
\end{center}
\caption{General junction} \label{F:figure 3}
\end{figure}
 The junction is usually  called  thin, if the diameter
of the quantum well $\Omega_{int}$ strongly dominates the width
$\delta$ of the wires $\omega$ attached  to it:  $\delta << $  diam
 $\Omega_{int}$.
Calculation of the scattering matrix  of a junction is a challenging
computational problem, yet  accessible for standard  commercial
programs, see  the discussion below. Physicists have  certain
preferences about  the boundary conditions at the vertices, see the
discussion below, Example 3.

{\bf Example 3: Thin symmetric T-junction. } For thin  symmetric
T-junction, with the ``bar'' orthogonal to the ``leg'',
\begin{figure}
[ht]
\begin{center}
\includegraphics [width=3in] {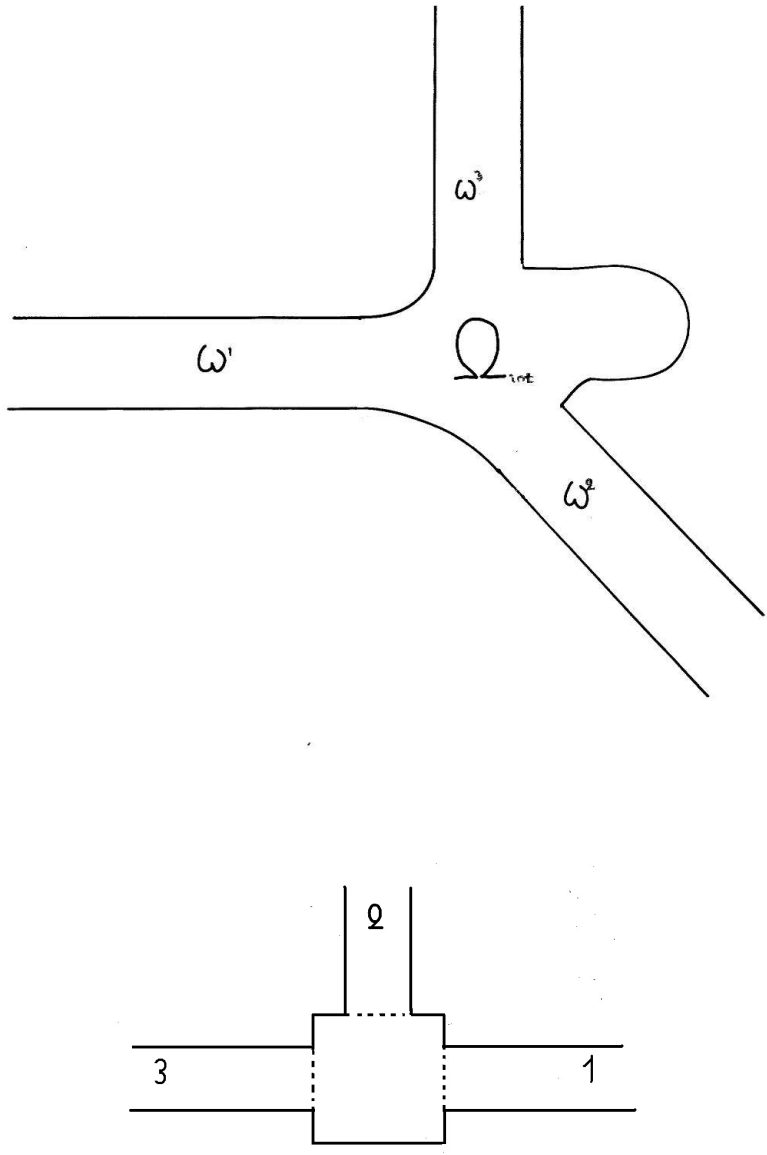}
\end{center}
\caption{ Simplest symmetric T-junction with a square vertex domain}
\label{F:figure 4}
\end{figure}
a  reasonably simple explicit formula for the  scattering matrix was
suggested in \cite{DattaAPL} based on reduction of the 2D scattering
problem on the junction   to the  corresponding  1D  scattering
problem on the corresponding quantum graph, see Fig. \ref{F:figure
5}
\begin{figure}
[ht]
\begin{center}
\includegraphics [width=3in] {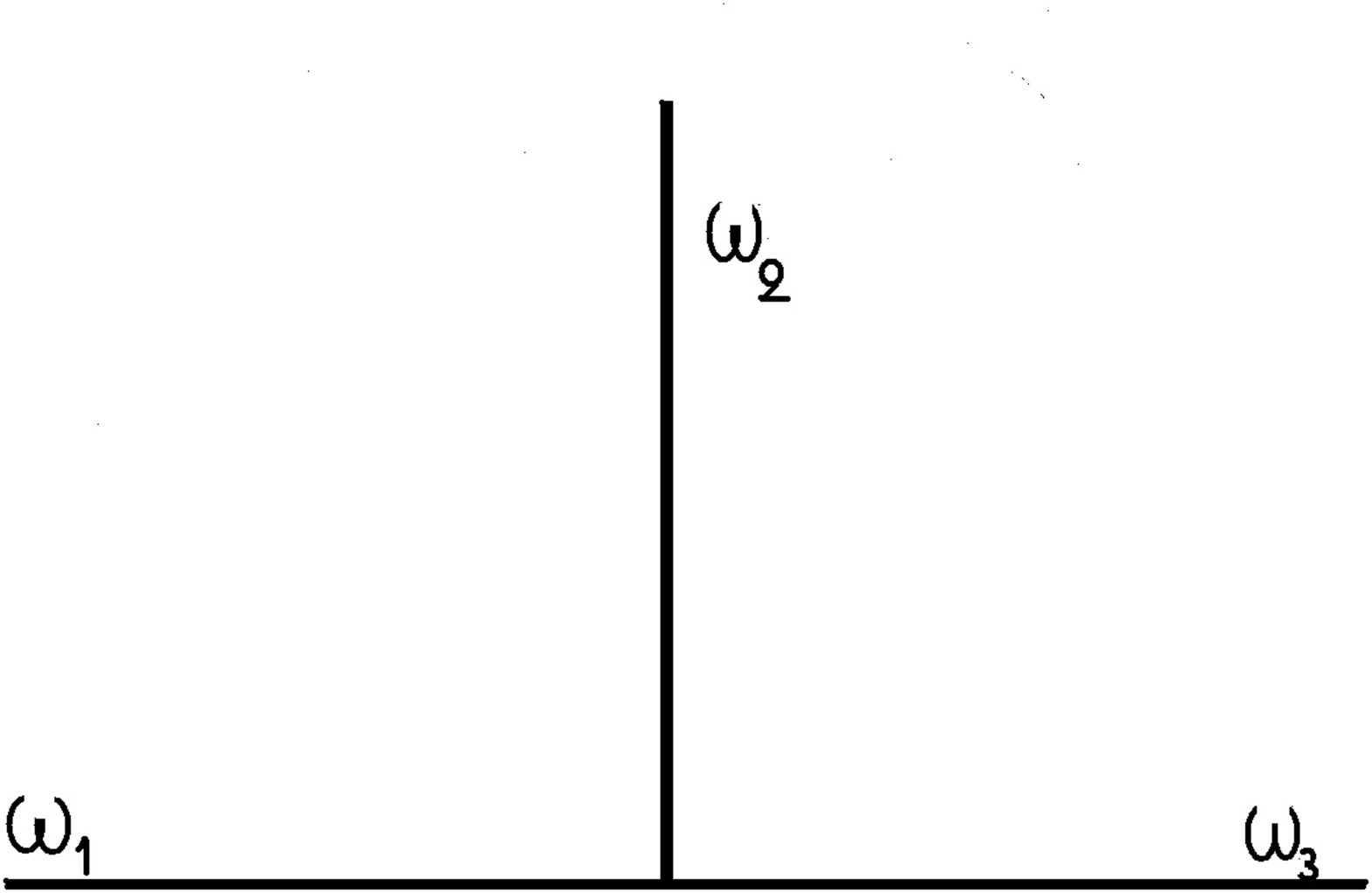}
\end{center}
\caption{ Model symmetric T-junction} \label{F:figure 5}
\end{figure}
The boundary conditions for the model T-junction suggested in
\cite{DattaAPL}, is presented in terms of limit values of the
wave-function on the 1D wires $\left\{ \psi_i \right\}^{3}_{i=1}:=
\vec{\psi}$ and the values of the corresponding outward derivative (
boundary currents) $\left\{ \psi^{\prime}_i \right\}^{3}_{i=1}:=
\vec{\psi'}$ at the node:
\begin{equation}\label{bndcnd}
 \psi_1 = \beta^{-1} \psi_2 = \psi_3 \, , \quad
\psi^{\prime}_{1} + \beta \psi^{\prime}_{2} + \psi^{\prime}_{3} = 0,
\end{equation}
or in the form
\begin{equation}\label{bndcnd2}
P_{\beta}^{\perp}\vec{\psi}=0 \, , \quad
P_{\beta}\vec{\psi}^{\prime}=0
\end{equation}
with the projection
\begin{equation}
\label{P_beta} P_{\beta} = \frac{1}{\beta^2 + 2} \left(
\begin{array}{ccc}
1 & \beta & 1 \\
\beta & \beta^2 & \beta \\
1 & \beta & 1
\end{array} \right).
\end{equation}
The scattering matrix of such a junction is constant $S_{\beta} = I
- 2 P_{\beta} $, with the phenomenological parameter $\beta$
responsible for connection between the bar and the leg. This formula
was intensely used, see for instance \cite{SGB,SGZ}, despite unclear
meaning of  the parameter $\beta$. See further  discussion of
transmission  across  the junction in \cite{Exner88,EP05,Grieser07}
 and find  more references therein.

In this paper, based on the resonance conception of conductance on
the junction, we suggest a semi-analytic procedure of calculation of
the scattering matrix and a method of reduction of a general thin
junction to a quantum graph. Moreover, we suggest a solvable model
of a  thin junction and reveal the meaning of the projection
$P_{\beta}$. In this paper we do not take into account the
spin-orbital interaction, just by disregarding the spin of  the
electron.

We consider a junction $\Omega$ constructed of a few straight leads
$\omega^m,\, \cup_{m=1}^M \omega^m = \omega$, width $\delta$,
attached orthogonally to the flat pieces $\Gamma_m$ of the
piecewise--smooth boundary of the vertex domain $\Omega_{int}$,
$\Omega = \Omega_{int} \cup \omega$. On smooth functions $u \in
 W^2_2 (\Omega)$ satisfying the homogeneous
Neumann boundary condition, we define the
 Schr\"{o}dinger operator
\[
-\bigtriangleup u + V u \ = : {\mathcal{L}}
\]
with the potential $V$ equal to the constant $V_{\delta}$ on the
leads and equal to a real bounded piecewise--continuous function on
$\Omega_{int}$ supplied  with Dirichlet boundary condition. The
operator ${\mathcal{L}}$ is essentially self--adjoint, and  it can
be considered as  a perturbation of the corresponding operator
${\mathcal{L}}_0$ defined by  the same differential expression and
an additional Dirichlet boundary condition on $\cup_{m=1}^M \Gamma_m
=: \Gamma$:
\[
{\mathcal{L}} \longrightarrow l^{\omega} \oplus L_{int} \ = \
{\mathcal{L}}_0\ .
\]
The  spectrum  $\sigma(L_{int})$ of $L_{int}$ is  discrete, and  the
 spectrum $\sigma^{\omega}$ of  $L^{\omega}$ is  absolutely continuous,
consists of a  countable set of branches $\sigma^{\omega} =
\cup_{m=1}^M\,\cup_{l=1}^{\infty} \sigma^m_l$ corresponding to the
parts $l^{m}_l$ of $l^{\omega}$
\[
l^{m}_l \ = \ -\frac{d^2}{d x^2} + \frac{\pi^2 l^2}{\delta^2} +
 V_{\delta},\quad l\geq 1\ ,
\]
with the homogeneous Dirichlet  boundary condition $u \big|_{\Gamma}
= 0$ at the bottom sections $x^m \big|_{\Gamma_m} =0$. The operators
$L^{m}_l$ on the wires are obtained from $L^{\omega}$ via separation
of variables, with the basis of cross--section eigenfunctions
$\left\{e^m_l \right\} = \left\{\sqrt{\frac{2}{\delta}}
\sin\frac{\pi l y}{\delta} \right\},\,l=1,2,\dots,\, m = 1,2,\dots
M$. Here the local transversal coordinate on $\omega^m$ is denoted
by $y$. The eigenfunctions of $L^{\omega}$ are scattered waves on
each lead $\omega^m$:
\[
\psi^m_l (x)\ = \ \chi^l_{+} (x)\quad - \quad\chi^l_{-} (x),\quad x
= x_m\geq 0\ ,
\]
represented as linear combinations of oscillating exponential modes
\begin{equation}
\label{chi+} \displaystyle \chi^{m,l}_{\pm} \ = e^{\pm i\,
\,\sqrt{\lambda - \lambda_l}\,\,\,x} \,\, e^m_l (y):=\  e^{\pm i
K^{^{m,l}}_{_{+}}\,x}\, e^m_l (y),\quad\lambda > \lambda_l =
\pi^2\,l^{2} \delta^{-2},
\end{equation}
with the reflection coefficients $S_{l} = 1$. The  perturbed
operator ${\mathcal{L}}$ is obtained  from ${\mathcal{L}}_0$ by
replacement of the homogeneous Dirichlet boundary condition on  the
bottom sections $\Gamma$ by the  smooth matching condition. The
corresponding scattered  waves are obtained via matching on $\Gamma$
a  solution of  the Schr\"{o}dinger  equation on the vertex domain
with the scattering Ansatz (see for instance \cite{ML71,
Mittra_77}):\,\,\,$
 \psi^m_l (x) =$
\begin{equation}
\label{Sansatz} \left\{
\begin{array}{c}
\chi^l_{+} (x) + \sum_{\pi^2 r^2/\delta^2 <\lambda }
S^{m,m}_{l,r}\chi^r_{-} (x) + \sum_{\pi^2 r^2/\delta^2 >\lambda }
s^{m,m}_{l,r}\xi^r (x), x\in \omega^m\\
\sum_{\pi^2 r^2/\delta^2 <\lambda } S^{m,n}_{l,r}\chi^r_{-} (x) +
\sum_{\pi^2 r^2/\delta^2 >\lambda } s^{m,n}_{l,r}\xi^r (x), x\in
\omega^n,\, n\neq m\ ,
\end{array}
\right.
\end{equation}
composed, for given  $\lambda$, of the above  oscillating modes
$\chi^r_{\pm}$ in the open channels, with the thresholds below
$\lambda,\,\, \lambda^m_r <\lambda $, and the exponentially
decreasing (``evanescent'') modes in the closed channels
\begin{equation}
 \label{Kplus} \displaystyle
\xi^{m,s}_{-} \ = \ e^m_s (y)\,\,e^{-\ \sqrt{ \lambda_s - \lambda
}\,\,\,x} := \ e^{- K^{{m,s}}_{_{-}}\,x}\, e^m_s (y),\quad\lambda <
\lambda_s ,
\end{equation}
associated  with the thresholds $ \lambda_s^m = \pi^2 \,s^2
\,\delta^{-2}$  of the closed channels in the  leads -- see
\cite{MathNach07} for  details.

The  quantum wells and the quantum wires are  usually manufactured
as a certain relief of the surface of the semiconductor. We  assume
in this paper that the scaled Fermi  level $\Lambda = 2 m^*E_F
\hbar^{-2}$ of the semiconductor is situated in the middle of the
first spectral band $\Delta_1 = [V_{\delta} +
\frac{\pi^2}{\delta^2},\, V_{\delta} + 4 \frac{\pi^2}{\delta^2}]$ of
the  wire, $\Lambda = V_{\delta} + \frac{5}{2}
\frac{\pi^2}{\delta^2}$. Then the first spectral band plays the role
of the conductivity  band and the junction has metallic  properties.
At low temperature $T$, the scattering processes are observed only
on the essential spectral interval
\begin{equation}
\label{T_interval}
 \Delta^T = [ \Lambda - 2 m^* \kappa T
\hbar^{-2},\,\Lambda - 2 m^* \kappa T \hbar^{-2}]\subset \Delta_1.
\end{equation}
If the electron's density is low, the main contribution to the
scattering picture is defined by one-body processes. In this paper
we  focus on one-body scattering on the essential spectral interval.
We disregard the spin-orbital interaction and neglect all effects
connected with electrons spin. Hence  we study the scattering on the
first spectral band $\Delta_1 =
[\pi^2\,\delta^{-2},\,4\pi^2\,\delta^{-2}]$ of the open channel, and
represent the cross-section space $L_2 (\Gamma)=:E$ as an orthogonal
sum of the entrance subspaces $E_{\pm}$ of the open and closed
spectral channels respectively:
\begin{equation}
\label{Eq:Eplus_minus} E_{+} \ = \ \bigvee_{m = 1}^M e^m_1,\quad
E_{-} \ = \ \bigvee_{m=1}^M\bigvee_{l=2}^{\infty}e^m_l, \quad
P_{E_{\pm}}=:P_{\pm}\ .
\end{equation}
The infinite linear system for the coefficients of the scattering
Ansatz, obtained from the matching conditions,can be solved, if the
Green functions $G^D_{ \Gamma} = G_{int}$ of the Schr\"{o}dinger
operators $L^D_{\Gamma} = L_{int}$  in $L_2(\Omega_{int})$, with
Dirichlet boundary conditions  is constructed. The operator
$L^D_{\Gamma}$ is defined on $W_2^2$-functions in $\Omega_{int}$,
with the Meixner conditions at the inner  corner points:
\begin{equation}
\label{L_int} L_{int}u = - \Delta u  + V u = \lambda u,\,\,
 u\big|_{\partial \Omega_{\int}} = 0.
\end{equation}
The Green function is  found from the equation:
\begin{equation}
\label{G:D} L^D_{ \Gamma} G^D_{\Gamma}\ = \ -\bigtriangleup
G^D_{\Gamma} + V G^D_{\Gamma} \ = \ \lambda G^D_{\Gamma} + \delta
(x-y),\,\,\,G^D_{\Gamma} \big|_{ \partial{\Omega_{int}}}\, = \, 0.
\end{equation}
Hereafter  we denote by  $\sigma^{D}$ the spectrum  of $  L^D_{
\Gamma}$. According to the general theory of second--order elliptic
equations, the solution $u$ of the  boundary problem
\begin{equation}
\label{relative_Dirichlet}
 - \Delta u  + V u = \lambda u,\,\,
u\big|_{\Gamma} = u_{\Gamma},\,\, u\big|_{\partial \Omega_{\int}
\backslash \Gamma} = 0.
\end{equation}
is represented by the  Poisson map
\[
u(x)\ = \ \int_{\Gamma} {\mathcal{P}}_{\Gamma} (x,\gamma,\lambda)
u_{\Gamma}(\gamma)\ d\gamma\ ,
\]
involving the kernel ${\mathcal{P}}_{int} (x,\gamma)= -
\partial G^D_{\Gamma}(x,\gamma)/\partial n_{\gamma} $. The
corresponding boundary current on $\Gamma$ is calculated as
\[
\frac{\partial u}{\partial n}\bigg|_{x\in \Gamma} \, = \, -
\int_{\Gamma} \frac{\partial^2
G^D_{\Gamma}(x,\gamma,\lambda)}{\partial n_x
\partial n_{\gamma}} u_{\Gamma}(\gamma)\  d\Gamma\ =:
{\mathcal{DN}}_{\Gamma} (\lambda)u_{\Gamma}\ .
\]
This  formal integral operator is restriction onto $\Gamma$ of the
Dirichlet-to-Neumann map, see \cite{SU2,Gezstezy05,Gezstezy06}.  For
the sake of brevity we call it here ``relative DN-map''. The
relative DN- map  is also a Nevanlinna class function
${\mathcal{DN}}(\lambda)$ for  Im $\,\lambda \leq 0$, with poles at
the eigenvalues of the corresponding Schr\"{o}dinger operator
${{L}}^D_{\Gamma} = L_{int}$. The relative  DN--map is a
pseudo--differential operator of order 1: for $W_2^2(\Omega)$
solutions $u$ the  DN--map acts from $W^{3/2}_2 (\Gamma)$  to
$W^{1/2}_2 (\Gamma)$ and for $W_2^{3/2}(\Omega)$ generalized
solutions the D--map acts from $W^{1}_2 (\Gamma)$ to $L_2 (\Gamma)$.

We  consider also the boundary problem
\begin{equation}
\label{relative_Neumann}
 -  \Delta u  + V u = \lambda u,\,\,
\frac{\partial u}{\partial n}\big|_{\Gamma} = \rho_{\Gamma},\,\,
u\big|_{\partial \Omega_{int} \backslash \Gamma} = 0.
\end{equation}
and the  operator
\begin{equation}
\label{L:N} L_{\Gamma}^N = -  \Delta u  + V u ,\,\, \frac{\partial
u}{\partial n}\big|_{\Gamma} = 0,\,\, u\big|_{\partial \Omega_{int}
\backslash \Gamma} = 0.
\end{equation}
with  the   relative  Neumann Green function $G^N_{\Gamma}$:
\begin{equation}
\label{G:N} L^N_{ \Gamma} G^N_{\Gamma}\ = \ -\bigtriangleup
G^N_{\Gamma} + V G^N_{\Gamma} \ = \ \lambda G^N_{\Gamma} + \delta
(x-y),\,\,\,G^N_{\Gamma} \big|_{ \partial{\Omega_{int}}\backslash
\Gamma}\, = \, 0,\, \frac{\partial G^N}{\partial n_x}\big|_{
\partial{\Omega_{int}}\backslash \Gamma} = 0.
\end{equation}
The map
\[
u(x) \, = \, \int_{\Gamma} G^N_{\Gamma} (x,\gamma,\lambda)
\rho_{\Gamma}(\gamma) d \,\,\Gamma\ =: Q^{\Gamma}_1
\rho_{\Gamma},\quad x\in \Omega_{int}\ ,
\]
gives  a solution of  the relative Neumann boundary problem
(\ref{relative_Neumann}). The trace of the solution on $\Gamma$
\[
u(x)\big|_{\Gamma} \ = \ \int_{\Gamma} G^N_{\Gamma} (x,\gamma)
\rho_{\Gamma} d \,\,\Gamma \ =:{\mathcal{ND}}_{\Gamma}^{int}
\frac{\partial \psi}{\partial n}\bigg|_{\Gamma},\,\, {x\in \Gamma}\
,
\]
defines the relative  Neumann--to--Dirichlet map which is  inverse
to the relative  Dirichlet--to--Neumann map defined above,
\[
{\mathcal{ND}}_{\Gamma}\ {\mathcal{DN}}_{\Gamma}\ = \ I_{\Gamma} .
\]
For $W_2^2 $ solutions $u$ the corresponding  DN--map  acts, on the
set of all regular spectral points  $\lambda$ of the Neumann
Schr\"{o}dinger, from $W^{1/2}_2 (\Gamma)$ onto $W^{3/2}_2
(\Gamma)$. For $W_2^{3/2} $ solutions the ND--map acts  acts from
$L_2 (\Gamma)$ onto $W^1_2 (\Gamma)$.

The coefficients of the scattering Ansatz (\ref{Sansatz}) can be
found, in principle, from the infinite linear system which is
obtained by substitution of the scattering Ansatz into the matching
condition (see \cite{MathNach07}). An important part of the
calculation is the proof of the formula for the DN--map in terms of
the $G^D_{\Gamma}$ (see \cite{MathNach07}), or, respectively, a
similar formula for the ND--map in terms of $ G^N_{\Gamma}$.
Selecting $E_{\pm}$ as  indicated above, (\ref{Eq:Eplus_minus}),
represent the ND--map of $L^N_{\Gamma}$ by $2\times 2$ operator
matrix with elements ${\mathcal{ND}}_{\pm, {\pm}} \ = \ P_{\pm}
{\mathcal{ND}}_{\Gamma} P_{{\pm}}$
\begin{equation}
\label{ND_matrix} {\mathcal{ND}}_{\Gamma} = \left(
\begin{array}{cc}
{\mathcal{ND}}_{+ +} & {\mathcal{ND}}_{+ -}\\
{\mathcal{ND}}_{- +}& {\mathcal{ND}}_{- -}
\end{array}
  \right).
\end{equation}
The similar decomposition of the DN--map of the  Schr\"{o}dinger
operator $L^{D}_{\Gamma}$ on $\Omega_{int}$
\begin{equation}
\label{DN_matrix} {\mathcal{DN}}_{\Gamma} \, = \, \left(
\begin{array}{cc}
{\mathcal{DN}}_{+ +} & {\mathcal{DN}}_{+ -}\\
{\mathcal{DN}}_{- +}& {\mathcal{DN}}_{- -}
\end{array}
  \right)
\end{equation}
was  used  in \cite{MathNach07}  in the course  of construction of a
convenient representation  for the  scattering matrix on the open
spectral bands. We set, in agreement with the above notations in
(\ref{chi+},\ref{Sansatz},\ref{Kplus}):
\[ \begin{array}{rcccl}
K_+ & = & \sum_{m}\sum_{open}\sqrt{\lambda -
\lambda_l}\,e^m_l\rangle \langle e^m_l & = & \sum_{m}\sqrt{\lambda -
\frac{\pi^2}{\delta^2}}
\,e^m_1\rangle \langle e^m_1\ ,\\
 K_- & = & \sum_{m}\sum_{closed}\sqrt{ \lambda_l -
\lambda}\,e^m_l\rangle \langle e^m_l & = & \sum_{m}\sum_{l\geq
2}\sqrt{ \lambda_l - \lambda}\,e^m_l\rangle \langle e^m_l\ .
\rule{0mm}{1.5em}
\end{array}
\]
Hereafter we  use the  standard  bra/ket notations, $e><e'\, : u \to
e \,<e',\,u> $, with the bar on the first factor of the dot--product
in $E = L_2 (\Gamma)$. The exponents of oscillating and decreasing
modes on the first spectral band spanned by the vectors $e_{\pm}\in
E_{\pm}$ are represented as:
\[
\chi_{\pm} e_+ = e^{\pm i K_+ x} e_+\ ,\qquad \xi_-  e_- = e^{- K_-
x} e_-\ .
\]
The matrices  $S^{m,n}_{l,r}$ and $ s^{m,n}_{l,r}$, which are
defined by the matching of the  scattering Ansatz to the solution of
the homogeneous equation  on $\Omega_{int}$, constitute respectively
the {\it scattering matrix} -- the square table of amplitudes in
front of the oscillating modes in open channels ($l = 1$):
\[
S \ = \ \sum_{m,n = 1}^M \,\,\,\,\sum S^{m,n}_{1,1} e^m_1
\rangle\,\, \langle e^n_1\ ,
\]
and the table of  amplitudes in  front of the evanescent modes
\[
s \ = \ \sum_{m,n = 1}^M \,\,\,\,\sum_{1,\, r\geq 2} s^{m,n}_{1,r}
e^m_1 \rangle\,\, \langle e^n_r\ .
\]
The  scattering matrix of the junction is represented (see
\cite{MathNach07} and  Theorem \ref{T:matching} below) in terms of
the matrix elements ${\mathcal{DN}},\, {\mathcal{ND}}$  combined in
aggregates
\begin{equation}
\label{DN_Gamma} {\mathcal{M}} \ = \ {\mathcal{DN}}_{++}-
{\mathcal{DN}}_{+-} \frac{I}{ {\mathcal{DN}}_{--} +
K_-}{\mathcal{DN}}_{-+}\,
\end{equation}
\begin{equation}
\label{ND_Gamma}
 {\mathcal{N}} \ = \ {\mathcal{ND}}_{++}-
{\mathcal{ND}}_{+-} K_- \frac{I}{ I_- + {\mathcal{ND}}_{--}K_-}
{\mathcal{ND}}_{-+}\ ,
\end{equation}
The width $\delta$ of the leads can serve as a small parameter in
the course of calculation of the scattering matrix. Thin networks,
with small $\delta $, are characterized  by large distance between
the neighboring spectral thresholds:
\[ \frac{\pi^2 (l+1)^2}{\delta^2} - \frac{\pi^2
l^2}{\delta^2} \ = \ \frac{(2 l + 1)\pi^2}{\delta^2}\ .
\]

One can prove  following \cite{MathNach07} that, for a ``thin
junction'', the denominator $ {\mathcal{DN}}_{--} + K_-$ is
invertible on a major part of a properly selected auxiliary spectral
interval $\Delta$, where the  DN-map is represented as a sum of a
rational function  and a  regular correcting term:
\begin{equation}
\label{rational_DN} {\mathcal{DN}}_{\Gamma} = \sum_{\lambda_s \in
\Delta} \frac{ \frac{\partial \varphi_s}{\partial n}\rangle \langle
{\frac{\partial \varphi_s}{\partial n}}}{\lambda_s - \lambda} +
{\mathcal{K}}^{\Delta_1} =: {\mathcal{DN}}^{\Delta} +
{\mathcal{K}}^{\Delta},
\end{equation}
The zeros of the denominator ${\mathcal{DN}}_{--} + K_-$ on $\Delta$
have an important operator--theoretic meaning: they are eigenvalues
of the {\it Intermediate Hamiltonian}. Hereafter we consider the
rational approximation (\ref{rational_DN}) and  the  corresponding
rational approximation of  ${\mathcal{DN}}_{--} = P_- {\mathcal{DN}}
P_-$:
\begin{equation}
\label{rational_DN_minus} {\mathcal{DN}}_{--}=
{\mathcal{DN}}^{\Delta}_{--} + {\mathcal{K}}^{\Delta}_{--},
\end{equation}
with a regular ``error'' ${\mathcal{K}}^{\Delta}_{--}$ on a complex
neighborhood $G(\Delta)$ of  $\Delta$.

{\it We  call the  junction $\Omega$ thin in closed channels, either
in $W_2^{1}(\Gamma)$ or in $W_2^{3/2}(\Gamma)$, if, respectively,
\begin{equation}
\label{Thin} \sup_{\Delta}\parallel K_-^{-1}
{\mathcal{K}}^{\Delta}_{--}\parallel_{W_2^{1}(\Gamma)} <  1\ ,\qquad
\mbox{or} \qquad \sup_{\Delta}\parallel K_-^{-1}
{\mathcal{K}}^{\Delta}_{--}\parallel_{W_2^{3/2}(\Gamma)} <  1
\end{equation}
}
This implies the following statement (see \cite{MathNach07}):
\begin{lemma}
\label{invertibility} {If the junction is  thin on closed  channels,
then the  denominator of  (\ref{DN_Gamma}) is  invertible
\[
\left[ {\mathcal{K}}^{\Delta}_{--} + K_- \right]^{-1}:
L_2(\Gamma)\to W_2^1(\Gamma)\ ,
\]
on  a corresponding ``major part of the essential spectral
interval'' - the complement of  the set of zeros $Z_{\Delta} \subset
\Delta$ of the determinant of the finite-dimensional matrix-function
:
\[
Z_{\Delta} = \left\{ \lambda: {\mbox{det}} \left[ I  + \left(
{\mathcal{K}}^{\Delta}_{--} + K_- \right)^{-1} \,
{\mathcal{DN}}^{\Delta}_{--}(\lambda) \right] = 0 \right\}
\]
A similar statement holds  for the above  denominator as an operator
from $ W_2^{1/2}(\Gamma)$ to $W_2^{3/2}(\Gamma)$.}
\end{lemma}

\begin{theorem}
\label{T:matching}{ The  substitution of the  scattering Ansatz
(\ref{Sansatz}) into the matching conditions on $\Gamma$ gives the
following formulae  for the  scattering matrix on $\hat{\Delta}$
\begin{equation}
\label{Eq:smatrix_M} S = \left[ i K_+ + {\mathcal
M}\right]^{-1}\left[  i K_+ - {\mathcal M}\right]\ ,
\end{equation}
\begin{equation}
\label{Eq:smatrix_N} S = \left[ {\mathcal N} i K_+ +
1\right]^{-1}\left[ {\mathcal N} i K_+ - 1\right]\ .
\end{equation}
 }
\end{theorem}
{\it Proof}. The scattering Ansatz generated  by the entrance vector
$e\in E_+$  is  constituted  by the  incoming wave $e^{iK_+\,x} e$,
 the  transmitted/reflected  wave $e^{-iK_+ \,x} S e$ and the
evanescent  wave $ e^{-K_- x} s e$:
\[
\Psi_{e} \ = \ e^{iK_+\,x} e +  e^{-iK_+ \,x} S e + e^{-K_- x} s e\
.
\]
The  boundary data of the  Scattering  Ansatz  at the  bottom
sections $\Gamma$  should match on $\Gamma$  the boundary data  of
the solution of  the  homogeneous  Schr\"{o}dinger equation inside
$\Omega_{int}$:
\[
L_{int} \psi\ = \ \lambda \psi,\qquad  \psi\bigg |_{\partial
\Omega_{int}\backslash \Gamma} \ = \ 0\ ,
\]
\begin{equation}
\label{bound_data} \psi\big|_{\Gamma} = \psi_{e}(0) =  e + S e + se\
,\qquad\frac{\partial\psi}{\partial n}\bigg|_{\Gamma} \ = \
\psi'_{e}(0) \ = \ iK_+ e - iK_+ Se - K_- se\ .
\end{equation}
Using the matrix representations (\ref{DN_matrix}, \ref{ND_matrix})
 for  ${\mathcal{DN}},\,{\mathcal{ND}} $, we  obtain from
(\ref{bound_data})  two equivalent linear systems  which describe
matching conditions on $\Gamma$
\begin{eqnarray}\label{Eq:DN}
\nonumber  iK_+ (1-S)e & = & {\mathcal{DN}}_{++}(1+S)e
 + {\mathcal{DN}}_{+-}s e\ ,\\
 \label{DNS}-K_- s e & = & {\mathcal{DN}}_{-+}(1+S)e +
{\mathcal{DN}}_{--}s e\ ,
\end{eqnarray}
and
\begin{eqnarray}\label{Eq:ND}
 \nonumber
(I+S)e & = & {\mathcal{ND}}_{++}\, i \,K_+ (I-S)e - {\mathcal{ND}}_{+-}K_- se, \\
\label{NDS} [I + {\mathcal{ND}}_{--} K_-] se & = &
{\mathcal{ND}}_{-+}\, i\,K_+ (I-S)e {\mathcal{ND}}_{--}se.
\end{eqnarray}
Eliminating  the component $se$ from them and using the former
notations ${\mathcal{M}},\, {\mathcal{N}}$ we  obtain the announced
representation for  the scattering matrix
(\ref{Eq:smatrix_M},\ref{Eq:smatrix_N}).

{\it The  end of the proof}

Consider the operator ${\mathcal{L}}$ defined by the above
Schr\"{o}dinger differential expression on the junction $\Omega =
\Omega_{int}\cup \omega$\ , with  zero Dirichlet  condition on the
boundary $\partial \Omega$. It is  essentially self--adjoint  on the
domain of smooth functions $u$, subject to  the Meixner restriction
$u \in W_2^1 (\Omega)$. Assume that the  entrance space $E =
L_2(\Gamma)$ on the cross--sections $\Gamma$ is decomposed as  $E_+
\oplus E_-$. We use the former  notations $P_{\pm}$ for  the
orthogonal projections in $E$ onto $E_{\pm}$. Consider the Glazman
splitting ${\mathcal{L}}_{\Gamma}$ obtained from ${\mathcal{L}}$ by
imposing an additional {\it partial zero boundary condition} on the
bottom sections $\Gamma$ of the leads:
\begin{equation}
\label{partial_bc} P_+ u\big|_{\Gamma}\ = \ 0\ ,
\end{equation}
complemented by the  standard  smooth matching condition on $\Gamma$
in  closed  channels. The operator ${\mathcal{L}}$ is  split by this
boundary conditions into  an orthogonal sum of two operators:
\[
{\mathcal{L}} \longrightarrow L_{\Lambda} \oplus l_{\Lambda} \ = \
{\mathcal{L}}_{\Lambda}\ .
\]
Here  $l_{\Lambda} = - {\displaystyle\frac{d^2}{dx^2}} +
\frac{\pi^2}{\delta^2} + V_{\delta}$ in $L^2(0,\infty)\times E_+ )
=: {\mathcal{H}}_+$, with  zero boundary condition at  the origin
$u(0)= 0$, and $L_{\Lambda}$ is  defined in  the orthogonal sum of
the channel space   $L^2(0,\infty)\times E_-  =: {\mathcal{H}}_-$ of
the closed channels and $L_2(\Omega_{int})$ on $W_2^2$ - smooth
functions, subject to the Meixner condition and the matching
condition on $\Gamma$ in closed channels:
\[
L_{\Lambda} : D({\Lambda})\ \longrightarrow \, L_2(\Omega_{int})
\oplus {\mathcal{H}}_-\, .
\]
\begin{theorem}\label{T:spectrum} {The operators $L_{\Lambda},\,l_{\Lambda} $ are
essentially self--adjoint. The absolutely continuous components of
spectra of the corresponding self--adjoint extensions  are
\begin{eqnarray}
\sigma_a (l_{\Lambda}) & = &[\lambda_1,\,\infty),\qquad\mbox{with\
multiplicity} \ M,
\nonumber \\
\sigma_a (L_{\Lambda}) & = &
\bigcup_{l=2}^{\infty}\left[\lambda_l,\infty \right)\ =:
\bigcup_{l\geq 2} \sigma_a^l\ .
\end{eqnarray}
where each branch $\sigma_a^l$ has  multiplicity $M$, and the total
multiplicity is  growing stepwise on the thresholds $\lambda_l$
separating the spectral bands $\Delta_l = \left[ \lambda_l,\,
\lambda_{l+1}\right]$. The spectral multiplicity of the
absolutely--continuous spectrum of $L_{\Lambda}$ on the spectral
bands $\Delta_l$ is equal to $M l(l+1)/2$. The discrete spectrum of
$L_{\Lambda}$ consists of a  countable set of eigenvalues
$\lambda^{\Lambda}_s$ accumulating at infinity. The singular
spectrum of $L_{\Lambda}$ is  empty.}
\end{theorem}
\begin{figure} [ht]
\begin{center}
\includegraphics [width=4.5in] {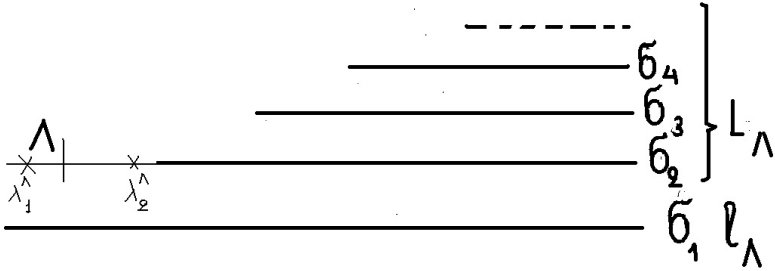}
\end{center}
\caption{ The intermediate Hamiltonian  $L_{\Lambda}$ inherits  the
closed branches of the continuous spectrum of the  unperturbed
operator. The  part  $l_{\Lambda}$  of  the split  operator inherits
the first - open - branch of the spectrum  of the  split  operator.
The resonance eigenvalues of the intermediate Hamiltonian define the
resonance conductance of the junction \label{F:figure 6}}
\end{figure}

The relation ${\mathcal{M}} {\mathcal{N}} = I$ observed from
comparison of the formul{\ae} (\ref{Eq:smatrix_N},
\ref{Eq:smatrix_M}) has an important operator--theoretic meaning. It
is  derived from the fact that  ${\mathcal{M}},\, {\mathcal{N}}$ are
respectively DN and ND--maps of the Intermediate Hamiltonian --- the
part $L^0$ of  the Glazman splitting
\begin{equation}
\label{Glazman_splitting} {\mathcal{L}}\longrightarrow
{\mathcal{L}}_{\Lambda} \ = \ L_{\Lambda} \oplus l_{\Lambda}.
\end{equation}
defined by the partial boundary condition $ P_+ u\big|_{\Gamma} = 0$
--- see \cite{ML71}.  Contrary to the standard
splitting ${\mathcal{L}}\longrightarrow L_{int} \oplus L^{\omega} $,
this splitting (\ref{Glazman_splitting}) is finite--dimensional ---
see \cite{Glazman}. The poles of ${\mathcal{M}}$ on the first
spectral band, below $\lambda_{min}$, are the eigenvalues of
$_{\Lambda}$.

 \vskip0.3cm
\section{Krein formulae for the intermediate DN-map \\ and ND-map,
with the compensated singularities}

Expressions ${\mathcal{M}},\,\mathcal{N}$ in the formulae
(\ref{DN_Gamma},\ref{ND_Gamma}) contain, at least formally, the
singularities  at the eigenvalues  of the operators
$L^D_{\Gamma},\,L_{\Gamma}^N$. Presence of these singularities  in
the conditions of the last theorem of the previous  section looks
strange. In fact we were able to prove, see \cite{APY_09}, that the
singularities in the first and second terms of the above Krein
formula for ${\mathcal{M}}$, inherited from $L^D_{\Gamma}$,
compensate each other, so that only the  singularities of the
denominators of (\ref{DN_Gamma})  play a role. Similar statement can
be proved, see the theorem \ref{T:comp_N} below, for
${\mathcal{N}}$. But in fact the compensation of singularities
permits to obtain more convenient representations for
${\mathcal{M}},\,{\mathcal{N}}$ - that is for DN and  ND maps of the
intermediate Hamiltonian $L_{\Lambda}$. These new representations
imply also the corresponding exact formulae for the scattering
matrix of ${\mathcal{L}}$, and convenient approximate expressions
for the scattering matrix as well.  This approximate expressions can
serve a base for construction of a fitted  solvable model of the
junction in form of star-shaped 1D quantum graph, and for derivation
of the boundary condition at the vertex.

We  begin with the discussion of  compensation of singularities in
the formula (\ref{DN_Gamma})   for the   DN map ${\mathcal{M}}$ of
the Intermediate Hamiltonian. It appeared, that the singularities of
the first and second term at the eigenvalues of $L_{int}$ compensate
each other, so that only the zeros of the denominator $
\mathcal{DN}_{--} + K_-$ arise as singularities of
$\mathcal{DN}^{\Lambda}$ on $\Delta$. A one-dimensional version of
the statement can be  found in \cite{Ring} and a rescription of the
classical Krein formula with compensated singularities  is given in
\cite{MP_08}. In this paper we review the  compensation
singularities in Theorem \ref{T:comp_M}, following  \cite{APY_09}
for a general thin junction and prove a similar statement, see
Theorem \ref{T:comp_N} for the Intermediate ND-map. We also obtain,
in course of calculations, an important ``byproduct'': a version of
analytic perturbation procedure for groups of eigen-pairs. Note that
the standard analytic perturbation procedure is  aimed on
calculation of an individual perturbed eigenvalue. Usually the
convergence of the corresponding perturbation series  is limited by
the condition of non- intersection  of corresponding terms
$\lambda_s(\varepsilon)\neq \lambda_t(\varepsilon)$. Our technique,
based on compensation of singularities, can be  used even to  study
overlapping terms and to study the transformation, under small
perturbations, of intersections of terms into quasi-intersections.
This  technique is aimed not on the calculation of an individual
perturbed eigenvalue, but rather on derivation of an approximate
algebraic equation for the perturbed eigenvalues and calculation of
the corresponding residues at the poles of the perturbed  DN-map. We
also  calculate, based on (\ref{Eq:smatrix_M}), the scattering
matrix of a ``relatively thin'' junction in the quantum network. We
also develop similar technique for  ND-map. In fact our technique
can be modified to calculate the scattering matrix for arbitrary
junction, see \cite{APY_09}.

For given temperature $T$  we  consider  an {\it essential spectral
interval} centered at the scaled Fermi level $\Lambda:  \Delta_T =
[\Lambda - 2 m^* \, T\,\hbar^{-2},\, \Lambda + 2 m^*\,  T\,
\hbar^{-2}]$. We assume that the temperature is {\it low}, so that
$\Delta_T$  is situated inside the auxiliary  spectral interval or
an open spectral set  $\Delta $
\[
\Delta_T \subset \Delta \subset  \left( \frac{\pi^2}{\delta^2} +
V_{\delta},\, \frac{4\pi^2}{\delta^2} + V_{\delta}\right) =:
\Delta_1.
\]
Our prime  aim is: to  construct on $\Delta_T$ a  convenient local
``quasi-one-dimensional'' representation for the intermediate DN-map
and one for the scattering matrix \footnote{Compare with the popular
one-dimensional formula for the scattering matrix in terms of the
Weyl function, derived in \cite{Pavlov_73} and intensely used by B.
Simon and F. Gezstezy in their approach to the spectral inverse
problem, see \cite{GS_97}.} of the junction {\it with compensated
singularities} inherited from the $L_{int}$, to substitute previous
formula (\ref{Eq:smatrix_M}).

Selecting an appropriate spectral interval (or just a  spectral set
) $\Delta : \Delta_T \subset \Delta $, we represent the DN-map
$\mathcal{D N}$ of $L_{int}$ on the essential spectral interval
$\Delta_T$  as a sum
\begin{equation}
\label{DN_local_spectral}
 \mathcal{D N}_{int} =  \sum_{\lambda_s \in \Delta} \frac{
\frac{\partial \varphi_s}{\partial n}\bigg|_{\Gamma}\rangle \langle
\frac{\partial \varphi_s}{\partial n}\bigg|_{\Gamma}}{\lambda -
\lambda_s }  +  \mathcal{K}^{\Delta} =:  \mathcal{D N}^{\Delta}\,\,
+ \mathcal{K}^{\Delta}.
\end{equation}
of the  rational  expression  constituted by the  polar terms with
singularities  at the   eigenvalues  $\lambda_s \in \Delta,\, s=
1,2,\dots N$ of the operator $L_{int}$ and the  analytic
operator-function ${\mathcal{K}}^{\Delta}$ on a complex neighborhood
$G_{\Delta_T}$ of $\Delta_T$.

We will also use the operators obtained from $\mathcal{D N}_{int}$
via framing it by the projections $P_{\pm}$, for instance:
\[
P_+{\mathcal{DN}}_{int}P_- =  P_+{\mathcal{DN}}^{\Delta}P_- +
 P_+{\mathcal{K}}^{\Delta} P_- = {\mathcal{D N}}^{\Delta}_{+-}\,\,+\,\,
 {\mathcal{K}}^{\Delta}_{+-}.
 \]
We introduce also the  linear hull  $E^{\Delta} = \bigvee_{s = 1}^N
\left\{\varphi_s \right\}$ - an invariant subspace  of $L_{int}$,
dim $E^{\Delta} = N$, corresponding to the spectrum of $L_{int}$
contained in $\Delta$ and the part $L^{\Delta}:= \sum_{\lambda_s \in
\Delta} \lambda_s \varphi_s \rangle \langle \varphi_s$ of $L_{int}$
in it. To calculate the intermediate DN-map  ${\mathcal{M}}$ in
terms of the standard DN - map  of $L_{int}$ we  have  to solve, see
(\ref{T:comp_M}) the equation:
\begin{equation}
\label{Eq5} [ {\mathcal{D N}}_{--} + K_- ] u = {\mathcal{D N}}_{-+}g
\end{equation}
on the  essential spectral interval $\Delta_T$.  It can be solved
based on Banach principle if $K_- $ can play a role of a large
parameter.

{\bf Definition} \label{Def: thin} {\it  The  junction, for  which
the operator}
\begin{equation}
\label{Large_parameter}
 \left[{\mathcal{K}}^{\Delta}_{--} + K_-\right]^{-1}
\end{equation}
{\it exists on  $\Delta_T$ is called hereafter ``relatively thin
junction'', for the selected  spectral set $\Delta$  and given
temperature $T$. A general network is called thin, if all junctions
of the network are thin.}

For a relatively thin junction, due to continuity of
${\mathcal{K}}_{--}^{\Delta}, K_-$ there exist also a complex
neighborhood  $G_{\Delta_T}$ of $\Delta_T$, where
${\mathcal{K}}^{\Delta}_{--} + K_-$ is invertible.

Hereafter we  assume that the  junction is thin. The case of an
arbitrary junction is considered in \cite{APY_09}.

The above definition of thin junctions (and networks) is based on
the following motivation. The DN-map of $L_{int}$ is homogeneous
degree $-1$. It acts from $W_2^{3/2}(\Gamma)$ to
$W_2^{1/2}(\Gamma)$, see \cite{SU2}. If $\Omega_{int}$  has a small
diameter $d$ then, the norm of the correcting term $\mathcal{K}$ is
estimated, generically, on the complement of the spectrum, as Const
$1/d$. The same estimate remains true for $P_-\mathcal{K}^{\Delta}
P_- := \mathcal{K}_{--}^{\Delta}$. The exponent $K_-$ on the
essential spectral band $\Delta$  acts from $W_2^{3/2}(\Gamma)$ to
$W_2^{1/2}(\Gamma)$ and the norm of its inverse is estimated  as
Const $\delta$. Then the $W_2^{3/2} -$ norm of
$K_-^{-1}\,\,\mathcal{K}_{--}^{\Delta} $ is estimated generically as
Const $\delta/d$. Hence, in particular, $ K_- +
\mathcal{K}_{--}^{\Delta} = K_- \left[I +
K_-^{-1}\,\,\mathcal{K}_{--}^{\Delta}\right]$ is invertible if
$\delta/d << 1$, see  more comments  in \cite{MathNach07}. Notice,
that for {\it arbitrary junction} an auxiliary  Fermi level
$\Lambda_1^F := \Lambda_1$ can be selected, see \cite{APY_09}, such
that the condition (\ref{Large_parameter}) is fulfilled. Now we
proceed  assuming that (\ref{Large_parameter})  is fulfilled.

Consider  the part $L_{int}^{\Delta}$  of  $L_{int}$ in the subspace
$E^{\Delta} = \oplus \sum_{\lambda_s \in \Delta} \varphi_s$:
\[
L_{int}^{\Delta} = \sum_{\lambda_s \in \Delta} \lambda_s
\,\,\varphi_s \rangle\,\langle \varphi_s: E^{\Delta} \to
E^{\Delta},\,\, \mbox {dim} E^{\Delta} = N.
\]

Assume that $\left\{\frac{\partial \varphi_s}{\partial
n}\right\}\bigg |_{\Gamma}$ are  linearly independent. Then dim
$\bigvee_{\lambda_s \in \Delta}\,\, \varphi_s  = $ dim $
\bigvee_{\lambda_s \in \Delta} \frac{\partial \varphi_s}{\partial
n}\bigg |_{\Gamma} = N $. Denote by ${\mathcal{T}}$ the map
\[
{\mathcal{T}} = \sum_{\lambda_s \in \Delta} \varphi_s \rangle
\langle \frac{\partial \varphi_s}{\partial n}\bigg|_{\Gamma},
\]
and  introduce
\[
\left( P_+ - {\mathcal{K}}^{\Delta}_{+-}
\frac{I}{{\mathcal{K}}^{\Delta}_{--} + K_-} P_- \right ) :=
{\mathcal{J}}(\lambda): E\to E_+.
\]
It is   obvious that  dim  $\left\{ {\mathcal {J}} \frac{\partial
\varphi_s}{\partial n}\bigg|_{\Gamma} \right\}_s  \leq $    dim
  $E^{\Delta}$. Later we  will  utilize  a  stronger

{\bf Assumption\,\,\, 1} {\it The \,\, vectors \,\,  ${\mathcal {J}}
\frac{\partial \varphi_t}{\partial n}\bigg|_{\Gamma}$ \,\, are\,\,
linearly\,\, independent\,\, in \,\, $E_+\,\,$ for \,\, any \,\,
$\lambda\,\, \in \Delta$, hence both :  dim $\left\{ \frac{\partial
\varphi_t}{\partial n}\bigg|_{\Gamma} \right\} = $ dim $E^{\Delta} =
N $ and }

\begin{equation}
\label{non_degenerate} W_J (\lambda):=\mbox{det} \left\{ \langle
{\mathcal {J}} \frac{\partial \varphi_s}{\partial
n}\bigg|_{\Gamma},\, {\mathcal {J}} \frac{\partial
\varphi_t}{\partial n}\bigg|_{\Gamma}\rangle\right\}_{s,t=1}^N
(\lambda) > 0.
\end{equation}
The  above condition (\ref{non_degenerate})  is  equivalent to the
pair of conditions:

1. The functions  $\frac{\partial \varphi_s}{\partial
n}\bigg|_{\Gamma},\,\, s = 1,2,3\dots N$ are linearly independent.

2. The operator  ${\mathcal {J}}^+ {\mathcal {J}}$ is  invertible in
the linear hull  $\bigvee_{s=1}^{s=N} \frac{\partial
\varphi_s}{\partial n}\bigg|_{\Gamma}$ for any  $\lambda \in
\Delta$.

Hereafter  we reduce the problem of compensation of singularities to
the  spectral analysis of the  Schr\"{o}dinger - type equation
\[
\left[ L^{\Delta} - Q(\lambda)\right] \psi = \lambda \psi
\]
in $E^{\Delta}$ with the $\lambda$-dependent ``potential''
\[
Q(\lambda) := {\mathcal{T}} \frac{I}{{\mathcal{K}}^{\Delta}_{--} +
K_-}{\mathcal{T}}^+ : E^{\Delta} \to E^{\Delta}.
\]
\begin{lemma}\label{L:positivity}{ For thin junction, on  the
essential spectral interval  $\Delta$ the  derivative
$\frac{\partial Q}{\partial \lambda}$ is  a positive matrix $N\times
N $. }
\end{lemma}
{\it Proof}.\,\,Recall that the branch  of the  square root
$\sqrt{\pi^2 \,\delta^{-2} + V_{\delta} -  \lambda }$ is defined
such that $\frac{ d K_-}{d\lambda} < 0$ on the conductivity band.
The correcting term ${\mathcal{K}}^{\Delta}$ is a meromorphic
operator - function with a negative  imaginary part in the upper
half-plane $\Im \lambda
> 0$ and a  positive imaginary part in the lower half-plane, hence
$\frac{d {\mathcal{K}}^{\Delta}_{--}}{d\lambda} <0$ on the
conductivity band  $\Delta_1$. This implies:
\[
\frac{\partial Q}{\partial \lambda} = -
{\mathcal{T}}\frac{I}{{\mathcal{K}}^{\Delta}_{--} + K_-}\left[
\frac{ d K_-}{d\lambda} + \frac{d
{\mathcal{K}}^{\Delta}_{--}}{d\lambda} \right]
\frac{I}{{\mathcal{K}}^{\Delta}_{--} + K_-}{\mathcal{T}}^+ >0,\,\,\,
\mbox {for}\,\,\, \lambda\in \Delta_T \in \Delta_1.
\]

{\it The end of the proof}

Based  on  some cumbersome calculation, we are able to derive, see
\cite{APY_09}, that all singularities in the Krein formula,
inherited from the eigenvalues $\lambda_s$ of the unperturbed
operator $L_{int}$, are compensated.
\begin{theorem}[{\bf Compensation of Singularities M}.]\label{T:comp_M}
{The Krein formula  (\ref{DN_Gamma}) for  the  intermediate  DN -
map, can be  re-written, for a thin junction,  on the spectral
interval  $\Delta$, as:
\[
 {\mathcal{M}}= {\mathcal{D N}}^{\Lambda}
={\mathcal{M}}_{reg} +
 {\mathcal{J}}  {\mathcal{T}}^+\rangle \frac{I}{\lambda
I^{\Delta} - L^{\Delta} + Q(\lambda )} \langle {\mathcal{T}}
{\mathcal{J}}^+ =:\]
\begin{equation}
\label{DN_polar} k (\lambda) +  {\mathcal{J}} {\mathcal{T}}^+\rangle
\frac{I}{\lambda I^{\Delta} - L^{\Delta} + Q(\lambda )} \langle
{\mathcal{T}} {\mathcal{J}}^+,
\end{equation}
where ${\mathcal{K}}^{\Delta}_{++} - {\mathcal{K}}^{\Delta}_{+-}
\frac{I}{{\mathcal{K}}^{\Delta}_{--} + K_-}
{\mathcal{K}}^{\Delta}_{-+} =: k(\lambda)$ is  a  regular part of
${\mathcal{M}}$ on $\Delta_T$. The representation (\ref{DN_polar})
remains valid on a complex neighborhood $G_{\Delta}$ of the
 spectral interval $\Delta$.}
\end{theorem}
{\bf Remark 1} The announced  rescription  (\ref{DN_polar}) of  the
Krein formula (\ref{DN_Gamma}) for the DN-map of the intermediate
Hamiltonian, has on the essential spectral interval only
non-compensated singularities, at the eigenvalues of the
intermediate Hamiltonian,  calculated as  zeros $\lambda^{Q}_s$ of
the denominator $\lambda I^{\Delta} - L^{\Delta} + Q(\lambda ): =
{\bf d}(\lambda)$:
\[
{\bf d} (\lambda^{Q}_s)\,\, \nu^{Q}_s = 0.
\]
These singularities  coincide with the  eigenvalues of the
intermediate Hamiltonian. We call the  above formula
(\ref{DN_polar}) for $\mathcal{DN}^{\Lambda}$ {\it the modified
Krein formula}. Inserting (\ref{DN_polar}) into  the above formula
(\ref{Eq:smatrix_M}) gives a convenient representation for the
scattering matrix of the relatively thin junction, which permits, in
particular, to calculate sharp resonances, situated near the
continuous spectrum,  based on eigenvalues of the intermediate
operator.

In the case of one-dimensional zeros  of the denominator ${\bf
d}(\lambda^{Q}_s) (\nu^{Q}_s)= 0, {\mathcal{J^+}}$ $ {\mathcal{T}}^+
\nu^{\Lambda}_s \neq 0 $, the corresponding residues are calculated
as projections onto the subspaces
\[
{\mathcal{E}}_s^{Q}= {\mathcal{J}}(\lambda^{Q}_s){\mathcal{T}}^+
\,\, \nu_s^{Q}.
\]
For  multidimensional zeros of the  denominator, ${\bf
d}(\lambda^{Q}_s) N_s^{Q} = 0 $, dim $N_s^{Q} \,\,
> 1$ the residues are  projections onto the images of the
corresponding null-spaces $N_s^{Q} = \bigvee_s \nu_s^{Q}$
\[
{\mathcal{E}}_s^{Q} = {\mathcal{J}}(\lambda^{Q}_s)
{\mathcal{T}}^+\,\, N_s^{Q}.
\]

The  above  expression  (\ref{DN_polar}) is  analytic in
$\Omega_{\Delta}$ on the complement of  the set of zeros of the
denominator  ${\bf d}(\lambda)$. This means that the eigenvalues of
the  Intermediate Hamiltonian are selected  from the set. Assume
that $\lambda^{\Lambda}_s, \,s=1,2,3,\dots N$ are simple  zeros of
the denominator.

\begin{theorem}\label{T:1}{If  the  Wronskian (\ref{non_degenerate})
does not vanish  at the algebraically  simple (first order ) zero
$\lambda^{Q}_1$ of the denominator, $W_J (\lambda^{Q}_1)\neq 0$,
then the zero is an eigenvalue of the Intermediate Hamiltonian, with
the same spectral multiplicity.}
\end{theorem}
{\it Proof}. It is  sufficient to prove, that the zero is a  first
order pole of the intermediate DN-map, with a finite-dimensional
residue having the same dimension as  the zero of the denominator.
Consider the equation
\begin{equation}
\label{Eq:12} {\bf d}(\lambda) u = \left[\lambda I^{\Delta} -
L^{\Delta} + Q(\lambda)\right] u = f, \,\,  \lambda \in E^{\Delta}.
\end{equation}
Assume that  $ {\bf d}(\lambda_1^Q) e_s^Q = 0$, and denote by
$P_1^Q$ an orthogonal projection onto the multiple eigen-space
$\bigvee_s e_s^{Q}$ of the operator $L^{\Delta}
 - Q(\lambda_1^Q) =: L^{\Delta}_Q$, and  by  $R^Q_{\lambda}$  the
corresponding  resolvent:
\[
\left[L^{\Delta} - Q(\lambda_1^Q) - \lambda I^{\Delta}\right]^{-1} =
R^Q_{\lambda} (\lambda).
\]
Then, for $\lambda$ close to  $\lambda_1^Q$ we can  substitute the
``potential'' $Q$ in the above equation by the Taylor expansion:
\[
Q(\lambda) = Q (\lambda_1^Q) + (\lambda - \lambda_1^Q)\frac{d Q}{d
\lambda} (\lambda_1^Q) +  \frac{(\lambda - \lambda_1^Q)^2}{2}
\frac{d^2 Q}{d \lambda^2} (\lambda_1^Q)+  \dots
\]
and represent  the resolvent near the pole $ \lambda_1^Q$  as
\begin{equation}
\label{Eq:pol} R^Q_{\lambda} =  \frac{P_1}{(\lambda_1^Q - \lambda)}
+ R^Q_{\bot,\lambda_1^Q},
\end{equation}
where  $R^Q_{\bot,\lambda_1^Q}$ is the part of the resolvent in the
complementary  invariant subspace  $E^{\bot}_1 =  (I - P_1)E$.
Replacing  $Q(\lambda)$  by the corresponding  first order Taylor
formula, we rewrite the  equation  ${\bf d} (\lambda)u =  f$ by  the
equation
\begin{equation}
\label{Eq:13} u + R^Q_{\lambda} (\lambda_1^Q - \lambda)\frac{d
Q}{d\lambda} u = - R^Q_{\lambda} f.
\end{equation}
To calculate the residue of the solution $u$ at the pole
$\lambda_1^Q$ we multiply (\ref{Eq:13}) by $\frac{d Q}{d\lambda}$
and by the spectral projection $P^Q_1\, $ of $L^{\Delta}_Q$ at
$\lambda_1^Q$ and take into account that the resolvent
$R^Q_{\lambda}$ of $L^{\Delta}_Q$ has a  simple pole and  neglect
the terms  vanishing  at  $\lambda_1^Q $, in particular all terms
arising  from the above Taylor expansion beginning from the second
$(\lambda - \lambda^Q_1)^2\frac{d^2 Q}{d\lambda^2}(\lambda^Q_1)$.
Due to positivity of $\frac{d Q}{d \lambda}$ the operator $I + P^Q_1
\frac{d Q}{d \lambda} P^Q_1$ is invertible. Then, due to
(\ref{Eq:pol})
\[
P^Q_1 \frac{d Q}{d \lambda} u = \frac{I}{I + P^Q_1 \frac{d Q}{d
\lambda} P^Q_1 } P^Q_1 \frac{d Q}{d \lambda} R^Q_{\lambda} f
\]
has the polar part at $\lambda_1^Q $
\begin{equation}\label{Eq:main}
u =  - \frac{P^Q_1 \left[I + P^Q_1 \frac{d Q}{d \lambda} P^Q_1
\right]^{-1} P^Q_1}{\lambda_1^Q - \lambda}f + \dots.
\end{equation}
The operator in the square bracket is positive and hence has  the
spectral form :
\[
P^Q_1 \left[ I + P^Q_1 \frac{d Q}{d \lambda} P^Q_1 \right]^{-1}P^Q_1
= \sum_{r} \alpha_r \,\, \nu_r \rangle \,\langle \nu_r.
\]
Then the polar term of ${\bf d}^{-1} = - \left[L^{\Delta}_Q
(\lambda)\right]^{-1}$ at $\lambda^Q_1$ is $(\lambda -
\lambda^Q_1)^{-1}{\sum_{r} \alpha_r \,\, \nu_r \rangle\, \langle
\nu_r}$ and the pole part of the intermediate DN-map at
$\lambda^Q_1$ is
\[
{\mathcal{DN}}^{\Lambda}_{pole} = \frac{\sum_{r}\,{\mathcal{J}}
{\mathcal{T}}^+\,\,\nu_r \rangle \, \alpha_r\,\langle
{\mathcal{J}}{\mathcal{T}}^+\,\,\nu_r }{\lambda - \lambda^Q_1}.
\]
Due to the orthogonality of  $\nu_s$ and  non-degeneracy of  the
Wronskian  $W^Q_{\Delta}$, the vectors ${\mathcal{J}} T^+\,\,\nu_r $
are linearly independent, hence $\lambda^Q_1$ is a simple pole of
the intermediate DN- map, with the spectral multiplicity dim
$\bigvee_s \,\, e^Q_s$.

{\it The end of the proof.}

 The scattering matrix of the original problem on the
essential spectral interval can  be obtained via replacement in
(\ref{Eq:smatrix_M}) the intermediate  DN-map by the  expression
(\ref{DN_polar}) with compensated singularities. This substitution
is  possible for  thin junctions, when the exponent $K_-$ in closed
channels can play a role of a large parameter, compared with the
error ${\mathcal{K}}^{\Delta}_{--}$ of the rational approximation
${\mathcal{D N}}^{\Delta}$ of $\mathcal{DN}$.

This condition may be not satisfied, for given  quantum network, at
the scaled Fermi level $\Lambda$. In that case  another
representation of the scattering matrix (\ref{Eq:smatrix_N}) can
help. We consider now the problem of compensation singularities for
the intermediate ND-map $ {\mathcal{N}}$ on the essential spectral
interval.

Denote by $\psi_s, \lambda^N_s$ the eigen-pairs of the  operator
$L^N_{\Gamma}$, see  (\ref{L:N}). Select the eigenvalues from the
spectral interval $\Delta_2$, to be  defined later, and introduce
$E^N_{\Delta_2} := \bigvee_{\lambda^N_s \in \Delta_2} \psi_s$ and
$E^N_{\Gamma}:= \bigvee_{\lambda^N_s \in \Delta_2}
\psi_s\big|_{\Gamma}$, and consider the map
\begin{equation}
\label{Eq:T_N} \tilde{\mathcal{T}}: \sum_{\lambda^N_s \in \Delta_2}
\psi_s \big|_{\Gamma} \rangle \,\langle \psi_s : \,\, E_{\Delta_2}
\to E^N_{\Gamma}, s= 1,2,\dots \tilde{N}
\end{equation}
{\bf Assumption 2}, see the paragraphs  1  and  2  below:

 1.{\it We  assume  that
the families $\left\{\psi_s\right\}_{s=1}^{\tilde{N}},\,\left\{
\psi_s\right\}_{s=1}^{\tilde{N}} \big|_{\Gamma}$ are linearly
independent, thus are bases in their linear hulls
$\tilde{E}^N_{\Delta_2},\, \tilde{E}^N_{\Gamma}$, dim
$\tilde{E}^N_{\Delta_2} = $ dim $ \tilde{E}^N_{\Gamma} =
\tilde{N}$}.

Represent the  compact in $L_2(\Gamma)$ relative ND-map
${\mathcal{ND}}^{\Gamma}_{int} =:{\mathcal{ND}}^{\Gamma}$ as
\[
{\mathcal{ND}}^{\Gamma}_{int} = \left(
\begin{array}{cc}
{\mathcal{ND}}_{+ +} & {\mathcal{ND}}_{+ -}\\
{\mathcal{ND}}_{- +}& {\mathcal{ND}}_{- -}
\end{array}
  \right)= \sum_{\lambda^N_s \in
\Delta_2}\frac{\psi_s \big|_{\Gamma}\rangle \langle \psi_s
\big|_{\Gamma}}{\lambda^N_s - \lambda}  +
\tilde{\mathcal{K}}^{\Delta_2} = : {\mathcal{ND}}^{\Delta_2}  +
\tilde{\mathcal{K}}^{\Delta_2},
\]
and  consider the corresponding matrices
\[
{\mathcal{ND}}^{\Delta_2} =  \left(
\begin{array}{cc}
{\mathcal{ND}}^{\Delta_2}_{+ +} & {\mathcal{ND}}^{\Delta_2}_{+ -}\\
{\mathcal{ND}}^{\Delta_2}_{- +}& {\mathcal{ND}}^{\Delta_2}_{- -}
\end{array}
  \right),\]
  and
  \[
 \tilde{\mathcal{K}}^{\Delta_2} =  \left(
\begin{array}{cc}
\tilde{\mathcal{K}}^{\Delta_2}_{+ +} & \tilde{\mathcal{K}}^{\Delta_2}_{+ -}\\
\tilde{\mathcal{K}}^{\Delta_2}_{- +}&
\tilde{\mathcal{K}}^{\Delta_2}_{- -},
\end{array} \right) =:
\left(
\begin{array}{cc}
\tilde{\mathcal{K}}_{+ +} & \tilde{\mathcal{K}}_{+ -}\\
\tilde{\mathcal{K}}_{- +}& \tilde{\mathcal{K}}_{- -},
\end{array} \right) =: \tilde{\mathcal{K}},
\]
in  the basis $E_{+},\, E_{-}$ of  $E = L_2(\Gamma)$. Hereafter we
omit the upper index $\Delta_2$ on matrix elements
$\tilde{\mathcal{K}}_{\pm\,\, \pm}$. To represent
${\mathcal{ND}}^{\Lambda} =: {\mathcal{N}}$ in the form with already
compensated singularities inherited from  the resolvent of $L^N$ ,
we  have to solve the equation
\begin{equation}
\label{Eq:1} \left( I + {\mathcal{ND}}_{--}\,K_- \right) u =
{\mathcal{ND}}_{-+}f
\end{equation}
Our second basic assumption is the  following:

2. {\it We  assume, that  the width $\delta$ of the leads, the
essential spectral interval $\Delta_T =: \Delta \subset\Delta_2$ and
the rational approximation $\tilde{\mathcal{K}}$ are selected such
that
\begin{equation}
\label{Neumann_thin}  I + \tilde{\mathcal{K}}^{\Delta_2}_{--}\,K_-
\end{equation}
is invertible  for $\lambda \in \Delta$}.

For  low temperature (that is for a relatively small essential
spectral interval  $\Delta_T$) this condition is equivalent to the
corresponding condition imposed just at the scaled Fermi level
$\Lambda$. It is satisfied, if
\begin{equation}
\label{Eq:N_thin} \mbox{sup}_{\lambda\in\Delta}
\parallel\tilde{\mathcal{K}}_{--}K_-
\parallel < 1.
\end{equation}

We  will not give here  a formal condition which guarantees 2, but
just notice that due to compactness of  the resolvent  on
$L^N_{\Gamma}$ for any $\Delta_T $ there exist ${\Delta_2}\supset
\Delta_T$ and the corresponding number $\tilde{N}_{\Delta_2} =:
\tilde{N}$ such that the error of the finite rational approximation
of  the resolvent is  small:
\[
G^N (x,s, \lambda) = G^N (x,s, \mu) + (\lambda -\mu )Q^{\Delta_2}_2
=
\]
\[
G^N (x,s, \mu) + (\lambda -\mu )\sum_{l=1}^{\tilde N}
\frac{\varphi_l(x)\rangle\,\langle\varphi_l(s)}{(\lambda_l -\mu)^2}
+ (\lambda -\mu )^2 \sum_{l=\tilde{N}+1}^{\infty}
 \frac{\varphi_l(x)\rangle\,\langle\varphi_l(s)}{(\lambda_l -\lambda)
(\lambda_l -\mu)^2} =
\]
\begin{equation}
\label{Eq:green} G^N (x,s, \mu) + \tilde{Q}_2 (x,s,\lambda,\mu)+
\tilde{\mathcal{K}}^{\Delta_2}(x,s,\lambda,\mu) =: \tilde{Q}
(x,s,\lambda,\mu) + \tilde{\mathcal{K}} (x,s,\lambda,\mu).
\end{equation}
Here we  choose $\mu$ large negative, so that $G^N (x,s, \mu)$ is a
kernel of a small integral operator and  denote hereafter
\[
G^N (x,s, \mu) + \tilde{\mathcal{K}}^{\Delta_2}(x,s,\lambda,\mu) =:
\tilde{\mathcal{K}} (x,s,\lambda,\mu),\,\,\tilde{Q}_2
(x,s,\lambda,\mu) = \tilde{\mathcal{ND}} (x,s,\lambda,\mu).
\]
Then  for the error $P_-{\tilde{\mathcal{K}}}P_-
=:\tilde{\mathcal{K}}_{--}$  of the rational approximation
$\tilde{Q}$ framed by the projections  onto $E_{-}$ the
corresponding estimate (\ref{Eq:N_thin})is valid. Denote
\[
\tilde{\mathcal{K}}_{++}-\tilde{\mathcal{K}}_{+-}K_-( I +
\tilde{\mathcal{K}}_{--}K_{-})^{-1}\tilde{\mathcal{K}}_{-+} =:
\tilde{\mathcal{N}}_{reg},
\]
\[
 \left\{ P_+ -
\tilde{\mathcal{K}}_{+-} K_-( I +
\tilde{\mathcal{K}}_{--}K_{-})^{-1}\right\} \tilde{\mathcal{T}}^+ =:
\tilde{\mathcal{J}}\tilde{\mathcal{T}}^+,
\]
\[
\tilde{\mathcal{T}} K_-( I +
\tilde{\mathcal{K}}_{--}K_{-})^{-1}\tilde{\mathcal{T}}^+ =:
V(\lambda)
\]
\[
L^{\Delta_2} - \lambda I^{\Delta_2} + V(\lambda)=:
L^{\Delta_2}(\lambda),
\]
\begin{equation}
\label{notations} \tilde{\mathcal{T}}\left\{P_+ -  K_-( I +
\tilde{\mathcal{K}}_{--}K_{-})^{-1}\tilde{\mathcal{K}}_{-+}\right\}
=:\tilde{\mathcal{T}} \tilde{\mathcal{J}}^+.
\end{equation}
\begin{theorem}
\label{T:comp_N} ({\bf Compensation of Singularities N}) {
\begin{equation}
\label{Eq:7} {\mathcal{N}} = {\mathcal{N}}_{reg} -
\tilde{\mathcal{J}}\tilde{\mathcal{T}}^+\frac{I}{L^{\Delta_2}(\lambda)}\tilde{\mathcal{T}}
\tilde{\mathcal{J}}^+
\end{equation}
}
\end{theorem}

{\it Proof} We treat the equation (\ref{Eq:1}) with use of the
Banach principle under assumption (\ref{Eq:N_thin}):
\begin{equation}
\label{Eq:2}
 K_- u + K_-( I + \tilde{\mathcal{K}}_{--}K_-)^{-1}
{\mathcal{ND}}^{\Delta_2}_{--}K_- u = K_-( I +
\tilde{\mathcal{K}}_{--}K_-)^{-1} {\mathcal{ND}}^{\Delta_2}_{-+}f.
\end{equation}
Recall  that ${\mathcal{ND}} = {\mathcal{ND}}^{\Delta_2} +\tilde
{\mathcal{K}}$ and notice that ${\mathcal{ND}}^{\Delta_2}$ is
connected with the part $L^{\Delta_2}$ of  $L^N$ in $E^{\Delta_2}$
as
\[
{\mathcal{ND}}^{\Delta_2} = \tilde{\mathcal{T}}^+
\frac{I}{L^{\Delta_2} - \lambda I^{\Delta}}\tilde{\mathcal{T}}.
\]
Then, denoting
\[
\frac{I}{L^{\Delta_2} - \lambda I^{\Delta_2}} \tilde{\mathcal{T}}K_-
u =:v
\]
we  rewrite the above equation (\ref{Eq:2}) as an equation for $v$
and obtain the solution of it  in terms of the inverse of the matrix
$L^{\Delta_2} - \lambda I^{\Delta_2} + {\mathcal{T}} K_-( I +
{\mathcal{K}}^{\Delta_2}_{--}\tilde{K_{-+}})^{-1}\tilde{\mathcal{T}}
=: L^{\Delta_2}(\lambda) $:
\[
v = [L^{\Delta_2}(\lambda) ]^{-1} V \frac{I}{L^{\Delta_2} - \lambda
I^{\Delta_2}} \tilde{\mathcal{T}}P_+ f.
\]
Substituting the result into (\ref{Eq:2})
\[
K_- u =  K_-( I +
\tilde{\mathcal{K}}_{--}K_{-})^{-1}\left\{\left[\tilde{\mathcal{T}}^+
\frac{I}{L^{\Delta_2} - \lambda I^{\Delta_2}} \tilde{\mathcal{T}} +
\tilde{\mathcal{K}}_{- +}\right]P_+ f -\right.
\]
\[
\left. \tilde{\mathcal{T}}^+
[L^{\Delta_2}(\lambda)]^{-1}\tilde{\mathcal{T}} K_-( I +
\tilde{\mathcal{K}}_{--}K_{-})^{-1}\left[\tilde{\mathcal{T}}^+
\frac{I}{L^{\Delta_2} - \lambda I^{\Delta_2}} \tilde{\mathcal{T}} +
\tilde{\mathcal{K}}_{-+}\right]\right. P_+ f.
\]
Then $ {\mathcal{N}}f = $
\[ P_+ \tilde{\mathcal{T}}^+
\frac{I}{L^{\Delta_2} - \lambda I^{\Delta_2}}\tilde{\mathcal{T}}P_+
f  + \tilde{\mathcal{K}}_{++} f - P_+
\left[\tilde{\mathcal{T}}^+\frac{I}{L^{\Delta_2} - \lambda
I^{\Delta_2}}\tilde{\mathcal{T}}P_- +
\tilde{\mathcal{K}}_{+-}\right]\times
\]
\[
 K_-( I +
\tilde{\mathcal{K}}_{--}K_{-})^{-1}\left\{\left[\tilde{\mathcal{T}}^+
\frac{I}{L^{\Delta_2} - \lambda I^{\Delta_2}} \tilde{\mathcal{T}} +
\tilde{\mathcal{K}}\right]P_+ f -\right.
\]
\[
\left.
 \tilde{\mathcal{T}}^+
 [L^{\Delta_2}(\lambda)]^{-1}\tilde{\mathcal{T}} K_-( I +
\tilde{\mathcal{K}}_{--}K_{-})^{-1}\left[\tilde{\mathcal{T}}^+
\frac{I}{L^{\Delta_2} - \lambda I^{\Delta_2}} \tilde{\mathcal{T}} +
\tilde{\mathcal{K}}_{-+}\right]P_+ f\right\}.
\]
Leading terms  inside  parentheses  give:
\[
 K_-( I +
\tilde{\mathcal{K}}_{--}K_{-})^{-1} \tilde{\mathcal{T}}^+ \left[
\frac{I}{L^{\Delta_2} - \lambda I^{\Delta_2}} -
\frac{I}{L^{\Delta_2}(\lambda)} V \frac{I}{L^{\Delta_2} - \lambda
I^{\Delta_2}}\right] =
\]
\begin{equation}
\label{Eq:3} K_-( I + \tilde{\mathcal{K}}_{--}K_{-})^{-1}
\tilde{\mathcal{T}}^+ \frac{I}{L^{\Delta_2}(\lambda)}.
\end{equation}
Taking into account the  leading term of the first addendum, we
obtain:
\[
P_+ \tilde{\mathcal{T}}^+ \frac{I}{L^{\Delta_2} - \lambda
I^{\Delta_2}}\tilde{\mathcal{T}} P_+ f - \tilde{\mathcal{T}}^+
\frac{I}{L^{\Delta_2} - \lambda I^{\Delta_2}}V\,
\frac{I}{L^{\Delta_2}(\lambda)}(\lambda)\tilde{\mathcal{T}} P_+ f =
\]
\begin{equation}
\label{Eq:4} P_+ \tilde{\mathcal{T}}^+
\frac{I}{L^{\Delta_2}(\lambda)}\tilde{\mathcal{T}} P_+ f.
\end{equation}
Lower  order  terms containing $\frac{I}{L^{\Delta_2} - \lambda
I^{\Delta_2}}$ in parentheses give:
\[
 -\tilde{\mathcal{K}}_{+-} K_-( I +
\tilde{\mathcal{K}}_{--}K_{-})^{-1} \tilde{\mathcal{T}}^+ \left[
\frac{I}{L^{\Delta_2} - \lambda I^{\Delta_2}} -
\frac{I}{L^{\Delta_2}(\lambda)} V \right] \tilde{\mathcal{T}} P_+ f
=
\]
\begin{equation}
\label{Eq:5} -\tilde{\mathcal{K}}_{+-} K_-( I +
\tilde{\mathcal{K}}_{--}K_{-})^{-1} \tilde{\mathcal{T}}^+
\frac{I}{L^{\Delta_2}(\lambda)}\tilde{\mathcal{T}} P_+ f,
\end{equation}
and the adjoint expression.  The  term which contains only the  main
singularity $[L^{\Delta_2}(\lambda)]^{-1}$ is
\[
\tilde{\mathcal{K}}_{+-} K_-( I +
\tilde{\mathcal{K}}_{--}K_{-})^{-1}\,\,\tilde{\mathcal{T}}^+
\frac{I}{L^{\Delta_2}(\lambda)}\tilde{\mathcal{T}}\,\, K_-( I +
\tilde{\mathcal{K}}_{--}K_{-})^{-1}\tilde{\mathcal{K}}_{-+}
\]
The terms which do not contain singularities result in:
\begin{equation}
\label{Eq:reg} \tilde{\mathcal{K}}_{++}-\tilde{\mathcal{K}}_{+-}K_-(
I +
\tilde{\mathcal{K}}_{--}K_{-})^{-1}\tilde{\mathcal{K}}^{\Delta}_{-+}
=: {\mathcal{N}}_{reg}
\end{equation}
Note that the operator $K_-( I +
\tilde{\mathcal{K}}_{--}K_{-})^{-1}$ is  self-adjoint on the first
spectral band. Then, collecting  all terms we  obtain the  announced
result.

{\it The end of the proof}

 Based  on the theorems \ref{T:comp_M},
\ref{T:comp_N} we can calculate the scattering matrix either in the
form (\ref{Eq:smatrix_M}) or in the form (\ref{Eq:smatrix_N}). One
of these formulae can be more  convenient  on the
 essential spectral interval, than another, depending on
 localization of  singularities  of  ${\mathcal{DN}}^{\Lambda} $ and
 ${\mathcal{ND}}^{\Lambda} $. Luckily, due to to
 ${\mathcal{DN}}^{\Lambda}\,\,\,
 {\mathcal{ND}}^{\Lambda} =  I_+ $, the singularities of  the factors
do not overlap, hence for any point $\lambda_0\in \Delta$ one can
select an interval centered at $\lambda_0$ where at least one of the
factors ${\mathcal{DN}}^{\Lambda} $ or ${\mathcal{ND}}^{\Lambda} $
can be substituted by the corresponding approximate expression based
on the bi-linear formulae suggested in \cite{PRRS08}.
 \vskip0.3cm

\section{Approximate Scattering Matrix and  the  boundary
condition at the  vertex of the quantum graph}

The standard method of calculation of the scattering matrix requires
solving of an infinite algebraic system anyway, though practically
admits a certain simplification in closed channels, see \cite{ML71}
\footnote{  The author is grateful to V. Katsnelson for important
comments in that connection.}, where it was done for  waveguides
with simple geometry. The approach based on the intermediate DN-map
gives a finite linear system for the Scattering matrix derived from
matching of the component  $\Psi_+ $ of the scattering Ansatz  on
the first (open) channel in wires with $p = \sqrt{\lambda -
V_{\delta} - \frac{\pi^2}{\delta^2}}$, on the first spectral band
$\Delta_1$:
\begin{equation}
\label{SansatzF}\Psi_+\, e_{_{+}} :=  e^{^{i p\xi}} e_{_{+}} +
e^{^{- i p\xi}} S(p) e_{_{+}}
\end{equation}
to the limit values on the spectrum, $\Im \lambda \to 0$, of the
solution of an intermediate boundary problem  with the boundary data
on $\Gamma$ defined by the  scattering Ansatz $\Psi$. The boundary
data $\Psi\big|_{\Gamma},\, \frac{\partial \Psi}{\partial
n}\big|_{\Gamma}$ are connected by the  intermediate DN-map, hence
the required matching gives a finite linear system for $S$:
\begin{equation}
\label{F_matching} ip\left[ e_{_{+}} - S(p) e_{_{+}}\right] =
{\mathcal{M}} (\lambda)\, \left[e_{_{+}} +  S(p) e_{_{+}}\right].
\end{equation}
Solving this  equation we  obtain  the formula for the  scattering
matrix  of the operator $\mathcal{L}$ on the first spectral band
$\Delta_1 $ in terms of $\mathcal{L}_{\Lambda}$ by the formula, see
(\ref{Eq:smatrix_M}). Due to Theorem \ref{T:1} one can substitute,
for a thin junction, the intermediate DN-map  ${\mathcal{M}}
(\lambda)$ in (\ref{Eq:smatrix_M}) by  an approximate  expression
for ${\mathcal{M}} =
 {\mathcal{M}}^{\Delta} + {\mathcal{K}}^{\Delta}$, where
${\mathcal{M}}^{\Delta} =: {\mathcal{M}}_{approx}$ is a rational
approximation ${\mathcal{DN}}^{\Delta}$ of  ${\mathcal{M}}$ on the
essential spectral interval $\Delta_T$, containing only polar terms
with poles on an auxiliary spectral interval $\Delta$, and
${\mathcal{K}}^{\Delta}$ is a regular part of ${\mathcal{M}}$ on
$\Delta$.

\begin{theorem}\label{T:2}
{ The resulting approximate  expression for the  scattering matrix
\begin{equation}
\label{T:sapprox}
 S\approx [ip P_+ + {\mathcal{M}}^{\Delta}]^{-1}
 [ip P_+ - {\mathcal{M}}^{\Delta}]=: S_{approx},
\end{equation}
with $p = \sqrt{\lambda -V_{\delta} - \pi^{2}\,\,\delta^{-2}}$, can
be  used as a  first step for the calculation of the exact
scattering matrix  via an analytic perturbation procedure.}
\end{theorem}

{\it Proof}\,\,\, Indeed, due to the above theorem \ref{T:1} the
error ${\mathcal{M}} - {\mathcal{M}}_{approx}
={\mathcal{K}}^{\Delta} $, with ${\mathcal{K}}^{\Delta}$  containing
the regular term ${\mathcal{M}}_{reg}$ too, is real and estimated by
$O (\delta \, d^{-2}_{int})$. Then, due to \ref{T:comp_M} we can
represent the exact scattering matrix in form of a product:
\begin{equation}
\label{S_anal_approx} S = \left( I + [ipP_+ +
{\mathcal{M}}^{\Delta}]^{-1} {\mathcal{K}}^{\Delta }\right)^{-1}\,\,
S_{approx} \,\, \left( I - [ipP_+ - {\mathcal{M}}^{\Delta}]^{-1}
{\mathcal{K}}^{\Delta}\right).
\end{equation}
Here  ${\mathcal{M}}^{\Delta},\, {\mathcal{K}}^{\Delta } $ are
hermitian on $\Delta_1$, hence  $\parallel [ipP_+ \pm
{\mathcal{M}}^{\Delta}]^{-1}\parallel << \delta$, for thin
junctions. Hence the analytic perturbation procedure of the
calculation the left and right factors of the  expression in
(\ref{S_anal_approx}) is geometrically convergent due to
$\delta\,\,\, O(\delta \, d^{-2}_{int}) << 1$. Thus  the exact
scattering matrix can be obtained  from  $S_{approx}$ by an analytic
perturbation procedure.

{\it The end of the proof}

One can construct various approximations for the  scattering matrix
based on  $S_{approx} $, replacing $M$ by various approximate
expressions, with controllable errors, see  for instance
(\ref{M_approx},\ref{M_thin}).

\subsection{ Simple resonance eigenvalue of the Intermediate Hamiltonian }
\begin{figure} [ht]
\begin{center}
\includegraphics [width=1.5in] {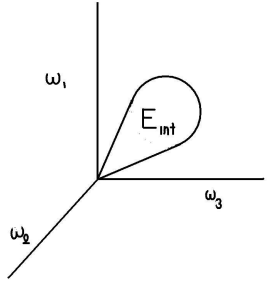}
\end{center}
\caption{1-d model of T-junction} \label{F:figure 8}
\end{figure}

The  simplest approximate formula for the  scattering matrix  can be
obtained in the case when there exist a single simple  eigenvalue
$\lambda^{\Lambda}_1$ of the  Intermediate Hamiltonian on  the
auxiliary spectral interval $\Delta$. Indeed, substituting the
intermediate DN-map  ${\mathcal{M}} = {\mathcal{K}}^{\Lambda} +
\alpha_1^2\,\,\frac{P_1^\Lambda}{\lambda -\lambda_1^\Lambda}$  by
the corresponding polar approximation  generated  by  the resonance
eigenvalue  $\lambda_1^{\Lambda}$ of the Intermediate Hamiltonian
and the boundary current of the corresponding  eigenfunction
\[
\alpha_1^2 P^\Lambda_1 = P_+ \frac{\partial
\varphi^{\Lambda}_1}{\partial n}\bigg|_{\Gamma}\rangle \,\,\langle
P_+ \frac{\partial \varphi^{\Lambda}_1}{\partial n}\bigg|_{\Gamma}
=: \vec{\psi^{\Lambda}_1}\rangle \,\langle \vec{\psi^{\Lambda}_1},
\]
we are able to obtain, due to preceding  theorem (\ref{T:2}), the
scattering matrix of a thin junction via an analytic perturbation
procedure based on  the jump-start
\begin{equation}
\label{jump-start} S_{jump-start} = \left[iK_+ + k(\lambda)+
\alpha_1^2\frac{P_1^\Lambda}{\lambda
-\lambda_1^\Lambda}\right]^{-1}\left[iK_+ -  k (\lambda)-
\alpha_1^2\frac{P_1^\Lambda}{\lambda -\lambda_1^\Lambda}\right].
\end{equation}
In the case when  the  first spectral band  $\Delta_1$ is the
conductivity band , the above approximate expression for the
scattering matrix can be represented, with $P_1^{\bot} = P_+ \ominus
P^\Lambda_1$ and $p_1 = \sqrt{\lambda - \pi^2 \delta^{-2} -
V_{\delta}} $ and $\lambda^\Lambda_1\approx \Lambda$ as:
\begin{equation}
\label{S_jump_start} S_{jump-start} =
 P_1^{\bot} + \frac{ip_1 - k (\lambda) - \alpha^2_1 \frac{1}{\lambda -\lambda^\Lambda_1}
}{ip_1 +  k (\lambda) + \alpha^2_1 \frac{1}{\lambda
-\lambda^\Lambda_1} } P^\Lambda_1.
\end{equation}
It corresponds  to the one-dimensional solvable model of the
junction, obtained  via  attachment an appropriate inner structure
to the vertex, see Fig. \ref{F:figure 8}.
This approximate  expression for the  scattering matrix can be
obtained via  imposing  on the  Scattering Ansatx  a
$\lambda$-dependent boundary condition at the vertex, see a
discussion in \cite{MathNach07}. Unfortunately  this boundary
condition does not correspond to  a self-adjoint operator, so that
it can't be interpreted  in terms of  Quantum Mechanics, the same as
prominent Wigner boundary condition, see \cite{Wigner51}.  We are
also able to represent the scattering matrix with use of a s1ngle
Blaschke-factor: $ S_{approx_1}(\lambda) = $
\begin{equation}
\label{approxS} P^{{\bot}}_{1} + \left[\frac{ip(\lambda -
\lambda^\Lambda_{1}) - \alpha^2_1 }{ip (\lambda -
\lambda^\Lambda_{1}) + \alpha^2_1}\right] P^\Lambda_{1}
 \equiv P^{{\bot}}_{1} + \Theta^\Lambda_1(\lambda) \, P^\Lambda_{1}.
\end{equation}
Notice that  the scalar Blaschke-factor  $\Theta^\Lambda_1$  is
close to $ -1$ on the  essential spectral interval
\[
\Delta_T : \left\{ \lambda :|\lambda - \lambda^\Lambda_{1}|\leq
2m^*\, \kappa T \,\hbar^{-2} < \alpha^2_1 \,\, p^{-1}
(\lambda^\Lambda_1)\right\},
\]
for low temperature $ T$, and it is close to $1$ on  the complement.
For thin junction and low temperature the boundary condition can be
reduced to Datta-type boundary condition,\cite{DattaAPL}, see below,
formula (\ref{Datta_1}) represented  in terms of  boundary currents
of the resonance eigenfunction of the  Intermediate Hamiltonian.

\begin{figure}
[ht]
\begin{center}
\includegraphics [width=3in] {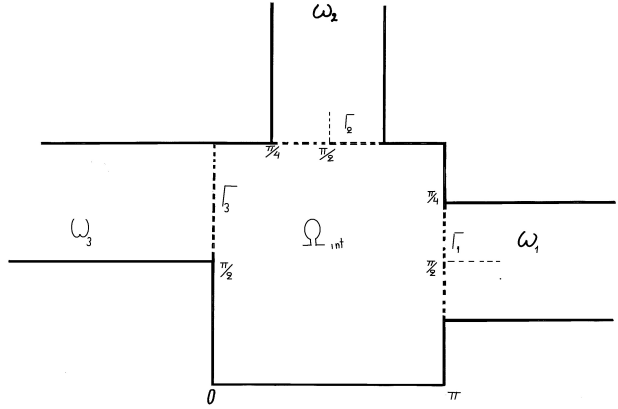}
\caption{ Simplest asymmetric T-junction} \label{F:figure 9}
\end{center}
\end{figure}

Once we already developed the  compensation procedure based on the
representation of  the  Intermediate  DN-map in terms of classical
DN-map,  we  can do one more step, expressing the approximate
scattering matrix  (\ref{jump-start}) in spectral terms of the
unperturbed  operator  $L_{int}$ on the vertex domain
$\Omega_{int}$, under assumption that it has a single resonance
eigenvalue  $\lambda_1 \in \Delta_T \subset \Delta$,
\[
L_{int}\varphi_1 = \lambda_1 \varphi_1.
\]
Then, for thin junction, the  intermediate Hamiltonian also has a
simple eigenvalue  near to $\lambda_1$.  We assume that the  major
part  of the correcting term ${\mathcal{K}}^{\Delta}$ in the
corresponding rational approximation of the  DN-map of $L_{int}$ is
defined  by a  finite sum of polar terms and a regular term
${\mathcal{M}}_{reg} = {\mathcal{K}}^{\Delta}_{++} -
{\mathcal{K}}^{\Delta}_{+-} \frac{I}{{\mathcal{K}}^{\Delta}_{--} +
K_-} {\mathcal{K}}^{\Delta}_{-+}\,\,$, see  (\ref{T:comp_M})
\[
{\mathcal{K}}^{\Delta} = \sum_{s=2}^{s=M} \frac{I}{\lambda
-\lambda_s}\frac{\partial \varphi_s}{\partial n} \rangle \langle
\frac{\partial \varphi_s}{\partial n} + {\mathcal{M}}_{reg},
\]
\[
{\mathcal{DN}}(\lambda) = \frac{I}{\lambda -\lambda_1}\frac{\partial
\varphi_1}{\partial n} \rangle \langle \frac{\partial
\varphi_1}{\partial n} + \sum_{s=2}^{s=M} \frac{I}{\lambda
-\lambda_s}\frac{\partial \varphi_s}{\partial n} \rangle \langle
\frac{\partial \varphi_s}{\partial n} =:  {\mathcal{DN}}^{\Delta} +
{\mathcal{K}}^{\Delta},
\]
\[
{\mathcal{T}} = \varphi_1 \rangle \langle \frac{\partial
\varphi_1}{\partial n},\,\,\, {\mathcal{J}} = P_+ -
{\mathcal{K}}_{+-}K_-^{-1} \left[I +
{\mathcal{K}}_{--}K_-^{-1}\right]^{-1} P_-.
\]
Hereafter we neglect the contribution from  higher terms of the
geometrically convergent series
\[
\left[I + {\mathcal{K}}_{--}K_-^{-1} \right]^{-1} P_- \approx I_- =
P_- ;
\]
\[
{\mathcal{J}} = P_+ - {\mathcal{K}}_{+-}K_-^{-1}\approx
 P_+ - \sum_{s=2}^{s=M} \frac{I}{\lambda
-\lambda_s}\,P_+\frac{\partial \varphi_s}{\partial n} \rangle
\langle \,\, K_-^{-1}\,\,\frac{\partial \varphi_s}{\partial n} .
\]
\[
Q(\lambda) = \varphi_1 \rangle \langle \frac{\partial
\varphi_1}{\partial n} K_-^{-1} \left[I + {\mathcal{K}}_{--}K_-^{-1}
P_-\right]^{-1} \frac{\partial \varphi_1}{\partial n} \rangle
\langle \varphi_1 \approx \varphi_1 \rangle \langle \varphi_1\,\,\,
\langle \frac{\partial \varphi_1}{\partial n} K_-^{-1}\frac{\partial
\varphi_1}{\partial n} \rangle.
\]
Then, with only  terms containing  $K_-^{-1}$  taken into account ,
we  obtain for ${\mathcal{M}}$, based  on Theorem \ref{T:comp_M}, an
approximate expression for
\[
{\mathcal{M}}_{approx} ={\mathcal{K}}^{\Delta}_{++} -
{\mathcal{K}}^{\Delta}_{+-}K_-^{-1} {\mathcal{K}}^{\Delta}_{-+}\,\,
+
\]
\begin{equation}
\label{M_approx}
 \left(P_+ - {\mathcal{K}}_{+-}K_-^{-1}\right)
\frac{\partial \varphi_1}{\partial n}\bigg|_{\Gamma}\rangle
\frac{I}{\lambda - \lambda_1 + \langle \frac{\partial
\varphi_1}{\partial n}\bigg|_{\Gamma} K_-^{-1}\frac{\partial
\varphi_1}{\partial n}\bigg|_{\Gamma}\rangle} \,\,\langle \,
\left(P_+ - {\mathcal{K}}_{+-}K_-^{-1}\right) \frac{\partial
\varphi_1}{\partial n}\bigg|_{\Gamma}.
\end{equation}
For ``very thin'' junction one can neglect even first order terms
containing $K_-^{-1}$, everywhere, except the expression staying in
the denominator, and obtain from (\ref{M_approx}) a simpler
approximate formula:
\[
 {\mathcal{M}}_{thin} =
{\mathcal{K}}^{\Delta}_{++} +
 P_+  \frac{\partial \varphi_1}{\partial
n}\bigg|_{\Gamma}\rangle \frac{I}{\lambda - \lambda_1 + \langle
\frac{\partial \varphi_1}{\partial n}\bigg|_{\Gamma}
K_-^{-1}\frac{\partial \varphi_1}{\partial n}\bigg|_{\Gamma}\rangle}
\,\,\langle \, P_+ \frac{\partial \varphi_1}{\partial
n}\bigg|_{\Gamma} =:
\]
\begin{equation}
\label{M_thin}
 {\mathcal{K}}^{\Delta}_{++} +
\alpha_1^2\frac{P_1^Q}{\lambda -\lambda_1^Q}
\end{equation}
where \[ P^Q_1 = e^Q_1\rangle \,\langle e^Q_1,\, e^Q_1 =
\alpha_1^{-1} P_+ \frac{\partial \varphi_1}{\partial
n}\bigg|_{\Gamma},
\]
 \[ \alpha_1 =
\parallel P_+ \frac{\partial \varphi_1}{\partial
n}\bigg|_{\Gamma}\parallel,\,\, \lambda_1^Q = \lambda_1 - \langle
\frac{\partial \varphi_1}{\partial n}\bigg|_{\Gamma}
K_-^{-1}\frac{\partial \varphi_1}{\partial
n}\bigg|_{\Gamma}\rangle.\]
This implies an approximate formula for
the  scattering matrix on the major part of the essential spectral
interval $\Delta_T$ \footnote{Roughly speaking, on a  complement of
a certain small neighborhood of  the zero $\lambda_1^Q $ of the
denominator. }, for low temperature, once the junction is
 {\it thin on the open channel}
\begin{equation}
\label{Thin_open} p^2(\Lambda) = |\Lambda - (\pi^2\,\delta^{-2} +
V_{\delta}) |
>> \parallel K^{\Delta}| \parallel.
\end{equation}
Notice that  arising of  the shape of the resonance eigenfunction of
the unperturbed operator $L_{int}$ in the corresponding approximate
jump-start formula (\ref{S_thin}) for the  scattering matrix
corresponds  to  folklore observation of physicists, that the
eigenfunctions react to perturbation  slower that the eigenvalues.
To derive the jump-start approximation for the  scatterin matrix,
denote $\sqrt{\lambda - \pi^2\,\delta^{-2} - V_{\delta}} =: p $.
Then we obtain on $\Delta_T$, similarly  to Theorem \ref{T:2}:
\begin{equation}
\label{S_thin} S (\lambda) \approx  P_1^{\bot}  +  \frac{i p - k -
\frac{\alpha_1^2}{\lambda-\lambda^Q_1}}{i p + k +
\frac{\alpha_1^2}{\lambda-\lambda^Q_1}} P_1 =: S^{Q}_{jump-start}.
\end{equation}
In the case when $\lambda^Q_1\approx \Lambda$  we  can replace on
$\Delta_T$ the Blaschke factor in front of $ P_1$ by  $-1$, which
implies on $\Delta_T$, for sufficiently low temperature:
\begin{equation}
\label{S_Thin_low} S^{Q}_{jump-start} (\lambda) \approx P_1^{\bot}
-  P_1 \equiv S_{Datta}
\end{equation}
When modeling the quantum network by  a one-dimensional graph, one
can attempt to define a boundary condition at the vertex which
implies  the scattering matrix  (\ref{S_Thin_low}). Indeed, forming
the component $\Psi_+$ of the scattering Ansatz  based on
(\ref{S_Thin_low}), we  see the the boundary values of the Ansatz
\[
\Psi_+ (x)\nu =  e^{iK_+ x} \nu  +  e^{-iK_+ x}\,\,S \nu
\]
satisfy the following boundary condition, similar \footnote{but not
yet equivalent, see an extended  discussion below, in next
subsection.} to one suggested in \cite{DattaAPL}:
\[
\Psi_+ (0)\,\,\nu =  2 P_1^{\bot} \nu,\,\,\frac{d}{dx}\,\Psi_+
(0)\nu = 2 P_1 \nu.
\]
In terms of the components $\Psi^n_+ (0)\nu,\,\frac{d}{dx}\,\Psi_+^n
(0)\nu $ of the boundary values of the Ansatz on the  bottom
sections $\Gamma_n$ of the wires and with use of the components of
the boundary currents $\vec{\psi}_1 = \left\{ \psi^n_1
\right\}_{n=1}^N$
\begin{equation}
\label{bound_currents} P_+ \frac{\partial \varphi_1}{\partial
n}\bigg|_{\Gamma_n} \equiv \psi^n_1
\end{equation}
we  obtain: $\langle \vec{\psi},\Psi_+(0)\nu\rangle = 0,\,
\frac{d}{dx} \Psi_+(0) \nu  \parallel \vec{\psi}$, or
\[
\langle \vec{\psi},\Psi_+ (0)\nu \rangle = \sum_{n}^N\Psi_+^n
\,\,\nu\,\, \bar{\psi^n} = 0, \]
\begin{equation}
\label{Datta_1}
 \frac{\frac{d}{dx}\,\Psi_+^1 (0)\nu}{\psi^1_1} =
\frac{\frac{d}{dx}\,\Psi_+^2 (0)\nu}{\psi^2_1}= \dots =
\frac{\frac{d}{dx}\,\Psi_+^n (0)\nu}{\psi^n_1} = \dots=
\frac{\frac{d}{dx}\,\Psi_+^N (0)\nu}{\psi^N_1}
\end{equation}
  \vskip0.3cm
{\bf Example 4: asymmetric T-junction} Consider a  two-dimensional
quantum network  $\Omega$ constructed as a simplest asymmetric
T-junction of three straight semi-infinite quantum wires width
$\pi/2$ attached as shown in Fig. \ref{F:figure 9} to the quantum
well - the square $\Omega_{int}: 0< x< \pi,\, 0< y< \pi$ on
$x,y$-plane.
The role of the one-electron Hamiltonian on $\Omega$ is played the
Dirichlet Laplacian. The spectrum of the corresponding unperturbed
Hamiltonian on the quantum well is discrete and the eigen-pairs e.g.
\[
 \lambda_1 = 2:  \varphi_{1,1} = \frac{2}{\pi} \sin x  \times \sin y;
\]
\vskip0.3cm
\[
\lambda_2 = \lambda_3 = 5;  \varphi_{1,2} = \frac{2}{\pi}  \sin x
\,\times\,\sin 2y \,\,\mbox{and}\,\, \varphi_{2,1} = \frac{2}{\pi}
\sin 2x \,\times\,\sin y;
\]
\[
P_5 =  P_{1,2} +  P_{ 2,1},\, P_{1,2} = \varphi_{1,2} \rangle
\,\,\langle \varphi_{1.2}: P_{ 2,1} = \varphi_{2,1} \rangle
\,\,\langle \varphi_{2,1};
\]
\vskip0.3cm
\[
\lambda_4 = 8; \varphi_{2,2} = \frac{2}{\pi} \sin 2x  \times \sin
2y; P_8 = P_{2,2} = \varphi_{2,2} \rangle \,\langle \varphi_{2,2};
\]
\vskip0.3cm
\[
\lambda_5 = \lambda_6 = 10; \varphi_{1,3} = \frac{2}{\pi}  \sin x
\,\times\,\sin 3y \,\,\mbox{and}\,\, \varphi_{3,1} = \frac{2}{\pi}
\sin 3x \,\times\,\sin y;\dots
\]
\vskip0.3cm
\[
\lambda_6 = \lambda_7 = 13; \varphi_{2,3} = \frac{2}{\pi} \sin 2x
\,\times\,\sin 3y \,\,\mbox{and}\,\, \varphi_{3,2} = \frac{2}{\pi}
\sin 3x \,\times\,\sin 2y;\dots
\]
\vskip0.3cm
\begin{equation}
\label{eigenfunctions} \dots  \dots  \dots
\end{equation}
are obtained  via separation of variables. We choose the basic
spectral  interval $\Delta =  [4,6]$, so that there is only one
multiple eigenvalue  $\lambda_2= \lambda_3 =  5$ of $L_{int}$ on
that interval, and use the approximate  formula (\ref{M_thin}) for
${\mathcal{M}}$. The
 normal boundary current $J_{1,2}$ of the resonance eigenfunctions $\varphi_{1,2}$
 and $\varphi_{2,1}$ is calculated  as
\[
 J_{1,2} =
 \left(
\begin{array}{c}
 \frac{\partial
\varphi_{1,2}}{\partial n}\bigg|_{\Gamma_1}\\
 \frac{\partial
\varphi_{1,2}}{\partial n}\bigg|_{\Gamma_2}\\
 \frac{\partial
\varphi_{1,2}}{\partial n}\bigg|_{\Gamma_3} \end{array}
 \right) =
 \left(
 \begin{array}{c}
- \frac{2}{\pi}\,\, \sin 2y \\\\
\frac{4}{\pi}\sin x\\\\
- \frac{2}{\pi}\,\, \sin 2y
\end{array}
 \right),\,\,
 \]
 \begin{equation}
\label{res_bound_current}
 J_{2,1} =
 \left(
\begin{array}{c}
 \frac{\partial
\varphi_{2,1}}{\partial n}\bigg|_{\Gamma_1}\\
 \frac{\partial
\varphi_{2,1}}{\partial n}\bigg|_{\Gamma_2}\\
 \frac{\partial
\varphi_{2,1}}{\partial n}\bigg|_{\Gamma_3} \end{array}
 \right) =
 \left(
 \begin{array}{c}
\frac{4}{\pi}\,\, \sin y \\\\
- \frac{2}{\pi}\sin 2x\\\\
- \frac{4}{\pi}\,\, \sin y
\end{array}
 \right).
\end{equation}
The spectrum of  the unperturbed Hamiltonian on the  wires is
absolutely- continuous and has  a band structure, with thresholds  $
\left\{l^2\right\},\, l=1,2,3\dots$ separating the spectral bands.
The multiplicity of the continuous spectrum jumps up by three units
on each threshold. We assume that  the first spectral band $\Delta_1
= [4,16] $ is  the conductivity band. The entrance subspace of the
open channel is spanned  by the inferior cross-section
eigenfunctions $e^{\bot}_{s,1} = (4/\pi)^{1/2}\sin 2 x^{\bot}_s,\,
s= 1,2,3$ on the bottom cross-section $\Gamma_s = [0< x^{\bot}_s <
\pi/2], s= 1,2,3$. The entrance subspace of closed channels is the
linear hull of the superior cross-section eigenfunctions
$e^{\bot}_{s,l} = (4/\pi)^{1/2}\sin 2 l\,x^{\bot}_s,\, s= 1,2,3,\,$
with $ l= 2,3,\dots$. We  will calculate the approximate scattering
matrix  (jump-start) of the thin junction based on the approximate
formulae for (\ref{M_approx}, \ref{M_thin}). The approximate
eigenvalues of the intermediate Hamiltonian are found as  zeros of
the denominator ${\bf d}(\lambda)$ represented as a $2\times 2$ -
matrix with respect to the basis $\varphi_{1,2},\varphi_{2,1}$;
\[
 {\bf
d}(\lambda) = \left(
\begin{array}{cc}
 \lambda - 5 & 0\\
0&\lambda - 5
\end{array}
\right) +
\]
\begin{equation}
\label{denominator}
\left(
\begin{array}{cc} \langle P_-\frac{
\partial \varphi_{1,2}}{\partial n}\big|_{\Gamma}\, K^{-1}_-\, P_-
\frac{\partial \varphi_{1,2}}{\partial n}\big|_{\Gamma}\rangle &
\langle P_- \frac{
\partial \varphi_{1,2}}{\partial n}\big|_{\Gamma}\, K^{-1}_-\, P_-
\frac{
\partial \varphi_{2,1}}{\partial n}\big|_{\Gamma}\rangle\\
\langle P_- \frac{
\partial \varphi_{2,1}}{\partial n}\big|_{\Gamma}\, K^{-1}_-\, P_-
\frac{\partial \varphi_{1,2}}{\partial n}\big|_{\Gamma}\rangle &
\langle P_- \frac{
\partial \varphi_{2,1}}{\partial n}\big|_{\Gamma}\, K^{-1}_-\, P_-
\frac{\partial \varphi_{2,1}}{\partial n}\big|_{\Gamma}\rangle
\end{array}
\right).
\end{equation}
Taking into account only components of the currents in the second
spectral channel
\[
K_- \frac{\partial \varphi_{1,2}}{\partial n}\bigg|_{\Gamma} \approx
e_{s,2} \langle \frac{\partial \varphi_{1,2}}{\partial
n}\bigg|_{\Gamma}, e_{s,2}\rangle
\]
and  introducing  the following notations for the integrals
\[
2 \int_0^{\pi/4} \sin 2x \sin 4x dx = 2/3 =: \alpha,\,\,
\]
\[
\int_0^{\pi/2} \sin x \sin 4x dx = - 4/15 =: \gamma,\,\, \alpha +
\gamma = 2/5 = :\beta,
\]
we represent the denominator (\ref{denominator}) and the  inverse
$[{\bf d}(\lambda)]^{-1}$  as
\[
{\bf
d}(\lambda) = \left(
\begin{array}{cc}
\lambda - 5& 0\\
0& \lambda - 5
\end{array}
\right) + \frac{\pi}{4 \sqrt{16 -\lambda}} \left(
\begin{array}{cc}
\alpha^2 & - \alpha \,\beta\\
- \alpha \,\beta & \beta^2
\end{array}
\right) =
 \]
\[
( \lambda - 5)\frac{1}{\alpha^2 + \beta^2} \left(
\begin{array}{c}
\beta\\
\alpha
\end{array}
\right)\rangle \,\, \langle \left(
\begin{array}{c}
\beta\\
\alpha
\end{array}
\right)
 +
 \]
 \[
 \left( \lambda - 5 + [\alpha^2 + \beta^2] \frac{\pi}{4  \sqrt{16
 -\lambda}}\right)\, \frac{1}{\alpha^2 + \beta^2}
 \left(
\begin{array}{c}
-\alpha\\
\beta
\end{array}
\right)\rangle \,\, \langle \left(
\begin{array}{c}
- \alpha\\
\beta
\end{array}
\right),
\]
\begin{equation}
\label{denominator_1} [{\bf d}(\lambda)]^{-1} =:
\frac{P_{5}}{\lambda - 5} + \frac{P_{5-\delta^Q}}{\lambda - 5 +
\delta^Q},
\end{equation}
with $\delta^Q  = \pi\,\, [\alpha^2 + \beta^2]/4 \,\,\,\sqrt{16
-\lambda} $. Here  $K_-$ is substituted  by  the contribution  $
4/\pi \,\ \sin 4x^{\bot} \rangle\,\, \langle \sin 4x^{\bot}$ from
the second spectral branch in the wires.

If the scaled Fermi-level is 5, then the corresponding multiple
resonance eigenvalue of $L_{int}$ is split into  pair  of
eigenvalues $\lambda^{\Lambda}_1 = 5,\,\lambda^{\Lambda}_2 \approx 5
- \frac{\pi\,\, [\alpha^2 + \beta^2]}{4  \sqrt{16
 - 5}}$, and  the jump-start approximation of the  scattering matrix
  can be calculated in terms
 of $P_+$ - projections of   the boundary currents of the  resonance
eigenfunctions  $\varphi_{1,2}, \varphi_{2,1} := $
\[
 P_+ \frac{\partial \varphi_{1,2}}{\partial n}\,
\bigg|_{\Gamma} =  P_+ J_{1,2} =
 \left(
\begin{array}{c}
 P_+\frac{\partial
\varphi_{1,2}}{\partial n}\bigg|_{\Gamma_1}\\
P_+ \frac{\partial
\varphi_{1,2}}{\partial n}\bigg|_{\Gamma_2}\\
P_+ \frac{\partial \varphi_{1,2}}{\partial n}\bigg|_{\Gamma_3}
\end{array}
 \right) =
 \]
 \[
 \left(
 \begin{array}{c}
- \frac{4}{\pi^2}\,\,\sin 2 x_1^{\bot}
 \int_{\Gamma_1} \sin 2 x_1^{\bot}\sin 2y d\Gamma_1\\\\
\frac{8}{\pi^2} \sin 2x_2^{\bot}\int_{\Gamma_2} \sin 2
x_2^{\bot}\sin x d\Gamma_3\\\\
- \frac{4}{\pi^2}\,\,\sin 2x_3^{\bot}\int_{\Gamma_3} \sin 2
x_3^{\bot}\sin 2y d\Gamma_3
\end{array}
\right) =
 \]
\[
\sin 2 x^{\bot}_2 \left(
\begin{array}{c}
0\\\\
16\sqrt 2 /3\pi^2\\\\
- 4/\pi
\end{array}\right) =:  \frac{2}{\sqrt{\pi}}\,\sin 2x^{\bot}_2\,\,
\vec{\psi}_{1,2}.
\]
\[
  P_+ \frac{\partial \varphi_{2,1}}{\partial n}\,
\bigg|_{\Gamma} = P_+ J_{2,1} =
 \left(
\begin{array}{c}
 P_+ \frac{\partial
\varphi_{2,1}}{\partial n}\bigg|_{\Gamma_1}\\
P_+  \frac{\partial
\varphi_{2,1}}{\partial n}\bigg|_{\Gamma_2}\\
P_+ \frac{\partial \varphi_{2,1}}{\partial n}\bigg|_{\Gamma_3}
\end{array}
 \right) =
 \]
 \[
 \left(
 \begin{array}{c}
\frac{8}{\pi^2}\,\,\sin 2 x^{\bot}_1\,\int_{\Gamma_1} \sin 2 x^{\bot}_1\sin y \, d\Gamma_1 \\\\
- \frac{4}{\pi^2} \,\sin 2 x^{\bot}_2\,\int_{\Gamma_2} \sin 2 x^{\bot}_2\sin 2x \, d\Gamma_2\\\\
- \frac{8}{\pi^2}\,\,\,\sin 2 x^{\bot}_3\,\int_{\Gamma_3} \sin 2
x^{\bot}_3\sin y \, d\Gamma_2
\end{array}
 \right)=
 \]
 \begin{equation}
\label{res_bound_ current_P} \sin 2x^{\bot}_2\left(
\begin{array} {c}
16/3 \pi^2 \\\\
0\\\\
- 16/3\pi^2
\end{array}
\right) = \frac{2}{\sqrt{\pi}}\,\sin 2x^{\bot}_2\,\,
\vec{\psi}_{2,1}.
\end{equation}
The  exponent  $K_+$ of  the first ( open) channel is represented as
$K_+ (\lambda) =  \sqrt{\lambda - 4} P_+ $, where the projection
$P_+$ onto the entrance subspace of open channels plays the role of
unity $I_+ = I_1 + I_2 + I_3$ in $E_+$ and is represented as
\[
P_+ = 4/\pi\left[ \sin 2 x^{\bot}_1 \rangle \langle \sin 2
x^{\bot}_1 +   \sin 2 x^{\bot}_2 \rangle \langle \sin 2 x^{\bot}_2 +
\sin 2 x^{\bot}_3 \rangle \langle \sin 2 x^{\bot}_3 \right].
\]
Then taking into account that the contribution ${\mathcal{K}}_{++}$
from the major polar part of  ${\mathcal{K}}\approx
\frac{P_8}{\lambda - 8}$ is
\[
{\mathcal{K}}_{++}\approx \frac{16}{\pi^3}\,\,\frac{\sin 2
x^{\bot}_3\rangle \,\langle \sin 2 x^{\bot}_3}{\lambda - 8} =
\frac{2}{\sqrt{\pi}}\sin 2 x_3^{\bot}\rangle \frac{4}{\pi^2 (\lambda
- 8)}\,\langle \frac{2}{\sqrt{\pi}}\sin 2 x_3^{\bot},
\]
and omitting the vector factors $ 2/\sqrt{\pi}\sin 2 x^{\bot}_1
\rangle$ and $ \langle 2/\sqrt{\pi} \sin 2 x^{\bot}_1$ on the right
and left side of the jump-start scattering matrix,
\begin{equation}
\label{jump-start_1} S_{jump-start} = \frac{ip I_+ - \frac{4
I_3}{\lambda - 8}
- \langle \left( \begin{array}{c} \vec{\psi}_{1,2}\\
\vec{\psi}_{2,1}
\end{array}\right) \left[ \frac{P_{5}}{\lambda - 5} + \frac{P_{5-\delta^Q}}{\lambda - 5 +
\delta^Q}\right] \left( \begin{array}{c} \vec{\psi}_{1,2}\\
\vec{\psi}_{2,1}
\end{array}\right)\rangle }{ip I_+ +\frac{4
I_3}{\lambda - 8} + \langle \left( \begin{array}{c}
\vec{\psi}_{1,2}\\ \vec{\psi}_{2,1}
\end{array}\right) \left[ \frac{P_{5}}{\lambda - 5} + \frac{P_{5-\delta^Q}}{\lambda - 5 +
\delta^Q}\right] \left( \begin{array}{c} \vec{\psi}_{1,2}\\
\vec{\psi}_{2,1}
\end{array}\right)\rangle}
\end{equation}
For better approximation of the  scattering matrix we  should use
better  approximation for $K_- , {\mathcal{K}}^{\Delta}$.

\subsection{Symmetric junction.}
The  Datta-type  boundary condition (\ref{Datta_1}) does not
coincide with the original  Datta-Das Sarma boundary condition
(\ref{bndcnd},\ref{bndcnd2}), suggested in \cite{DattaAPL}, because
the phenomenological Datta-Das Sarma boundary condition was
suggested for a  T-junction which is  symmetric with respect to the
left-right reflection. The resonance concept of the conductance
permits to derive, for a symmetric junction, the original Datta-
das- Sarma condition and interpret the phenomenological parameter
$\beta$.

Consider a symmetric junction $\Omega$ consisting of a square
$(0,\pi)\times(0,\pi)$, and the quantum wires width $\pi/2$,
attached  in the middle of  the sides $\Gamma_1,
\Gamma_2,\Gamma_3$,, see (\ref{F:figure 4}).  The role of the
one-electron Hamiltonian is  played by the Laplacian on $\Omega$
with zero boundary conditions. Similarly to above example we assume
that the electrons are supplied from the second wire, in the first
spectral channel. We assume that the scaled Fermi level is $\Lambda
=$10 and the conductivity band is $4\leq \lambda\leq 16$, the
eigenvalues of $L_{int}$ embedded into the conductivity band are
$\lambda_0 = 5,\,\lambda_1 = 8,\,\lambda_2 = 10,\, \lambda_3 = 13 $.
The corresponding eigenfunctions of $L_{int}$  are found in previous
subsection via separation of variables, see (\ref{eigenfunctions}).
The role of the resonance eigenfunctions  is  played by
$\varphi_{1,3},\varphi_{3,1}$, with the eigenvalue $\lambda_5 =
\lambda_6 = 10 $. We  also use the  symmetric and  antisymmetric
linear combinations  of them
\[
2^{-1/2}[\varphi_{1,3} + \varphi_{3,1}] =: \varphi_{s},\,\,
2^{-1/2}[\varphi_{1,3} - \varphi_{3,1}] =: \varphi_{a}.
\]
We denote the corresponding  boundary currents as
\[
\frac{\partial \varphi_{s}}{\partial n}\bigg|_{\Gamma} =: J_{s},\,\,
\frac{\partial \varphi_{a}}{\partial n}\bigg|_{\Gamma} =: J_{a}
\]
and consider  the projections of them $P_+ J_{sym},\,P_+ J_{asym}$
onto  the  entrance space of the first (open) channel. We assume
that the temperature is low, so that the role of an essential
spectral interval is played by $\Delta_T = [9,11]$. Then the
eigenfunctions and eigenvalues of the intermediate Hamiltonian can
be  found based on  Theorem \ref{T:comp_M}, taking into account the
approximate calculation of the  potential $Q(\lambda)$  of
$L^{\Delta}(\lambda)$ : \[
Q(\lambda)\approx \left(
\begin{array}{c}
\varphi_{1,3}\\
\varphi_{3,1}
\end{array}
\right)\rangle
 {\bf D}\langle\left(
\begin{array}{c}
\varphi_{1,3}\\
\varphi_{3,1}
\end{array}
\right),
\]
with
\[
{\bf D} = \left(
\begin{array}{cc} \langle P_-\frac{
\partial \varphi_{1,3}}{\partial n}\big|_{\Gamma}\, K^{-1}_-\, P_-
\frac{\partial \varphi_{1,3}}{\partial n}\big|_{\Gamma}\rangle &
\langle P_- \frac{
\partial \varphi_{1,3}}{\partial n}\big|_{\Gamma}\, K^{-1}_-\, P_-
\frac{
\partial \varphi_{3,1}}{\partial n}\big|_{\Gamma}\rangle\\
\langle P_- \frac{
\partial \varphi_{3,1}}{\partial n}\big|_{\Gamma}\, K^{-1}_-\, P_-
\frac{\partial \varphi_{1,2}}{\partial n}\big|_{\Gamma}\rangle &
\langle P_- \frac{
\partial \varphi_{3,1}}{\partial n}\big|_{\Gamma}\, K^{-1}_-\, P_-
\frac{\partial \varphi_{3,1}}{\partial n}\big|_{\Gamma}\rangle
\end{array}
\right),
\]
and thus neglecting $Q$ in the  case of thin networks.  Hence, in
the first order approximation, the perturbed eigenvalues of the
Intermediate Hamiltonian remain the same : $\lambda^Q_5= \lambda^Q_6
= 10$. Due to reflection symmetry of the junction there are two
eigenfunctions of the Intermediate Hamiltonian which correspond to
the  eigenvalue multiplicity 2 obtained  based on the Theorem
\ref{T:comp_M}. The corresponding eigenfunctions and the projections
of the normal currents onto $E_+$ are respectively symmetric and
anti-symmetric:
\[
\vec{\psi}_{a,s} = P_+ \frac{\partial \varphi_{a,s}}{\partial n}|P_+
\frac{\partial \varphi_{a,s}}{\partial n}|^{-1}
\]
\[
\vec{\psi}_{a} = \alpha_a\frac{1}{\sqrt{2}} \left(
\begin{array}{c} 1\\0\\
-1 \end{array}\right)= \alpha_a \,\, e_a,\,\,
 \vec{\psi}_{s} =\alpha_s \frac{1}{\sqrt{2+
\gamma^2}} \left(
\begin{array}{c} 1\\\gamma\\
1 \end{array}\right)= \alpha_s \,\, e_s,
\]
and the  orthogonal complement in $E_+$ is  spanned by the  vector
\[
\frac{\gamma}{\sqrt{4 + 2\gamma^2}} \left(
\begin{array}{c} 1\\ - 2 /\gamma  \\
1 \end{array}\right) =: \frac{1}{\sqrt{2 + \beta^2}} \left(
\begin{array}{c} 1\\\beta  \\
1 \end{array}\right),
\]
with $\beta = -2/\gamma$. In the first order approximation
${\mathcal{K}}$ can be  substituted by the contribution from the
nearest eigenvalue $\lambda_4 = 8$, and hence ${\mathcal{K}}_{++} =
k = 0$.

 The intermediate  DN-map  is represented, due to  Theorem
\ref{T:comp_M} by the formula
\[
{\mathcal{M}} =  \frac{ \alpha_a^2 P_a + \alpha_s^2 P_s}{\lambda -
\lambda_2^Q} + {\mathcal{K}}_{++}\approx   \frac{ \alpha_a^2 P_a +
\alpha_s^2 P_s}{\lambda - 10},
\]
where  $P_a = e_a\rangle \,\langle e_a,\,\, P_s = e_s\rangle
\,\langle e_s$. Then  denoting by $P^{\bot}: = P_+ - \left[P_a + P_s
\right] =: P_+ - P_{Q}$, we  represent the  scattering matrix of the
symmetric junction as

\[
S = P^{\bot} + \frac{ip P_Q - \frac{ \alpha_a^2 P_a + \alpha_s^2
P_s}{\lambda - \lambda_2^Q}}{ip P_Q + \frac{ \alpha_a^2 P_a +
\alpha_s^2 P_s}{\lambda - \lambda_2^Q}}.
\]
Here the role of the scalar Blaschke factor  $\Theta_1$ in
(\ref{approxS}) is played by the $2\times 2$ matrix
\[
\Theta = \frac{ip[\lambda - \lambda_2^Q] P_Q - [\alpha_a^2 P_a +
\alpha_s^2 P_s]}{ip[\lambda - \lambda_2^Q] P_Q + [\alpha_a^2 P_a +
\alpha_s^2 P_s]}.
\]
The matrix $\Theta$ for low temperature  is close  to  $-P_Q$ on the
corresponding small essential spectral interval  $\Delta_T$ centered
at $\lambda_2^Q$. Then the  scattering matrix is represented as
\[
S =  P^{\bot}_Q -  P_Q,\,\,
\]
with  the pair of  complementary projections $ P^{\bot}_Q$ and $
P_Q$, dim $ P^{\bot}_Q =  1$, dim  $ P_Q = 2$. Scattering Ansatz on
the model graph
\[
\Psi =  e^{ip P_+ x} e  +  e^{-ip P_+ x} Se
\]
satisfies  the following boundary condition at the vertex $x=0$:
\[
\Psi(0) = (I + S)e = 2 P^{\bot}_Q e,\, \Psi'(0) = ipP_+(I - S)e =
2ip P_Q e.
\]
Taking into account that  dim $P^{\bot}_Q = 1$,\, dim $P_Q = 2$, we
can re-write the  previous formulae as  boundary conditions imposed
on the Ansatz:
\[
P_Q^{\bot} \vec{\psi}' (0)= 0, \,\,\,
 \vec{\psi}(0) \mbox{ is \,\,parallel\,\,\, to}\,\,\, P_Q^{\bot}
e\,\,\,\mbox{or}\,\,\,
\]
\begin{equation}
\label{Datta_bc} \langle  e^{\bot}, \psi' (0) \rangle =  0,\,\,
\frac{\psi_1 (0)}{e^{\bot}_1} = \frac{\psi_2 (0)}{e^{\bot}_2}=
\frac{\psi_3 (0)}{e^{\bot}_3}.
\end{equation}
where $P_Q^{\bot} =: e^{\bot}\rangle \,\langle e^{\bot},\,\, e^{\bot
} =  (e^{\bot }_1, e^{\bot }_2,e^{\bot }_3) = (2 +
\beta^2)^{-1/2}(1, \beta, 1)$. This  condition coincides with the
original Datta-das-Sarma  boundary condition, see
(\ref{bndcnd},\ref{bndcnd2}).  Our analysis reveals the meaning of
the phenomenological parameter $\beta$.

\vskip0.3cm
 \section{A solvable model of a thin junction}
Generally, a Schr\"{o}dinger operator with non-constant coefficients
or one  in  a non-standard domain rarely admits spectral analysis in
explicit form. For qualitative analysis of quantum systems the
Schr\"{o}dinger operator often is substituted by  a solvable model,
constructed by the von Neumann operator extension technique,
\cite{Neumann},  see for instance \cite{BF61,Demkov,Albeverio} and
an extended list of references in \cite{AK00}. In particular, the
substitution of the network by a proper one-dimensional graph, with
special boundary conditions at the vertices, looks like a convenient
tool for the qualitative analysis of the Schr\"{o}dinger equation on
the network. Unfortunately the estimation of the error caused by the
substitution of the network by the corresponding graph is difficult.
Shrinking of a ``fattened graph'' $\Omega_{_{\delta}}$ to
one-dimensional graph was studied in numerous papers, see for
instance \cite{Kuch01,Kuch02}. The authors considered a compact
network $\Omega_{\delta}$ constructed of the vertex domains
$\Omega_{in}$, with the diameter proportional to $\delta^{\alpha},\,
0< \alpha < 1 $ and several finite leads $\omega^m$, width $\delta
$, joining them to each other. In \cite{RS01} they developed, based
on \cite{WF92,Schat96}, a variational technique for description of
the asymptotic behavior of the discrete spectrum of the
Schr\"{o}dinger operator on the quantum network $\Omega_{_{\delta}}$
of various grades  $\alpha$ of thinness. It appeared that the
(discrete) spectrum of the Laplacian on the compact shrinking
``fattened graph'' $\Omega_{_{\delta}}$ tends to the spectrum of the
Laplacian on the corresponding one-dimensional graph but with
different boundary conditions at vertices depending on the speed of
shrinking: the Kirchhoff boundary conditions at the nodes, in the
case of ``small protrusion" $1/2 < \alpha < 1 $, or the homogeneous
Dirichlet boundary conditions, in case of `` large protrusion'', $0<
\alpha< 1/2$, see \cite{Kuch02}, theorems 1,2,3.
\par
In this section we  consider a  thin quantum networks with small
protrusion $\alpha = 1$, assuming  diam $\delta<<$ diam $
\Omega_{in} $. We will construct a {\it quantitatively consistent}
solvable model of the quantum network, in  the form of a star-graph
with a vertex supplied with  inner space and appropriate vertex
Hamiltonian. The scattering matrix of the properly fitted model
serves as a local approximation - on a certain ``essential''
spectral interval $\Delta$ - of the scattering matrix  of the
original network. In contrast to the quoted above results for
compact networks, where the wave-functions are obtained based on the
variational approach, we use Dirichlet-to-Neumann map
${\mathcal{D}}{\mathcal{N}}^{\Lambda}$ of an intermediate
Hamiltonian $L_{\Lambda}$, to derive an explicit formula for the
scattered waves on the original network, see (\ref{Eq:smatrix_M}).
The scattering matrix of the star-graph model is obtained via
replacement of ${\mathcal{M}}$ in (\ref{Eq:smatrix_M}) by the
corresponding rational approximation on $\Delta$, based on Theorem
\ref{T:comp_M}. This defines all parameters of the model in terms of
spectral characteristics of $L_{\Lambda}$. In course of construction
and fitting  of the solvable star-graph model we also define the
energy-dependent boundary conditions at the vertex for the
Schr\"{o}dinger equation on the graph. In the simplest case when
only one resonance eigenvalue $\lambda_{0}$ of the intermediate
Hamiltonian is present on $\Delta$, this condition depends linearly
on the spectral parameter, and is parametrized by coordinates  $\Re
k_{0},\, \Im k_{0} $ of the corresponding resonance. \par Note that
in \cite{Schrader} an algorithm for construction of  the scattering
matrix of  the quantum graph of the scattering matrices of
star-shaped elements is described and, in \cite{Har2_00} a
convenient formula for the scattering matrix of the star-graph in
terms of the boundary parameters at the vertex is suggested.  For
extended discussion of properties of star graphs see
\cite{Keating03,Keating04}. \vskip10pt

The star-graph solvable model of the thin junction will be
constructed as a finite-dimensional perturbation of an orthogonal
sum of the non-perturbed Hamiltonian $l_{_{\Lambda}}$ in the  open
channels and a  finite matrix $A$ acting in the  {\it inner  space}
of the vertex. We will choose the parameters of the model such that
the model scattering matrix coincides with the few-pole
approximation ${\bf S}_{_{\Delta}}$ of the complete scattering
matrix ${\bf S}$. Then the constructed model will be  automatically
{\it fitted} (i.e.  quantitatively consistent). We assume that the
spectral variable is  scaled  such that original  the Schr\"{o}
dinger equation on the wires is  just $-\Delta u + V_{\delta}u
=\lambda u$. The scattering Ansatz of the model in open channels
satisfies on the wires $\omega$  the same equation as the scattering
Ansatz on the original network
\begin{equation}
\label{nonpert} l_{{\Lambda}} U := -\, \frac{d^{2} {U}^{{\omega}}}{d
x^{2}}
   +  \sum_{{s,m}}\frac{\pi^{{2}}\,
   s^{{2}}}{\delta^{2}}\,p_{s}^{m} U +
{V}_{\delta} {U}:=  \lambda \,\,,\,\, \lambda \in \Delta,\,\,U =
U^{\omega}= \left(u^1,u^2,\dots u^M\right).
\end{equation}
We assume that  the  Schr\"{o}dinger equation on the  quantum well
$\Omega_{int}= \Omega_0$, with the same spectral parameter, is
represented as
\[
L_0 u := -\Delta u + V u = \lambda u,
\]
with a corresponding effective mass $\mu_0$. We  assume that the
Intermediate relative  DN-map, $ P_+\frac{\partial u}{\partial
n}\bigg|_{\Gamma} =:{\mathcal{DN}}u$, with respect to $\Gamma$, is
calculated and the corresponding rational approximation is selected.
Then the construction of the vertex part of the model will be done
with a major change of the original Intermediate Hamiltonian.
\par
For thin networks an auxiliary spectral interval $\Delta$ is
selected  inside $\Delta_1 =
[\frac{\pi^2}{\delta^2},\,4\frac{\pi^2}{\delta^2} ]$, and  hence
does not overlap with the continuous spectrum of the intermediate
Hamiltonian $L_{_{\Lambda}}$. Only a finite number $ N $ of
eigenvalues of the intermediate operator are situated on $\Delta$.
Then substitution of $\mathcal{M}$ on $\Delta$ by  the rational
approximation ${\mathcal{M}}_{\Delta}$ may cause only a minor and
controllable error. Now we will prove that there exist a
finite-dimensional perturbation of the operator $l_{{\Lambda}}\oplus
{A}$ such that the scattering matrix of the perturbed operator
coincides with ${S}_{{\Delta}}$. The perturbation will be
constructed via operator restriction-extension procedure applied to
the orthogonal sum $l_{_{\Lambda}}\oplus A$, where  $A$ is an
$N\times N$ Hermitian matrix $: E_A \to E_A $, dim $E_A = N$. The
parameters of the model will be properly selected to fit the
spectral data of the Intermediate Hamiltonian on the original
quantum network, within the auxiliary spectral interval $\Delta$.

Assume that the positive matrix $A $ is defined by its spectral
decomposition
\[
A = \sum_{_{r}} \alpha^{^2}_{_r} P_{_{r}}.
\]
Here  $\alpha^{^2}_{_r} > 0$ are eigenvalues of $A$, and $P_{_r} =
\nu_{_{r}}\rangle\,\langle \nu_{_{r}}$ are the corresponding
orthogonal spectral projections. The eigenvalues and the boundary
parameters $\,\beta$ of the model, see below (\ref{one}), will be
defined later, based on comparison of the scattering matrix of the
model with the essential scattering matrix
\begin{equation}\label{S_essential}
S_{\Delta}= [iK_+ + {\mathcal{DN}}_{\Delta}]^{-1} [iK_+ -
{\mathcal{DN}}_{\Delta}].
\end{equation}
\par
Consider restrictions of both $l_{_{\Lambda}} $ and $A$ to symmetric
operators on the corresponding domains. The  restriction of  $
l_{_{\Lambda}} \big|_{D^{^{l}}_{_{0}}} = l_{_0} $  is defined on
functions vanishing near  $x = 0$. Then the adjoint operator
$l_{_{0}}^{^{+}}$ is  defined on  $W_{_{2}}^{^{2}} \left(E_{_{+}},\,
R_{_{+}}\right)$, and the boundary form of it is calculated via
integration by parts:
\begin{equation}
\label{boundform} {\mathcal{J}}_{_l} (U,\, V) =
  \langle l_{_{0}}^{^{+}} U,\,V\rangle
- \langle U ,\, l_{_{0}}^{^{+}} V\rangle = \langle U'(0),\, V
(0)\rangle  -  \langle U(0),\, V'(0)\rangle,
\end{equation}
where  $U (0),\,V(0) \in E_{_{+}}$ and  the  derivatives are taken
in the  outgoing direction on $\Gamma$ with respect to
$\Omega_{in}$.
\par
Restriction of the  matrix $A$ is  equivalent to selection of the
deficiency subspace  for the given value $i$ of the spectral
parameter. Choose a generating subspace $N_{_{i}}$, $
\overline{\bigvee_{_{k>0}} A^{^k} N_{_{-i}}} = E_{_{A}}$ such that
$\frac{A + i I}{A - i I} N_{{i}} \cap N_{{i}} = 0$, dim $N_{{i}} =
d$, set $D^{^A}_{{0}} = (A-iI)^{^{-1}} \left( E_{_{A}} \ominus
N_{_{i}} \right)$ and define the restriction of the inner
Hamiltonian  as $A\to A_{{0}} = A \big|_{D^{{A}}_{{0}}}$. We develop
the extension procedure for general  $N_{_{i}}$ and fit it later
based on spectral data of the intermediate operator, see Theorem
4.1, 4.2. In our construction $N_{_{i}} \subset E_{_A} $ will play a
role of the deficiency subspace at the spectral point $i$, dim
$N_{_i} = d,\,\, 2d \leq N$  and the dual deficiency subspace is ${
N}_{-i} = \frac{A + i I}{A - i I} N_{_{i}}$. The domain of the
restricted operator $A_{_{0}}$ is not dense  in $E_{_{A}}$, because
$A$ is bounded. Nevertheless, since the deficiency subspaces
$N_{_{\pm i}}$ do not overlap, the extension procedure for the
orthogonal sum $l_{0}\oplus A_{{0}}$ can be developed.  We will do
it here with use of the symplectic formalism, see for instance
\cite{Extensions}. In this case the ``formal adjoint'' operator for
$A_{_{0}}$ is defined on the defect $N_{_{i}} + N_{_{-i}}:=
{\mathcal{ N}} $ by the von Neumann formula : $A_{_{0}}^{^{+}} e \pm
i\,e = 0$ for $e\in N_{_{\pm i}}$. Then the extension is
constructed, see lemmas 3.1-3.4 below, via restriction of the formal
adjoint onto a certain plane in the defect where the boundary form
vanishes (a ``Lagrangian plane''). According to the classical  von
Neumann construction  all Lagrangian planes are parametrized by
isometries $V: N_{_{i}}\, \to \, N_{_{i}}$ in the form
\[
   {\mathcal{T}}_{_{V}} = \left( I -  V\right) N_{_{i}}.
\]
In case when the  deficiency  subspaces do not overlap, the
corresponding  isometry is  admissible, and, according to
\cite{Krasn} there exist a self-adjoint extension  $A_V$ of the
restricted operator $A_0$. We construct this  extension based on the
following
\begin{lemma} { The lagrangian plane ${\mathcal {T}}_V$ in the  defect
forms a non-zero angle with the domain $D^{{A}}_{{0}}$ of the
restricted operator $A_0$}.
\end{lemma}
{\it Proof}\,\, Indeed, if  $A_V$ is the  extension, then on the
${\mathcal {T}}_V$ it coincides with the restriction of the  formal
adjoint, and on the  domain $D^{{A}}_{{0}}$ it coincides with $A_0$.
Then  assuming that ${\mathcal {T}}_V$ and  $D^{{A}}_{{0}}$ overlap,
we obtain, for some $f^{\bot} \bot N_i,\,\, \nu \in N_i$
\[
\frac{1}{A - i I} f^{\bot} = \nu - V \nu.
\]
Applying $A_V -iI$ to both parts of this equation, we  obtain
\[
f^{\bot} = - 2i \nu,
\]
hence  $f^{\bot} = - 2i \nu = 0$.

{\it End of the proof.}
\par
It  follows  from the Lemma that, once the  extension is constructed
on the Lagrangian plane, the whole construction of the extended
operator can be accomplished in the form of  a direct sum of the
closure of the restricted operator and the extended operator on the
Lagrangian plane.
\par
Note that the  operator  extension procedure may be  developed
without  assumption of non-overlapping, see \cite{Krasn}. In
particular, the case  dim$E_{_{A}} = 1 $, which is not formally
covered by the above procedure, was analyzed in \cite{Shirokov80}
independently  of \cite{Krasn}. The relevant formulae for the
scattering matrix and scattered waves remain true and may be
verified  by the direct calculation. We will  use  this fact in
section 7 below.
\par
We  will use hereafter notations  and some  facts concerning the
symplectic  operator extension procedure, see  Appenxix and
references therein. Choose an ortho-normal basis in ${N}_i$ :
$\left\{f_{_s}\right\},\,s= 1,2,\dots ,d$, as a set  of deficiency
vectors  of the restricted operator $A_0$.  Then the vectors
$\hat{f}_{_{s}}= \frac{A + i I}{A - i I} f_s$ form an ortho-normal
basis in the dual deficiency subspace $N_{{-i}}$. Under the above
non-overlapping condition one can use the formal adjoint operator
$A_0^+$ defined on  the defect $N_{{i}} + N_{{-i}} = {\mathcal
{N}}$:
\begin{equation}
\label{formal_adj}
 u = \sum_{s=1}^{{d}} [x_{s}\, f_{s} +
\hat{x}_{s} \, \hat{f}_{s}] \,\in \, {\mathcal{N}},
\end{equation}
by the von Neumann formula, see  \cite{Glazman},
\begin{equation}
\label{von_neumann_form}
 A_0^+ u = \sum_{s=1}^{{d}} [ -i\,\,
x_{s}\, f_{s} + i\,\, \hat{x}_{s}\, \hat{f}_{s}].
\end{equation}
In order to use the symplectic version of the operator-extension
techniques we introduce in  the defect  a  new basis $W_{s}^{\pm}$,
on which the formal adjoint $A_0^+$ is correctly defined due  to the
above non-overlapping condition:
  \[
   W_s^{+}= \frac{f_{_{s}} + \hat{f}_{_{s}}}{2} = \frac{A}{A-iI} f_{_s},
   \,\,W_{s}^{-} = \frac{f_{_{s}} - \hat{f}_{_{s}}}{2 i}= - \frac{I}{A-iI}
   f_{_{s}},
   \]
    \[
   A_0^+ W_s^{+} = W_s^{-},\,\, A_0^+ W_{s}^{-} = - W_{s}^{+}.
   \]
It is  convenient to represent  elements  $u \in {\mathcal{N}}$ via
the new  basis as
\begin{equation}
\label{formal_adj_12}
 u = \sum_{s=1}^{{d}} [\xi^{^+}_{s}\,  W_{s}^{+} +
{\xi}_{s}^{^{-}} \,  W_{s}^{-}].
\end{equation}
Then, using  notations  $\sum_{s=1}^{{d}} \xi_{s,\pm}\, e_{_{s}}:=
\vec{\xi}_{_{\pm}} $ we re-write  the above von Neumann formula as
\begin{equation}
\label{formal_adj12} u = \frac{A}{A - iI} \vec{\xi}^{^{u}}_{_{+}} -
\frac{1}{A - iI} \vec{\xi}^{^{u}}_{_{-}},\,\,\,\, A^{^+}_{_{0}} u =
- \frac{1}{A - iI} \vec{\xi}^{^{u}}_{_{+}} - \frac{A}{A - iI}
\vec{\xi}^{^{u}}_{_{-}}
\end{equation}
The following formula of integration by parts for abstract operators
was proved in (\cite{Extensions}):

\begin{lemma}{Consider the  elements  $u,v$ from
the  domain of  the  (formal) adjoint  operator $A_0^+$:
$$ u =
\frac{A}{A-iI} \vec{\xi}^{^u}_{_{+}} - \frac{1}{A-iI}
\vec{\xi}^{^u}_{_{-}},\,\, v = \frac{A}{A-iI} \vec{\xi}^{^v}_{_{+}}
- \frac{1}{A-iI} \vec{\xi}^{^v}_{_{-}}
$$
with  coordinates  $ \vec{\xi}^{u}_{\pm},\,\vec{\xi}^{v}_{\pm}$:
 \[
\vec{\xi}^{^u}_{_{\pm}} = \sum_{s=1}^{d} \xi^{^u}_{_{s,\pm}}
f_{_{s,i}}\in {N_{_{i}}},\,\, \vec{\xi}^{^v}_{_{\pm}} =
\sum_{s=1}^{d} \xi^{^v}_{_{s,\pm}} f_{_{s}}\in {N_{_{i}}}.
\]
Then,  the boundary  form  of the  formal adjoint operator  is equal
to
\begin{equation}
\label{b_form} {\mathcal{J}}_{_{A}} (u,v) = \langle A_0^+ u,v
\rangle - \langle  u, A_0^+ v \rangle = \langle
\vec{\xi}^{^u}_{_{+}}, \vec{\xi}^{^v}_{_{-}}\rangle_{_N} - \langle
\vec{\xi}^{^u}_{_{-}}, \vec{\xi}^{^v}_{_{+}} \rangle_{_N}.
\end{equation}
}
\end{lemma}
One can see that the coordinates  $
\vec{\xi}^{u}_{\pm},\,\vec{\xi}^{v}_{\pm}$ of the  elements $u, v $
play  the role of the  boundary values $\left\{ U'(0),\, U(0),\,\,
V'(0),\, V(0) \right\}$. We  will call them  {\it symplectic
coordinates} of the element $u,v$. The next statement proved in
\cite{Extensions} is the core detail of the fundamental Krein
formula \cite{K}, for generalized resolvents of symmetric operators.
In our situation, it is used in course of calculation of the
scattering matrix.
\begin{lemma}{ The  vector-valued
function of the  spectral parameter
 \begin{equation}
 \label{incomp}
 u (\lambda) =  \frac{A +iI}{A - \lambda I}\,\,\, \vec{\xi}^{^u}_{_{+}} :=
 u_0 + \frac{A}{A-iI} \vec{\xi}^{^u}_{_{+}}
 - \frac{1}{A-iI} \vec{\xi}^{^u}_{_{-}},
 \end{equation}
satisfies the adjoint  equation  $ [A_0^+ - \lambda I] u =
 0$, and  the  symplectic coordinates
$\vec{\xi}^{^u}_{_\pm}\in N_{_{i}} $ of it are  connected  by  the
formula
 \begin{equation}
        \label{krein}
        \vec{\xi}^{^u}_{_{-}} = - P_{N_{i}}\frac{I + \lambda A}{A -
        \lambda} \vec{\xi}^{^u}_{_{+}}
        \end{equation}
        }
 \end{lemma}

 {\it Proof}\,\,\, see in  Appendix, subsection 9.1 or in \cite{Extensions}.

        Introduce the  map
        $$  P_{ N_{i}} \frac{I + \lambda A}{A - \lambda I}
P_{N_{i}} =: - {\mathcal{M}}
        : {N_{i}} \to { N_{i}}. $$
       The matrix-function
       ${\mathcal{M}} = P_{N_i} A P_{N_i} - P_{N_i} \frac{I+
        A^2}{A-\lambda I} P_{N_i}$
        has a  negative  imaginary  part  in the upper half-plane
        $\Im \lambda > 0$ and  serves an abstract analog
         of the celebrated  Weyl-Titchmarsh function.
          The operator-function ${\mathcal{M}}$ exists almost
        everywhere on the real axis $\lambda$, and  has a   finite  number
        of  simple  poles at  the  eigenvalues  $\alpha_{_r}^{^2}$ of $A$.
        This function  plays  an  important role in description of spectral
        properties of self-adjoint extensions of  symmetric operators, see
        \cite{K,Gorbachuk}.
        \par
        We  construct a solvable model of the quantum network as a
    self-adjoint  extension of the  orthogonal sum  $l_{_{0}} \oplus
    A_{_{0}}$. We consider
    the orthogonal sum of the  corresponding adjoint  $l^{^+}_{_{0}}$
    and  the formal adjoint:  $l^{^+}_{_{0}} \oplus A^{^+}_{_{0}}$, and
    calculate the  corresponding boundary form
    ${\bf J} ({\bf U},{\bf V}) := {\mathcal{J}}(U,V) + {\mathcal{J}}(u,v)$
    on elements $(U,u) := {\bf
     U}$ from the orthogonal sum  of the corresponding spaces. The
        self-adjoint extensions of  the operator $l_{_{0}} \oplus A_{_{0}}$
        are obtained, based on restrictions of  the
        adjoint operator  ${\bf A}_0^+= l^{^+}_{_{0}} \oplus A^{^+}_{_{0}}$ onto
        Lagrangian planes of the form  ${\bf J} ({\bf U},{\bf
        V})$. These
        planes may be defined by the boundary conditions  connecting
        symplectic coordinates $ U'(0),\,U (0) ,\,\, \vec{\xi}^u_{+},\,
         \vec{\xi}^u_{-} $ of components of  corresponding  elements in the
        deficiency subspaces. For instance, one may select a
        finite-dimensional operator
        $\beta : E_{_{+}} \oplus N_{_{i}} \to  E_{_{+}} \oplus N_{_{i}}$
         and define the Lagrangian plane
        ${\bf L}_{\beta}$ by the boundary condition
        \begin{equation}
        \label{one}
         \left(
        \begin{array}{c}
         U'(0)\\ \vec{\xi}_{+}
        \end{array}
        \right)
        =
         \left(
        \begin{array}{cc}
        \beta_{_{00}} & \beta_{_{01}}\\ \beta_{_{01}}^{^{+}} & 0
        \end{array}
        \right)
         \left(
        \begin{array}{c}
        U(0) \\ \vec{\xi}_{-}
        \end{array}
        \right).
        \end{equation}
The  extension  defined by (\ref{one}) on the Lagrangian plane is
continued  onto the whole  space  $L_{_{2}} (E_{_{+}},\, R_{_{+}})
\oplus E_{_{A}} $   by forming the direct sum with  the closure of
the restricted operator  $A_0$, see \cite{Krasn}. This construction
gives a self-adjoint extension ${\bf A}_{\beta}$ of $l_0 \oplus A_0$
in $L_2({E_{_{+}}},\, R_{_{+}}) \oplus E_{_{A}} $, defined by the
boundary condition (\ref{one}).

The absolutely continuous spectrum of the operator ${\bf A}_{\beta}$
coincides with  the spectrum of the exterior part  of the model, and
hence it coincides with the spectrum of the trivial component
$l_{_{\Lambda}}$ of the split operator ${\mathcal{L}}_{_{\Lambda}}$
(in the open channels). The corresponding eigenfunctions of ${\bf
A}_{_{\beta}}$ on the first spectral band
  ${\Delta}_1\supset \Lambda$  can be  found, see \cite{AK00},
via  substitution  into the above  boundary  condition for the
column, combined  of  the  Scattering Ansatz  in the open channels
with (\ref{incomp}), and, in the  outer space, with $K_+ =
\sqrt{\lambda - V_{\delta} -\pi^2\,\delta^{-2}}$:
  \begin{equation}
  \label{S_wave}
\Psi = \left(
\begin{array} {c}
e^{i K_+ x} \nu  + e^{ -i\, K_{+}x} {\bf S} \nu\\
\frac{A + iI}{A - \lambda I}\vec{\xi}^{u}_{+}
\end{array}\right),
  \end{equation}
with $\beta_{10} = \beta^+_{01}$. It  gives the linear equation for
the Scattering Matrix:
\[
\left(
\begin{array}{c}
i K_{_{+}} \left( \nu - S \nu\right)\\
\vec{\xi}_{_+}
\end{array}
\right) = \left(
\begin{array}{cc}
\beta_{_{00}}& \beta_{_{01}}\\
\beta_{_{10}}& 0
\end{array}
\right) \,\, \left(
\begin{array}{c}
\nu + S \nu\\
{\mathcal{M}}\vec{\xi}_{_+}
  \end{array}
  \right).
\]
Solving  this  equation we obtain  the  scattered  waves and the
scattering matrix:
\begin{lemma}
{The scattering matrix  for  the  constructed extension is an
analytic function of the  spectral parameter $\lambda$:
\begin{equation}
\label{model}
  {\bf S}(\lambda) = \frac{
  i K_{_{+}} - \left[\beta_{_{00}} +
\beta_{_{01}} {\mathcal{M}} \beta_{_{10}} \right]}{i K_{_{+}} +
\left[\beta_{_{00}} + \beta_{_{01}} {\mathcal{M}} \beta_{_{10}}
\right]},
\end{equation}
with the denominator of the  fraction preceding the  numerator. The
coordinate $\vec{\xi}_{_+}$ of the inner  component of the scattered
wave (\ref{S_wave}) is defined  as \[\vec{\xi}_{_+} = \beta_{10}
\frac{2 ip}{ip + [\beta_{00} + \beta_{_{01}} {\mathcal{M}}
\beta_{_{10}}]},
\]
with $p = \sqrt{\lambda -V_{\delta} + \pi^2\,\delta^{-2}}$.}
\end{lemma}
\vskip0.3cm
\section{Fitting of the solvable model} It remains to choose the
eigenvalues of $A$, the subspace $N_{_{i}}$ and the matrix parameter
$\beta$, such that the operator-function $\left[\beta_{_{00}} +
\beta_{_{01}} {\mathcal{M}} \beta_{_{10}} \right]$ acting  in
$E_{_{+}}$ coincides with the essential DN-map
${\mathcal{D}}{\mathcal{N}}^{^{\Lambda}}_{_{\Delta}}$ of the
intermediate Hamiltonian. Denote by $Q_{_{s}}$ the spectral
projection corresponding to the eigenvalue $k^{^2}_{_s}$ of $A$,
framed by the projections $P_i$ onto the deficiency subspace
$N_{_{i}}$
\[
Q_{_{s}} = P_{_{i}} P_{_{s}} P_{_{i}}.
\]
Then the  above expression takes the form :
\[
\left[ \beta_{_{00}} + \beta_{_{01}} {\mathcal{M}} \beta_{_{10}}
\right]=
\]
\begin{equation}
\label{KreinQ} \left[\beta_{_{00}} + \sum_{_{r=1}}^{^{N_{_{T}}}}
\alpha_{_r}^{^2} \beta_{_{01}} Q_{_{sr}} \beta_{_{10}}\right] -
\sum_{_r} \frac{1 + \alpha_{_{r}}^{^4}}{\alpha_{_{r}}^{^2} - \lambda
}\beta_{_{01}} Q_{_{r}}\beta_{_{10}}.
\end{equation}
We will define the boundary parameters
$\beta_{_{10}},\,\,\beta_{_{01}}  = \beta_{_{10}}^{^ +}$ later, but
once they are defined, we choose $\beta_{_{00}}$ such that the first
summand in the  right side of (\ref{KreinQ}) coincides with $k_M$
$\beta_{_{00}} + \sum_{_{r}} \alpha_{_r}^{^2}\beta_{_{01}} Q_{_{r}}
\beta_{_{10}} = - k_M$. Then the scattering matrix  takes the form:
\begin{equation}
\label{ModelSmatr} {\bf S}(k) = \frac{i K_{_{+}} - k_M +
\sum_{_{r=1}}^{^{N}} \frac{1 +
\alpha_{_{r}}^{^4}}{\alpha_{_{r}}^{^2} -\lambda} \beta_{_{01}}
Q_{_{r}}\beta_{_{10}}}{i K_{_{+}} + k_M - \sum_{_{r=1}}^{^N} \frac{1
+ \alpha_{_{r}}^{^4}}{\alpha_{_{r}}^{^2} - \lambda} \beta_{_{01}}
Q_{_{r}}\beta_{_{10}}},
\end{equation}
which  coincides  with  the  essential  scattering matrix  if  and
only  if  the  corresponding   Krein function
\begin{equation}
\label{Kreinfunk}
 k_M - \sum_{_{r=1}}^{^{N}} \frac{1 +
\alpha_{_{r}}^{^4}}{\alpha_{_{r}}^{^2} -\lambda} \beta_{_{01}}
Q_{_{r}}\beta_{_{10}}
\end{equation}
coincides with the essential  part ${\mathcal{D}}{\mathcal{
N}}^{^{\Lambda}}_{_{\Delta}}$ of the $DN$-map  of the Intermediate
Hamiltonian  on the essential spectral interval $\Delta_T$:
\begin{equation}
\label{polarterm} {\mathcal{DN}}^{\Lambda} \approx k(\lambda ) +
\sum_{_{r=1}}^{^{N}}\frac{P_{_{+}}\frac{\partial\,
\varphi_{_{r}}}{\partial n}\rangle \langle
P_{_{+}}\frac{\partial\,\varphi_{_{r}}}{\partial n}}{\lambda_{_r} -
\lambda}.
\end{equation}
Summarizing these results we  obtain the  following conditional
statement  for  the  extension constructed  based on the boundary
condition (\ref{one}) in case  when  $N_{_{i}} \cap N_{_{-i}} = 0$
or  dim $E_{_{A}} = 1$:
\begin{theorem}
{The  constructed  operator   ${\bf A}_{_{\beta}}$ is  a  solvable
model  of  the  Quantum network on the essential interval $\Delta$,
if and only  if  the dimension of  the  space $E_{_{A}}$ coincides
with the  number $N$ of  eigenvalues of the intermediate operator on
$\Delta\subset [\lambda_{max},\lambda_{min}]$, the eigenvalues
$\alpha^{^2}_{_{r}}$ of the inner Hamiltonian  $A =
\sum_{_{r=1}}^{^{N}} \alpha^{^{2}}_{_{r}} \nu_{_r}\rangle \, \langle
\nu_{_r}$ coincide with eigenvalues of the intermediate operator on
$\Delta$, there exists a deficiency subspace $N_{_{i}}$ of the inner
Hamiltonian such that $N_i \cap \frac{A+iI}{A-iI} N_i  = 0$ and the
operator $\beta_{_{01}}: N_{{i}} \to E_{_{+}}$ such that for the
ortho-normal basis $\left\{ e_{_s}\right\}_{_{s = 1}}^{^{N}}$ of
eigenvectors of $A$ in $E_{_{A}}$
\begin{equation}
\label{fitting}
  P_{_{+}}\frac{\partial\,
\Psi_{{r}}}{\partial n} = [1 + \alpha_{_{r}}^{^4}]^{^{1/2}}
\beta_{_{01}} P_{_{N_{i}}} \nu_{_r},\, r= 1,2,\dots,N.
\end{equation}
 }
\end{theorem}
Eliminating  the inner variables, we can  reduce the model to the
Schr\"{o}dinger equation with the constant potential on open
channels, and appropriate boundary conditions  on the bottom
sections:
\begin{equation}
\label{energy-dep} \frac{d
U^{^{\omega}}}{dx}\bigg|_{_{\Gamma}}=\left[ k_M  -
\sum_{_{r=1}}^{^{N}}\frac{P_{_{+}}\frac{\partial\,
\Psi_{{r}}}{\partial n}\rangle \langle P_{_{+}}
\frac{\partial\,\Psi_{{r}}}{\partial n}}{\lambda_{_r} - \lambda}\,
U^{^{\omega}}\right]\bigg|_{_{\Gamma}}.
\end{equation}
Unfortunately, this straightforward construction does  not  fulfil
basic requirements of  quantum mechanics, and hence we  proceed via
construction  a self-adjoint operator in
$\left[L_{{2}}(0,\infty)\times E_{_{+}}\right]\oplus E_{_{A}})$.
\par
Dr. M. Harmer suggested  an important strengthening  of the previous
conditional statement, by proving a general theorem of existence of
the subspace $N_{_{i}}$ and the projection $P_{N_{i}}$ which satisfy
the condition of Theorem 4.1. The proof we provide below only
slightly differs from the original proof in \cite{Harmer04}: we
added an explicit formulae  for $\beta_{_{01}},\, P_{N_{i}}$ in
terms of the corresponding Gram matrix. \par Denote  by
$L^{^{\Delta}}_{_{\Lambda}}$ the restriction of the  intermediate
operator $L_{_{\Lambda}}$ onto the invariant subspace
$E_{_{\Delta}}= E_A$ corresponding to  the part $\sigma_{_{\Delta}}
= \left\{\lambda_{_{1}},\, \lambda_{_{2}},\, \dots \lambda_{_{N}}
\right\}$ of its spectrum on the essential interval $\Delta$, and
consider the linear  map
\begin{equation}
\label{P} \sum_{_{s}} [1 + \alpha^{^{4}}_s]^{^{-1/2}}
P_{_{+}}\frac{\partial\, \varphi_{_{s}}}{\partial
n}\bigg|_{_{\Gamma}}\langle *,\varphi_{_{s}}\rangle
:=\Phi_{_{\Delta}}
\end{equation}
from  $E_{_{\Delta}}$ to  $ E_{_{+}}$,dim $E_+ = {\bf n}$ .
\begin{theorem}(M. Harmer) {\it
The map $\Phi_{_{\Delta}}$ defines a one-to one  correspondence
between  two  $d$ - dimensional subspaces, $2 d < N$:
\[
\Phi_{_{\Delta}}^{^{+}}\Phi_{_{\Delta}} E_{_{\Delta}}: =
N_{_{\Delta}}\subset E_{_{\Delta}}\,\,\,\mbox{and}\,\,\,
\Phi_{_{\Delta}}\Phi_{_{\Delta}}^{^{+}} E_{_{+}}:=
E^{^{\Delta}}_{_{+}}\subset E_{_{+}}
\]
If the  subspace  $N_{_{\Delta}}$ is a generating subspace of
$L^{^{\Delta}}_{_{\Lambda}} $ and
\begin{equation}
\label{non-overlap} N_{_{\Delta}} \cap
\left(L^{^{\Delta}}_{_{\Lambda}} - iI \right)^{^{-1}}
\left(L^{^{\Delta}}_{_{\Lambda}} + iI \right) N_{_{\Delta}} = 0,
\end{equation}
then there exist a unique  pair of the boundary operator
$\beta_{_{01}}: E_{_{A}}\to E_{_{+}}$ and the subspace
$N_{_{i}}\subset E_{_{A}}$, which satisfy the condition of the
previous theorem. }
\end{theorem}
{\bf Remark  2}  This  theorem  gives an interpretation of the
solvable model described in theorem 4.1, in  terms of the
intermediate Hamiltonian via selection of the inner  Hamiltonian $A$
as a part $L_{_{\Lambda}}^{^{\Delta}}$ of $L_{_{\Lambda}}$ in the
invariant subspace corresponding to the  essential spectral interval
$\Delta$. The subspace $N_{_{\Delta}} \subset E_{_{\Delta}} $ plays
the role of the deficiency  subspace $N_{_{i}}$ of the inner
Hamiltonian and $ \left(L^{^{\Delta}}_{_{\Lambda}} - iI
\right)^{^{-1}} \left(L^{^{\Delta}}_{_{\Lambda}} + iI \right)
N_{_{\Delta}} $ plays the role of the  dual subspace $ N_{_{-i}}$.
\par
\noindent {\it Proof}\,\, The map $\Phi_{_{\Delta}}$ is represented
by the ${\bf n} \times N $ matrix $\Phi$ of columns $\phi_{_s}$ with
respect to the orthogonal basis of cse-functions
$\left\{e_{_{t}}\right\}$ in $E_{_{+}}$. The condition
(\ref{fitting}) is equivalent to the representation of the operator
$\Phi_{_{\Delta}}$ in form $ \beta_{_{01}} \,\, P_{_{N_{i}}}$, where
$\beta_{_{01}}$ is a bounded  operator acting from $E_{_{A}}$ into
$E_{_{+}}$ and $P_{_{N_{i}}}$ is an orthogonal projection  in
$E_{_{A}}$ onto the deficiency subspace $N_{_{i}}$. We  will
construct both $\beta_{_{01}},\, P_{_{N_{i}}}$ from the data encoded
in $\Phi_{\Delta}$.
\par
The non-negative Gram operator
$\Phi_{\Delta}\,\,\Phi^{^{+}}_{\Delta}$ in $E_{_{+}}$ has the
spectral representation
\[
\Phi_{\Delta}\,\,\Phi^{{+}}_{\Delta} = U^{^{+}} D \,\,\,U.
\]
The non-negative diagonal  matrix  $D$ is  invertible on the
orthogonal complement $\hat{E}$ of the corresponding null-space
$\hat{E}_{_{0}}$. We denote the restriction $D$ onto $\hat{E}$ by
$\hat{D}$.  One  can assume that  the subspace $\hat{E}$ belongs to
some  extended space $\hat{E} \oplus\hat{E}_{_{0}}$ which contains
$E_{_{A}}$, and the operator  $U^{^{+}}$ acts from $\hat{E}
\oplus\hat{E}_{_{0}}$ onto $E_{_{+}}$ as an isometry. The operator $
U^{^{+}} \hat{D}^{^{1/2}}$ coincides with $\Phi_{\Delta}$. Hence the
operator $\Phi_{\Delta}$ is  presented as a  product $\beta
\hat{P}$, with $\beta = \beta_{_{01}} = U^{^{+}}\hat{D}^{^{1/2}}:
\hat{E}\to E_{_{+}}$ and $\hat{P} = P_{_{\hat{E}}}:=
P_{_{N_{i}}}\subset E_{_{A}}$, dim $N_{i} =$ dim $\hat{E} = d$ and
coincides with the dimension of the resonance entrance subspace of
the intermediate operator. Up to some non-essential isometry we  may
assume  that $E_{_A} = E_{_{\Delta}}$,\,$A =
L^{^{\Delta}}_{_{\Lambda}}$,\, $N_{_{i}} = N_{_{\Delta}}$. The
condition  (\ref{non-overlap}) guarantees that  $N_{_{i}} \cup
N_{_{-i}} = 0$.

{\it End of the proof}
\par
In the case when only one resonance eigenvalue $\alpha^{^2}_{_0}$ of
the intermediate operator sits  on the essential spectral band, the
obtained  model scattering matrix
\begin{equation}
\label{one_pole} {\bf S}(p) = \frac{i K_{+} - k_M + \frac{1 +
\alpha_{_{0}}^{^4}}{\alpha_{_{0}}^{^2} - \lambda }\beta_{_{01}}
Q_{_{0}}\beta_{_{10}}}{i K_{+} +  k_M - \frac{1 +
\alpha_{_{0}}^{^4}}{\alpha_{_{0}}^{^2} - \lambda} \beta_{_{01}}
Q_{_{0}}\beta_{_{10}}}
\end{equation}
is  a single-pole approximation of  the scattering matrix  of the
network. The  condition of the  above  theorem  is obviously
fulfilled for the single-pole  approximation, when
$P_{_{+}}\frac{\partial \varphi_{_{0}}}{\partial n} \neq  0$,  $d =
1,\,N = 1$, and $\beta_{_0}$ is a  one-dimensional operator mapping
the one-dimensional subspace $N_{_{i}}$ onto  the resonance entrance
subspace  in $E_{_+}$ spanned  by $ P_{_{+}}\frac{\partial
\varphi_{_{0}}}{\partial n}$. For thin or shrinking  networks  one
can estimate, (see \cite{BMPPY} and  more details in \cite{MPP04,
MathNach07}) the deviation of the single-pole and/or few-poles
approximations from the exact scattering matrix on the network, in
terms of the ratio $d/\mbox{diam}\,\Omega_{in}$.
\par
We postpone the discussion of the  non-stationary scattering matrix
for QN to  forthcoming publications. But  we  notice here that the
local wave  operators  ( see \cite{Birman_68}) and  the
corresponding scattering matrix on the essential spectral band can
be  defined for the  pair $\left( {\mathcal{L}}, {\bf
A}_{_{\beta}}\right)$. \vskip0.3cm
\section{ A solvable model as  a jump-start in  the\\
analytic perturbation procedure}

Recall that the  exact scattering matrix was  approximated by the
essential or approximate scattering matrix. In this section we
consider this phenomenon from the point of view of complex analysis,
for the simplest star-shaped network constructed of a single model
quantum well with one semi-infinite wire attached to it.
\par
Consider a thin quantum network constructed of a quantum well
$\Omega_{in}$ and a single quantum wire  of width $\delta$ attached
to it,\, $\delta/\mbox{diam}\,\Omega_{in} <<1$. Assume that the
Fermi level is situated on the first spectral band in the wire,
which has multiplicity $1$. Without loss of generality we may assume
that the component of the corresponding solvable model in the open
channel is presented by the Schr\"{o}dinger equation with
$2\mu^{^{\parallel}} = 2\mu^{^{\bot}} = I, \, V^{^{\omega}} = V,\,
K_{_{+}} = p = \sqrt{\lambda - \frac{\pi^{2}}{\delta^{{2}}}-V }$ and
one-dimensional subspace $E_{_{+}}$:
\begin{equation}
\label{singlechan} - u'' = p^{^2} u,\, \, 0< x < \infty.
\end{equation}
 Assume that the  model Hamiltonian  ${\bf A}_{_{\beta}}$  is constructed
as suggested  in the previous section based on the ``inner
Hamiltonian" $A$, the differential operator $l_{_{\Lambda}}$ and the
boundary parameters which are reduced to the  coupling constant
$\beta_{_{01}}:= \beta$. Hereafter we will use the re-normalized
eigenvalues $\alpha_{_{s}}^{{2}}-\frac{\pi^{2}}{\, \delta^{{2}}} - V
:= k^{{2}}_{{s}} > 0$. Introducing that notation into the Krein
function, and  submitting the  boundary parameter to the condition
$\beta_{_{00}} + \sum_{_{s = 1}}^{^{N}} \alpha_{_{s}}^{{2}}
\beta_{_{01}}\, q_{{s}}\, \beta_{_{10}} = 0$ we obtain the
corresponding few-pole scattering matrix  as a function of the
wave-number $p$, with physically meaningful limit behavior at
infinity ${\bf S}^{^\beta}(p) \to I$:
\begin{equation}
\label{scalarsmatr} {\bf S}^{\beta}(p) = \frac{ip - k -
  \beta^{^2} \sum_{s}
\frac{1 + \alpha_{{s}}^{4}}{ p^{^2} - k_{{s}}^{^2} } q_{{s}}}{ip + k
+ \beta^{2} \sum_{_s} \frac{1 + \alpha_{{s}}^{^4}}{ p^{2}
-k_{_{s}}^{2}} q_{{s}}}.
\end{equation}
Here $q_{_s} = |\langle e,e_{_s}\rangle|^{2}$. Zeros $p_{_{s}}
(\beta)$ of the Scattering matrix (\ref{scalarsmatr}) - the
resonances - sit in the upper half-plane $\Im p > 0$ and approach
the points $ \pm k_{_s},\, k_{_{-s}}: = -k_{_s}$, when $\beta \to
0$.
\par
Assume  that the resonance eigenvalue $\alpha^{^2}_{_{0}} =
k_{_{0}}^{2} + \delta^{{-2}}\,\pi^{{2}} + V$ is situated  close to
the  scaled Fermi-level $\Lambda$ and  the coupling constant $\beta
:= \beta_{_{01}}$ is relatively small, see below. Separating the
resonance term in the numerator and denominator of
(\ref{scalarsmatr})
\[
\left[ip - \beta^{^2} \frac{1 + \alpha_{_{0}}^{^4}}{k_{_{0}}^{^2} -
p^{^2}} q_{_{0}} \right] + \left[k + \beta^{^2} \sum_{_{s\neq
0}}\frac{1 + \alpha_{_{s}}^{^4}}{ p^{^2} - k_{_{s}}^{^2} }
q_{_{s}}\right],
\]
and multiplying by  $(ip)^{^{-1}}$ one can see that $-
\frac{\beta^{^2}}{ip}\left[ k + \sum_{_{s\neq 0}}\frac{1 +
\alpha_{_{s}}^{^4}}{ p^{^2} - k_{_{s}}^{^2}} q_{_{s}}\right]$ plays
the  role of the  small parameter. The resonance $ k_{_{0}}(\beta)$
originated from the eigenvalue $ k^{^2}_{_0} $ of the operator $A$
(more precisely : from the point $ + k_{_0}$ ) can be obtained as a
solution  $p = k_{_{0}}(\beta)$ of the equation
\begin{equation}
\label{disp}
  p = k_{_{0}} - \frac{\beta^{^2} (1+ \alpha^{^4}_{_{0}})
q_{_0} }{(p + k_{_{0}} )\left( ip + k + \beta^{^2} \sum_{_{s\neq 0}}
\frac{1 + \alpha_{_{s}}^{^4}}{p^{^2} - k_{_{s}}^{^2}}
q_{_{s}}\right)}.
\end{equation}
Another  resonance  originated  from the  point $- k_{_{0}}$
corresponds  to the same eigenvalue, and it sits at the  symmetric
point $ -\bar{k}_{_{0}}(\beta)$ with respect to  the  imaginary
axis. Remaining  resonances  $k_{_s} (\beta),\,\, s\neq 0$,  can be
found from similar equations. All  functions  $k_{_{s}} (\beta) $
are analytic functions  of $\beta,\, k$ in small neighborhoods of
$(0,\, \pm k_s)$. They sit in the upper  half-plane symmetrically
with respect to the imaginary axis  $k_{_{- s}} (\beta) = -
\bar{k}_{_{s}} (\beta) $. The scattering  matrix (\ref{scalarsmatr})
is unitary on the  real axis  $k$ and  has poles at  the
complex-conjugate  points $\bar{k}_{_{s}} (\beta)$ in the  lower
half-plane, and hence  it is presented  by  the finite Blaschke
product  which tends  to $1$ when  $|k|\to \infty$:
\begin{equation}
\label{SBlashke} {\bf S}^{^{\beta}} (p) =  \prod_{s}\frac{p -
k_{_s}(\beta)}{p - \bar{k}_{_s}(\beta)}.
\end{equation}
The  outer  component of the  scattered  wave  is  presented as
\begin{equation}
\label{F_ansatz} \Psi^{^{\beta}}_{_{0}} = e^{ - ipx}  +  {\bf
S}^{^{\beta}}(k) e^{ ipx}.
\end{equation}
It fulfills appropriate  boundary  condition  at the  place  of
contact with the model quantum dot. The  inner  component of the
 scattered wave can be obtained from  Lemma 3.4.
\par
We  explore  the  model scattering problem  for  small values of
$\beta$. Though the resonances depend analytically on $\beta$,
neither the scattering matrix (\ref{scalarsmatr},\ref{SBlashke}) nor
the scattered wave depend analytically of  $ \left(\beta, p\right)$
on the  product of a small neighborhood of the  origin in $\beta$ -
plane  and small neighborhoods  of $\pm k_{_{0}}$ in $p$-plane. The
analyticity is lost due to presence of the points $\pm k_{_{0}}$
where the resonances are created at $\beta = 0$: both $k_{_{0}}
(\beta)$ and $\bar{k}_{_{0}} (\beta)$ approach the same  point
$k_{_{0}}$ when $\beta \to 0$. The corresponding ``resonance''
factor of the scattering matrix
\begin{equation}
\label{onepole2} {\bf S}^{^{\beta}}_{_{0}} (p) = \frac{[p -
k_{_0}(\beta)]\, [p + \bar{k}_{_0}(\beta)]}{[p -
\bar{k}_{_0}(\beta)]\,[p + k_{_0}(\beta)]},
\end{equation}
is  non-analytic on $(\Omega_{{\beta}}\times\Omega_{_{k_{_0}}}\times
\Omega_{_{-k_{_0}}})$, though $k_{_{\beta}} = \Re k_{_{\beta}} + i
\Im k_{_{\beta}}$ is analytic function of $\beta$ due to
(\ref{disp}). But the complementary factor of the scattering matrix
\begin{equation}
\label{complement} {\bf S}_{_{\beta}}^{^{0}} (p) = \prod_{s\neq
0}\frac{p - k_{_s}(\beta)}{p - \bar{k}_{_s}(\beta)}.
\end{equation}
is  analytic on $(\Omega_{{\beta}}\times\Omega_{_{k_{_0}}}\times
\Omega_{_{-k_{_0}}})$ and can be expanded into the  power series
over $\beta^{^{m}},\, m=0,1,2,\dots$. Assume now that the function
$k_{_{0}} (\beta)$ is  known. Then the  following statement is true:

\begin{theorem} There  exists  a  one-dimensional perturbation
${\bf A}^{^{\beta}}_{_0}$  of  the  operator
\[
l_{_0} u =  -u'' ,\,\, u \bigg|_{_0} = 0
\]
with a  non-trivial  inner  component,  such that  the  scattering
matrix of the  pair  $({\bf A}^{^{\beta}}_{_0},l_{_0})$ coincides
with $-{\bf S}^{^{\beta}}_{_{0}} (p)$. Then the scattering matrix
${\bf S}$ of the complementary pair
  $({\bf A}_{_{\beta}},{\bf A}^{^{\beta}}_{_0})$
is  equal to  the  complementary  factor $ - {\bf
S}_{_{\beta}}^{^{0}} (p) $:
\[
{\bf S}^{^{\beta}} (p) = {\bf S}^{^{\beta}}_{_{0}} (p)\,\,{\bf
S}_{_{\beta}}^{^{0}} (p).
\]
The  complementary  factor is an analytic function  of $(\beta,\,
k)$ on the product $(\Omega_{{\beta}}\times\Omega_{_{k_{_0}}}\times
\Omega_{_{-k_{_0}}})$  of a  small neighborhood of the origin  in
$\beta$-plane  and
  a  small  neighborhood of the  pair
   $(k_{_{0}},\,-k_{_{0}} )$ in the  $p$-plane.
\end{theorem}
{\it Proof} is presented in \cite{Pavlov_C07}. \vskip0.3cm {\bf
Corollary 1} The exterior component of the scattered wave of the
operator ${\bf A}_{_0}^{^{\beta}}$ presented by the Ansatz
(\ref{F_ansatz}) with ${\bf S}_{\beta}$ taken in the form
(\ref{SBlashke}) is not analytic with respect to the coupling
constant $\beta_{_{01}}:= \beta$ near the origin. The
non-analyticity of the scattered  wave is  caused by the presence of
the non-analytic factor $- {\bf S}_{_0}^{^{\beta}}$ in the
scattering matrix. From theorem 4.1, we interpret this factor as the
scattering matrix for the  pair $\left({\bf A}_{_0}^{^{\beta}},\,
l_{_{0}}\right)$. The complementary factor $- {\bf
S}^{^0}_{_{\beta}}$ is  analytic with respect to the coupling
constant $\beta$. It can be interpreted as the scattering matrix for
the pair $\left({\bf A}^{^{\beta}},{\bf A}_{_0}^{^{\beta}}\right)$.
Summarizing our  observation we suggest, for  our  example, the
following  two-steps modification of the analytic perturbation
procedure  on continuous  spectrum:

a. First step is the construction of the  solvable model and
calculation of the corresponding (non-analytic with respect to the
coupling constant $\beta$ at the origin) scattering matrix. This is
``the jump-start'' of the analytic perturbation procedure.

b. Second step  is the calculation of the analytic factor of the
scattering matrix of the  model by the  standard analytic
perturbation procedure.  The  analytic factor is interpreted as the
scattering matrix between the constructed solvable model and the
perturbed operator ${\bf A}^{^{\beta}}$.
\par
The  obtained  connection between resonances and  analytic
perturbation series on the continuous  spectrum  recalls the
connection between  small denominators in celestial mechanics and
divergence of perturbation series, observed by  H. Poincar\'{e}, see
\cite{Poincare}. More historical comments about  intermediate
Hamiltonian and the jump-start may be found  in \cite{PA05}, where
similar modification of the analytic perturbation procedure  for the
Friedrichs model is suggested.
\par{\bf Remark 3}
Note that recovering {\it exact} information on the resonance
$k_{_{0}}(\beta)$ and on the corresponding residue for the perturbed
operator  ${\bf A}_{_{\beta}}$, which we need to develop the
``jump-start" procedure, may be a tricky problem almost equivalent
to the original spectral problem. On the other hand, if the {\it
approximate} resonance factor ${\bf S}_{_0}^{^{\beta}}$ is used
instead the the  exact factor, then the division of the scattering
matrix through ${\bf S}_{_0}^{^{\beta}}$ would not eliminate
singularity, hence the complementary factor of the scattering matrix
would be  still non-analytic at the  origin and hence could not  be
obtained via analytic perturbation procedure.

\section{Acknowledgement }
The author acknowledges support from the Russian Academy of
Sciences, Grant RFBR  03-01- 00090. The  author is grateful to V.
Katsnelson for important references and very interesting materials
provided, and to  M. Harmer for  deep remarks concerning  the
fitting of  the star-graph model.
 \vskip0.3cm
\section{Appendix: symplectic operator extension procedure}

John von Neumann in 1933 has found conditions which guarantee
existence of a self-adjoint extension of given unbounded symmetric
operator, and suggested a procedure of construction of the
extension, see symplectic version in \cite{Extensions}.  For given
symmetric operator ${\mathcal{A}}_0$ defined on $D_0$  in the
Hilbert space $H$, see \cite{Glazman}  and given  complex value
$\lambda,\,\Im \lambda \neq  0$ of the spectral parameter:

\begin{definition} \label{Def_supspace} Define  the {\bf deficiency
subspaces}
\[
N_{\lambda}:= H \ominus  \overline{ \left[{\mathcal{A}}_0 - \lambda
I\right]\,\, D_0},
\]
\[
N_{\bar{\lambda}}: = H \ominus  \overline{ \left[{\mathcal{A}}_0 -
\bar{\lambda} I\right]\,\, D_0}.
\]
\end{definition}

 The dimension of $N_{\lambda},\,N_{\bar{\lambda}}$ is constant on
the whole  upper and lower spectral  half-plane  $\Im\lambda
>0,\,\Im\lambda <0 $ respectively.

\begin{definition} \label{Def_index} Introduce the deficiency index
$\left( \mbox{dim} N_{\lambda},\,\mbox {dim} N_{\bar{\lambda}}
\right):= (n_+,\, n_-)$ of the operator ${\mathcal{A}}_0$.
\end{definition}
J. von Neumann  proved that
\begin{theorem}
\label{Extension}{ The hermitian operator ${\mathcal{A}}_0$  has  a
self-adjoint extension if and  only if $n_+ = n_-$ }
\end{theorem}

The  idea of  construction of the  extension is based on the
following theorems von Neumann, see for instance \cite{Glazman}:

{\begin{theorem} \label{Neumann1}{  The domain of the adjoint
operator is represented as a direct sum of the domain
$D_{\bar{A}_{0}}$ of the closure and the deficiency subspaces, in
particular:
\[
D_{A^+_0}= D_{\bar{A}_{0}} +  N_i + N_{-i}.
\]

The deficiency subspaces of the densely-defined operator are the
eigen-spaces of the adjoint operator:
\[
{\mathcal{A}}^+_{0} e_i = -i e_i,\,\, e_i \in N_i,\,\, {\mathcal{
A}}^+_{0} e_{-i} = i e_{-i},\,\, e_{-i} \in N_{-i}.
\]
}
\end{theorem}
\begin{theorem} \label{Neumann2} { If  ${\mathcal{A}}_0$
is an Hermitian operator with deficiency indices $\left(
n_+,\,n_-\right)$,\\ $n_- = n_+$ and $V$ is an isometry $V : N_i \to
N_{-i}$. Then the  isometry $V$ defines\\ a self-adjoint extension
${\mathcal{A}}_V$ of ${\mathcal{A}}_0$, acting on the domain
\[
D_{A_V} = D_{\bar{A}_{0}} + \left\{ e_i + V e_i ,\,\, e_i \in
N_i\right\}
\]
as a restriction of ${\mathcal{A}}^+_0$ onto $D_{A_V}$:
\[
A_{V}: u_0 + e_i + V e_i \to \bar{\mathcal{A}}_0 u_0 - i e_i + i V
e_i.
\]
}
\end{theorem}

J. von Neumann  reduced  the construction of the  extension of the
symmetric operator ${\mathcal{A}}_0$ to an equivalent problem of
construction of an extension of the corresponding isometrical
operator - the Caley transform  of  ${\mathcal{A}}_0$. It is  much
more convenient, for differential operators, to construct the
extensions based on so-called {\it boundary form}.

{\bf Example 5. Symplectic Extension procedure for the differential
operator} Consider the second order differential operator
\[
L_0 u = - \frac{ d^2 u}{d x^2},
\]
defined  on all square integrable functions, $u\in L_2(0,\infty)$,
with square- integrable derivatives of the first and second  order
and  vanishing near the origin. This  operator is  symmetric and
it's adjoint $L_0^+$ is  defined by the  same differential
expression on  all square integrable functions with square
integrable derivatives of the first and  second  order and no
boundary condition at the origin. This  operator is not symmetric:
its boundary form
\[
{\mathcal{J}} (u,v) = \langle L_0^+ u,v\rangle - \langle u,L_0^+ v
\rangle = u'(0)\bar{v}(0)- u(0)\bar{v}'(0),\,\, u,v \in D_{L_0^+}
\]
is generally non equal to zero for  $u,v \in D_{L_0^+}$.  But  it
vanishes on  a "Lagrangian plane" ${\mathcal{P}}_\gamma \subset
D_{L_0^+} $ defined by the boundary condition
\[
u'(0) = \gamma u(0),\, \gamma = \bar{\gamma}.
\]
The restriction $L_{\gamma}$ of the  $L_0^+$ onto the Lagrangian
plane ${\mathcal{P}}_\gamma $ is a  self-adjoint  operator in  $L_2
(0,\infty)$: it is  symmetric, and  the inverse of it $(L_{\gamma}-
\lambda I)^{-1}$, at each complex spectral point $\lambda$, exists
and is defined on the whole space  $L_2 (0,\infty)$.

The  operator extension procedure used  above for the differential
operator,  can be applied to general symmetric operators and serves
a convenient alternative for  construction  of solvable models  of
orthogonal sums of differential operators  and finite matrices. We
call the abstract analog of the extension procedure  the {\it
symplectic  extension procedure}. Let $A$ be  a  self-adjoint
operator in a finite-dimensional Hilbert space $E$, dim $E = d$, and
$N_i := N$ is  a subspace of $E$, dim $N = n < d/2$, which does not
overlap with $\frac{A + iI}{A - iI} N_i := N_{-i}$:
\[
N_i \cap N_{-i} = \left\{ 0\right\}.
\]
Define the  operator $A_0$ as a  restriction of $A$ onto  $D_0 :=
\frac{I}{A-iI} E \ominus N$. This  operator is  symmetric, and the
subspaces $N_{\pm i}$ play roles of it's deficiency subspaces. The
operator  can $A_0$ can be extended to the  self-adjoint operator
$A_{\Gamma} \supset A_0$ via  simplectic  extension procedure
involving the  corresponding boundary form: selecting a basis
$\left\{e^+_s\right\}_{s=1}^n := g_s\in N_i$, we consider the dual
basis  $\left\{ \frac{A + iI}{A - iI} g_s = g_s^-
 \right\}_{s=1}^n \in N_i$. Introduce, following
 \cite{Extensions}, another basis in
the  defect $N = N_i + N_{-i}$
\[
W^+_s =  \frac{1}{2}\left[ g_s +  \frac{A+iI}{A-iI} g_s \right],\,\,
W^-_s =  \frac{1}{2i}\left[ s_s - \frac{A+iI}{A-iI} g_s \right].
\]
Due to  $A_0^+ g_s  + i g_s = 0,\,\, [A_0^+  - iI] \frac{A+iI}{A-iI}
g_s = 0$ we  have,

\[
 A_0^+ W^+_s = W^-_s, \,\, A_0^+ W^-_s = - W^+_s.
\]
Following  [24] we  will use  the  representation of elements from
the domain of the adjoint operator by the expansion on
 the new basis:
\[
u = u_0 + \sum_{s} \xi^s_+ W^+_s  + \xi^s_- W^-_s,
\]
with  $u_0 \in D(A_0)$ and simplectic coordinates  $\xi^s_{\pm}$.

We also introduce the {\it boundary vectors } of  elements from
$D(A_0^+)$
\[
\vec{\xi}_{\pm}:=  \sum_{s} \xi^s_{\pm} g_s \in N_i,
\]
\[
 u = u_0 +  \frac{A}{A - iI}\vec{\xi}^u_{+} - \frac{I}{A -
iI}\vec{\xi}^u_{-}: = u_0 +  n^u,\, u_0 \in D(A_0) \, n^u \in N.
\]
Define  the formal adjoint operator $A_0^+$ on the  defect ${\bf N}=
N_i + N_{-i}$ as:
\[
A_0^+ e_+ = -i e_+,\,\mbox{for}\, e_+ \in N_i, \,\,
 A_0^+ e_- = i e_-, \,\mbox{for}\, e_+ \in N_{-i},
\]
\[
 A_0^+ ( e_+ + e_-) = - ie_+ + i e_-.
\]
Then we  have:
\[
 A_0^+ W^+_s = W^-_s, \,\, A_0^+ W^-_s = - W^+_s.
\]
Following \cite{Extensions},  we  will use  the  representation of
elements from the domain of the adjoint operator by the expansion on
 the new basis:
\[
 u = u_0 + \sum_{s} \xi^s_+ W^+_s  + \xi^s_- W^-_s,
\]
with  $u_0 \in D(A_0)$ and simplectic coordinates  $\xi^s_{\pm}$. We
also introduce the {\it boundary vectors } of  elements from
$D(A_0^+)$
\[
\vec{\xi}_{\pm}:=  \sum_{s} \xi^s_{\pm} g_s \in N_i,
\]
\[
 u = u_0 +  \frac{A}{A - iI}\vec{\xi}^u_{+} - \frac{I}{A -
iI}\vec{\xi}^u_{-}: = u_0 +  n^u,\, u_0 \in D(A_0) \, n^u \in N.
\]
Then the  boundary form of  $A_0^+$ is  calculated as
\[
\langle A_0^+ u, v \rangle - \langle u,A_0^+ v \rangle : =
{\mathcal{J}}(u,v) = \langle \vec{\xi}^u_+, \vec{\xi}^v_-\rangle -
\langle \vec{\xi}^u_-, \vec{\xi}^v_+\rangle
\]
\vskip0.3cm
\subsection{Operator Extensions: Krein formula }

{\bf Theorem}: {\bf Krein formula} {\it Consider a  closed symmetric
operator $A_0$ in the  Hilbert space $\mathcal{H}$, obtained via
restriction of the  self-adjoint operator $A$ onto  the dense domain
$D (A_0)$, with finite-dimensional deficiency subspaces $N_{\mp
i},\, P_{N_i}:= P_+$,  dim $N_i =$  dim $N_{-i}$. Then the resolvent
of the selfadjoint extension $A_M$ defined by the boundary
conditions
\begin{equation}
\label{M}
 \vec{\xi}_+ = M \vec{\xi}_-
\end{equation}
is  represented, at regular points of $A_M$, by the formula:
\[
\left( A_M - \lambda I \right)^{-1} = \frac{I}{ A - \lambda I} -
\frac{A+iI}{A -\lambda I} P_+  M \frac{I}{ I + P_+ \frac{I+ \lambda
A}{A - \lambda I} P_+ M} P_+  \frac{A-iI}{A -\lambda I}
\]
}
 {\it Proof}. For the convenience of the  reader we  provide below
the sketch of the proof of the  Krein formula via simplectic
operator extension procedure. Solution of the homogeneous equation
$\left(A^+ - \lambda I\right) u = f$ is reduced to finding $u_0,
\vec{\xi}_{\pm}$ from the equation
\begin{equation}
\label{u0}
 (A-\lambda I)u_0 - \frac{I + \lambda A}{A -
iI}\vec{\xi}^u_+  -  \frac{A-\lambda I}{A - iI}\vec{\xi}^u_- = f.
\end{equation}
Applying to  this  expression the operator $\frac{A - iI}{A-\lambda
I}$, due to  $(A-iI)u_0 \bot N_i$, we  obtain
\[
\vec{\xi}_- = - \frac{I}{I + \frac{I+ \lambda A}{A - \lambda I}P_+}
P_+  \frac{A-iI}{A -\lambda I}  f.
\]
Then, from the above equation (\ref{u0}) and from the boundary
condition (\ref{M}), we derive:
\[
u_0 = \frac{1}{A-iI}\left[ \frac{I + \lambda A}{A-\lambda I}
\vec{\xi}_+ + \vec{\xi}_- \right] + \frac{I}{A -\lambda I} f,
\]
and
\[
u = u_0 + \frac{A}{A -iI} \vec{\xi}_+ -  \frac{I}{A -iI} \vec{\xi}_-
=
\]
\[
\frac{I}{A - \lambda I} f - \frac{A+iI}{A -\lambda I} P_+ M
\frac{I}{ I + P_+ \frac{I+ \lambda A}{A - \lambda I}P_+ M} P_+
\frac{A-iI}{A -\lambda I}f.
\]

{\it The end of the proof}


\begin{thebibliography}{99}
\bibitem{AA_1965} V. Adamjan, D. Arov {\it  On a class of scattering
operators and characteristic operator-functions of contractions}.
(Russian) Dokl. Akad. Nauk SSSR 160 (1965) pp 9--12.

\bibitem{APY_09} V. Adamyan, B. Pavlov, A. Yafyasov {\it  Modified
Krein Formula and analytic perturbation procedure for scattering on
arbitrary junction.} International Newton Institute report series
NI07016, Cambridge, 18 April 2007, 33p. To be  published in the
proceedings of  M.G.Krein  memorial conference, Odessa, April 2007.

\bibitem{Glazman} N.I.Akhiezer, I.M.Glazman, {\it Theory of
Linear Operators in  Hilbert Space}, (Frederick Ungar, Publ.,
New-York, vol. 1, 1966) (Translated from  Russian by M. Nestel)

\bibitem{AK00}  S.Albeverio, P. Kurasov \textit{Singular Perturbations of
Differential Operators}, London Math. Society Lecture Note Series
271. Cambridge University Press (2000)

\bibitem{Albeverio} S. Albeverio, F. Gesztesy, R. Hoegh-Krohn, H.
Holden, {\it Solvable models in quantum
mechanics}.\,{Springer-Verlag},\,{New York},{1988}.

\bibitem{BMPPY} N.Bagraev, A.Mikhailova, B. Pavlov, L.Prokhorov,
A.Yafyasov {\it Parameter regime of a resonance quantum switch.} In:
Phys. Rev. B, 71, 165308 (2005), pp 1-16.

\bibitem{BF61} F.A.Berezin, L.D.Faddeev {\it A  remark on Schr\"{o}dinger
equation with a  singular potential}  Dokl. AN SSSR, {\bf 137}
(1961) pp 1011-1014

\bibitem{Birman_68} M. Birman {\it A local test for the existence of wave operators}.
 (Russian) Izv. Akad. Nauk SSSR Ser. Mat. 32 1968 914--942.

\bibitem{Ring} V.Bogevolnov, A.Mikhailova,B. Pavlov, A.Yafyasov
{\it About  Scattering  on  the  Ring} In: "Operator  Theory :
Advances  and Applications", Vol 124 (Israel Gohberg Anniversary
Volume), Ed. A. Dijksma, A.M.Kaashoek, A.C.M.Ran, Birkh\"{a}user,
Basel (2001) pp 155-187.

\bibitem{BP07} J. Bruening, B. Pavlov  {\it On calculation of Kirchhoff constants
of helmholtz resonator}. International Newton Institute report
series NI07060-AGA, Cambridge, 04 September 2007, 40p.

\bibitem{CH1}
R. Courant, D. Hilbert, {\it Methods of mathematical physics.} Vol.
II. {\it Partial differential equations}. Reprint of the 1962
original. Wiley Classics Library. A Wiley-Interscience Publication.
John Wiley \& Sons, Inc., New York (1989). xxii+830 pp.

\bibitem{Datt}
S.~Datta.
\newblock {\it Electronic Transport in Mesoscopic systems}.
\newblock Cambridge University Press, Cambridge (1995)

\bibitem{DattaAPL} S. Datta and B. Das Sarma  \textit{Electronic
analog of the electro-optic modulator}.
 Appl. Phys. Lett. {\bf 56},7 (1990) pp 665-667.

\bibitem{Demkov}  Yu.N.Demkov, V.N.Ostrovskij, {\it Zero-range potentials and
their applications in Atomic Physics}, Plenum Press, NY-London,
(1988).

\bibitem{Exner88}  P. Exner, P. \^{S}eba. \textit{A new type of quantum
interference transistor}. Phys. Lett. A 129:8,9, 477 (1988)

\bibitem{EP05} P.Exner, O.Post {\it Convergence of graph-like thin
manifolds} J. Geom. Phys. {\bf 54},1, (2005) pp 77-115.

\bibitem{Opening}  M. Faddeev, B. Pavlov. \textit{Scattering by resonator
with the small openning}. Proc. LOMI, v126 (1983). (English
Translation J. of Sov. Math. v27, 2527 (1984)

\bibitem{Fermi} E.Fermi {\it Sul  motto dei
neutroni nelle sostance idrogenate} (in Italian) Richerka
Scientifica {\bf 7} p 13 (1936)

\bibitem{GS_97} F. Gesztesy,\, B.Simon,
{\it Inverse spectral analysis with partial information on the
potential. I. The case of an a.c. component in the spectrum.} Papers
honouring the 60th birthday of Klaus Hepp and of Walter Hunziker,
Part II (Z\"{u}rich, 1995). Helv. Phys. Acta 70, no. 1-2, (1997) pp
66--71.

\bibitem{Gezstezy05} F. Gesztezy, Y. Latushkin, M. Mitrea and
M. Zinchenko {\it Non--selfadjoint operators, infinite determinants
and some applications}, Russian Journal of Mathematical Physics,
{\bf 12}, 443--71 (2005).

\bibitem{Gezstezy06} F. Gesztezy, M. Mitrea and M. Zinchenko, {\it  On
Dirichlet-to-Neumann maps and some applications to modified Fredholm
determinants} \textit{preprint}, (2006).

\bibitem{Gorbachuk}V.I. Gorbachuk, M.L. Gorbachuk. {\it Boundary value problems for
operator differential equations}. Translated and revised from the
1984 Russian original. Mathematics and its Applications (Soviet
Series), 48. Kluwer Academic Publishers Group, Dordrecht, 1991.
xii+347

\bibitem{Gramotnev_2000} D. Gramotnev, D. Pile {\it  Double resonant
 extremely asymmetrical scattering of electromagnetic waves in
 non-uniform periodic arrays} In:  Opt. Quant. Electronics., {\bf
 32}, (2000) pp 1097-1124.

\bibitem{Grieser07} D. Grieser {\it Spectra of graph neighborhoods and scattering}.
Proc. Lond. Math. Soc. (3) 97, no. 3 (2008) pp. 718--752.

\bibitem{Har2_00} M.~Harmer.
\newblock Hermitian symplectic geometry and extension theory.
\newblock {\em Journal of Physics A: Mathematical and General}, (\bf 33)
(2000)  pp 9193--9203.

\bibitem{Harmer04} M. Harmer {\it Fitting parameters for a Solvable Model of
a  Quantum  Network} The University of Auckland, Department of
Mathematics report series 514 (2004), 8 p.

\bibitem{HPY07} M. Harmer, B. Pavlov, A. Yafyasov {\it Boundary condition
at the  junction}, in: Journal of  Computational Electonics,{\bf
6}(2007)  pp 153-157.

\bibitem{Kato} T. Kato {\it Perturbation theory for linear
operators} Springer Verlag, Berlin-Heidelberg-NY, second edition
(1976)

\bibitem{Keating03} J.P.Keating, J. Marlof, B. Winn {\it Value  disttribution
of the  eigenfunctions and spectral determinants of quantum
star-graphs }  Communication of Mathematical Physics, {\bf 241}, 2-3
(2003) pp 421-452.

\bibitem{Keating04} J.P. Keating, B. Winn  {\it No quantum ergodicity for
star graphs}  Communication of Mathematical Physics, {\bf 250}, 2
(2004) pp 219-285.

\bibitem{Kirchhoff} G.R.Kirchhoff.{\it Gesammelte Abhandlungen}
Publ. Leipzig: Barth, 1882, 641p.

\bibitem{Kirchhoff_constants_08} J. Br\"{u}ning, B. Pavlov {On calculation of
Kirchhoff constants for  Helmholtz resonator} International Newton
Institute, report series NI07060-AGA, Cambridge, 04 September, 2007,
38 p.

\bibitem{Schrader}  V. Kostrykin and R. Schrader. \textit{Kirchhoff's rule
for quantum wires}. J. Phys. A: Math. Gen. {\bf 32}, 595 (1999

\bibitem{Krasn} M.A.Krasnosel'skij {\it On selfadjoint extensions of
Hermitian Operators} (in Russian) Ukrainskij Mat.Journal {\bf 1}, 21
(1949)

\bibitem{K} M. G. Krein {\it Concerning the resolvents of an Hermitian
operator with deficiency index (m,m)},Doklady  AN USSR ,{\bf 52}
(1946) pp 651-654.

\bibitem{Kuch01} P. Kuchment, H. Zeng {\it
 Convergence of Spectra of mesoscopic Systems
Collapsing  onto Graph} Journal of Mathematical Analysis and
Applications, {\bf 258}(2001) pp 671-700.

\bibitem{Kuch02} P. Kuchment {\it Graph models for  waves in thin structures}
Waves in Periodic and  Random Media, {\bf 12},1 (2002) R 1 - R 24

\bibitem{Marsden2003} S. Lall, P. Krysl, J. Marsden {\it
Structure-preserving model reduction for mechanical systems} In:
Complexity and nonlinearity in physical systems (Tucson, AZ, 2001),
Phys. D {\bf 184}, 1-4 (2003) pp 304-318.

\bibitem{Lax}  Lax, Peter D.; Phillips, Ralph S. {\it Scattering theory.}
 Second edition. With appendices by Cathleen S. Morawetz and Georg Schmidt.
 Pure and Applied Mathematics, 26.
 Academic Press, Inc., Boston, MA, (1989) xii+309 pp.

\bibitem{Livshits62} M.S.Livshits {\it  Method of nonselfadjoint operators
in the theory of waveguides}  In: Radio Engineering and Electronic
Physics. Pulb. by  American Institute of Electrical Engineers, {\bf
1} (1962) pp 260-275

\bibitem{Madelung} O. Madelung. {\it Introduction to solid-state
theory}. Translated from German by B. C. Taylor. Springer Series in
Solid-State Sciences, 2. Springer-Verlag, Berlin, New York ( 1978)

\bibitem{MPP04} A. Mikhailova, B. Pavlov, L. Prokhorov{\it
 Modeling of quantum networks}' arXiv
math-ph: 031238,\,2004,\,69 p.

\bibitem{MathNach07} A. Mikhailova, B. Pavlov, L. Prokhorov.
  {\it Intermediate Hamiltonian via  Glazman
splitting and analytic perturbation  for meromorphic
matrix-functions.} In: Mathematishe Nachrichten, {\bf 280}, 12,
(2007) pp 1376-1416

\bibitem{MP_08}A. Mikhailova,\,B. Pavlov {\it Remark on compensation of
singularities  in Krein formula} Accepted by: {\it Operator Theory:
 Advances and Applications} Proceedings of OTAMP06, Lund. Editors:
 S. Naboko, P. Kurasov. 16 p.

\bibitem{ML71}R. Mittra, S. Lee {\it Analytical techniques in the theory
of  guided waves} The  Macmillan Company, NY, Collier-Macmillan
Limited, London, 1971.

\bibitem{Mittra_77} L.Ko,\, R. Mittra,
{\it A new approach based on a combination of
integral equation and asymptotic techniques for solving
electromagnetic scattering problems} IEEE Trans. Antennas and
Propagation AP-25, no. 2, (1977) pp 187--197.

\bibitem{Neumann} J. von Neumann {\it Mathematical foundations
of quantum mechanics} Twelfth printing. Princeton Landmarks in
Mathematics. Princeton Paperbacks. Princeton University Press,
Princeton, NJ, (1996)

\bibitem{Newton} R.G. Newton {\it Scattering theory of waves
and particles}Newton, Reprint of the 1982 second edition [Springer,
New York; MR0666397 (84f:81001)], with list of errata prepared for
this edition by the author. Dover Publications, Inc., Mineola, NY,
2002.

\bibitem{NK_87} N. Nikol'skii, S. Khrushchev, {\it  A functional model and some
problems of the spectral theory of functions} (Russian) Translated
in Proc. Steklov Inst. Math. 1988, no. 3, 101--214. Mathematical
physics and complex analysis (Russian). Trudy Mat. Inst. Steklov.
176 (1987), 97--210, 327.

 \bibitem{Sz_N_F_1970} B.Sz.-Nagy, C., Foias,{\it  Harmonic analysis of
 operators on Hilbert space}. Translated from the French and revised North-Holland
 Publishing Co., Amsterdam-London; American Elsevier Publishing Co., Inc.,
 New York; Akadémiai Kiadó, Budapest 1970 xiii+389 pp.

  \bibitem{Novikov_99} S. Novikov,
  {\it Schrodinger operators on graphs and symplectic geometry}
  In : The Arnoldfest (Toronto, ON, 1997), Ed.: Fields Inst. Commun.,
  24, Amer. Math. Soc., Providence, RI, (1999)  pp 397--413.

\bibitem{Pavlov_73} B. Pavlov{\it On one-dimensional scattering of
plane waves on an arbitrary potential}, Teor. i Mat. Fiz.,v16, N1,
1973, pp 105-119.

\bibitem{Extensions} B. Pavlov {\it The theory of extensions and
explicitly solvable models} (In Russian) Uspekhi Mat. Nauk, {\bf
42}, (1987) pp 99-131

\bibitem{DN01} B. Pavlov. {\it S-Matrix and Dirichlet-to-Neumann  Operators} In:
\textit{ Encyclopedia of Scattering}, ed. R. Pike, P. Sabatier,
Academic Press, Harcourt Science and  Tech. Company (2001) pp
1678-1688

\bibitem{PA05} B. Pavlov, I. Antoniou
{\it  Jump-start in analytic perturbation procedure for Friedrichs
model.} In J. Phys. A: Math. Gen. 38 (2005) pp 4811-4823.

\bibitem{Hadr05} B. Pavlov, V. Kruglov {\it Operator Extension technique
for resonance scattering of neutrons  by  nuclei.} In: Hadronic
Journal {\bf 28} (2005) pp 259-268.

\bibitem{NZMJ05}B. Pavlov, V. Kruglov  {\it Symplectic  operator-extension
technique and  zero-range quantum models} In:  New Zealand
mathematical Journal 34,2 (2005) pp 125-142.

\bibitem{PY07} B. Pavlov, A. Yafyasov {\it Standing waves and
resonance transport  mechanism in quantum networks } With  A.
Yafyasov, Surface  Science 601 (2007), pp 2712 - 2716

\bibitem{Pavlov_C07} B. Pavlov {\it  A star-graph model via operator  extension}
Mathematical Proceedings of the Cambridge Philosophical Society,
Volume 142, Issue 02, March 2007, pp 365-384.

\bibitem{PRRS08}B. Pavlov. T. Rudakova, V. Ryzhii, I. Semenikhin
{\it Plasma waves in two-dimensional electron channels:
 propagation and trapped modes.} Russian Journal of mathematical
 Physics, {\bf 14},4 (2004)pp 465-487.

\bibitem{PP08} L. Petrova, B. Pavlov {\it Tectonic plate
under  a localized boundary stress: fitting of a zero-range solvable
model.} Jounal of Physics A, {\bf 41} (2008) 085206 (15 pp)

\bibitem{Poincare} H. Poincare {\it Methodes nouvelles de la
m\'{e}canique celeste} Vol. 1 (1892), Second edition: Dover, New
York (1957)

\bibitem{PS96} C. Presilla, J. Sjostrand {\it Transport properties in
resonance tunnelling heterostructures} In: J. Math. Phys. {\bf 37},
10 (1996), pp 4816-4844.

\bibitem{Prigogine73} I. Prigogine {\it Irreversibility as a
Symmetry-breaking Process} In : Nature, {\bf 246}, 9  (1973)

\bibitem{Rayleigh} Lord Rayleigh {\it The  theory of  Helmholtz resonator} Proc.
Royal Soc. London {\bf 92} (1916) pp 265-275.

\bibitem{RS01} J.Rubinstein, M.Shatzman {\it Variational approach  on
multiply connected  thin strips I : Basic estimates and convergence
of the Laplacian spectrum} Arch. Ration. Mech. Analysis {\bf 160},
4, 271 (2001)

\bibitem{Schat96} M. Schatzman {\it On the  eigenvalues of the
Laplace operator on a thin set with Neumann boundary conditions}
Applicable Anal. {\bf 61}, 293 (1996)

\bibitem{SGB}
I.~A. Shelykh, N.~G. Galkin, and N.~T. Bagraev.
\newblock {\it Quantum splitter controlled by Rashba
spin-orbit coupling}. {\em Phys. Rev.B} {\bf 72},235316 (2005)

\bibitem{SchA_2000} J.H. Schenker, M. Aizenman {\it The creation of
spectral gaps by graph decoration} Letters of Mathematical
Physics,{\bf 53}, 3, (2000) pp 253-262.

\bibitem{Shirokov80} J. Shirokov {\it Strongly singular potentials in
three-dimensional Quantum Mechanics} (In Russian) Teor. Mat. Fiz.
{\bf 42} 1 (1980) pp 45-49.

\bibitem{SGZ}J.~Splettstoesser, M.~Governale, and U.~Z{\"{u}}licke.
\newblock {\it Persistent current in ballistic mesoscopic rings with Rashba
spin-orbit coupling.}
\newblock {\em Phys. Rev. B}, {\bf 68}:165341, (2003).

\bibitem{StrSeba03} P. Streda, P. Seba  {\it Antisymmetric spin
filtering in one-dimensional electron systems via uniform spin-orbit
coupling} Phys. Rev. Letters {\bf 90}, 256601 (2003)

\bibitem{SU2}  J. Sylvester,\thinspace G. Uhlmann  {\it The Dirichlet to
Neumann map and applications.} In: \textit{ Proceedings of the
Conference '' Inverse problems in partial differential equations
(Arcata,1989)''}, SIAM, Philadelphia, 101 (1990)

\bibitem{WF92}A. Wentzel, M. Freidlin {\it Reaction-diffusion equations
with randomly perturbed boundary conditions} Annals of Probability
{\bf 20}, 2 (1992) pp 963-986.

\bibitem{Wigner51}  E.P.Wigner, {\it On a class of analytic functions
from the quantum theory of collisions} Annals  of  mathematics, {\bf
2}, N53, 36 (1951)

\bibitem{ScattXu02} H. Q. Xu  {\it Diode and transistor behaviour of
three-terminal ballistic junctions}  Applied Phys. Letters {\bf 80},
853 (2002)

\end{thebibliography}
\end{document}